\documentclass[reprint,amsmath,amssymb,aps,longbibliography]{revtex4-2}

\usepackage{graphicx}
\usepackage{dcolumn}
\usepackage{bm}
\usepackage{physics}
\usepackage{placeins}
\usepackage{subscript}
\usepackage[super]{nth}
\usepackage{afterpage}
\usepackage{xcolor}  

\usepackage{caption}
\usepackage{subcaption}
\usepackage{float}
\renewcommand{\thesubfigure}{\alph{subfigure}}
\begin{document}

\preprint{APS/123-QED}

\author{Eyal Uzner}
\affiliation{%
  Technion- Israel Institute of Technology, Schulich Faculty of Chemistry and Faculty of Physics, Haifa, 32000036, Israel
}%

\author{Ofer Neufeld}%
 \email{ofern@technion.ac.il}
\affiliation{%
  Technion- Israel Institute of Technology, Schulich Faculty of Chemistry, Haifa, 32000036, Israel
}%


\title{Ultrafast spectroscopy and role of interlayer coupling in high harmonic generation from layered solids
}

\begin{abstract}
High harmonic generation (HHG) in solids has recently emerged as a powerful all-optical approach for probing material properties and ultrafast electron dynamics in quantum systems. It has been widely applied for studying two-dimensional and layered solids of various kinds. In these studies, the laser is usually polarized within the layered planes, where most electron dynamics occurs, while out-of-plane hopping is commonly neglected. This is despite of interlayer hopping being ubiquitous in nano-systems. Here we develop theory for HHG in layered solids in presence of interlayer coupling and employ it for studying strong-field driven hexagonal BN, graphite, and the transition metal dichalcogenide WS$_2$. We show that sufficiently intense couplings can alter typical HHG emission characteristics such as angular or ellipticity dependence even when the driving laser is polarized in-plane. We develop an analytic perturbation theory for the laser-driven current expanded in the interlayer coupling parameter and explicitly show that HHG yields follow a 4'th order polynomial form, which is validated numerically. Our work should motivate experiments for probing interlayer coupling via HHG spectroscopy, as well as exploring its modulation as a control parameter for ultrafast dynamics and attopulse generation via laser driving and mechanical strain. 
\end{abstract}


\maketitle


Strong-field laser driving of solid-state has recently been heavily explored as an avenue for ultrafast spectroscopy and coherent control over quantum matter\cite{Ghimire2014cc,Ghimire2019,GaardeTutorialHHGSolids2021}. In this paradigm, a material system is irradiated by intense off-resonant femtosecond laser pulses, which initiate ultrafast electron dynamics within the band structure. Electronic currents generated by the laser can either be directly measured\cite{Schiffrin2013,Higuchi2017,Langer2020,Hanus2021,Feher2026}, or cause an optical emission of high harmonic generation (HHG) through interband and intraband mechanisms\cite{GaardeTutorialHHGSolids2021}. Both optical and electronic signals were shown to carry unique information about the solid's intrinsic properties, as well as the out-of-equilibrium ultrafast dynamics. For instance, HHG has been employed for probing band structure\cite{PhysRevLett.115.193603,Lanin2017} and Berry curvature\cite{Luu2018b,Lv2021}, Berry phase\cite{Uzan-Narovlansky2024}, topological aspects of bands\cite{Bauer2018,Silva2019a,Chacon2020, Bai2021,Schmid2021,Baykusheva2021a,Heide2022a,OferPRX2023,Chen2025a}, electron-phonon interactions\cite{Bionta2021,Neufeld2022g,DixitHHGGraphenePRA2022,Zhang2024,Zhang2024a}, valley excitations\cite{Jimenez-Galan2020,Mitra2024,Tyulnev2024,Lively2024,Ofer2025SolidsNatCommun}, and more. Similarly, nonlinear photocurrent generation was used to study light-dressed phases\cite{Weitz2024,Galler2023}, coherence\cite{Heide2021}, band structure\cite{Galler2025}, and light-matter symmetries\cite{Neufeld2021a,Sharma2022,Sharma2023,Bharti2024,Lesko2025,Kanega2025}. 

In the realm of two-dimensional and layered solids, strong-field physics is especially promising for light-manipulation of valley degrees of freedom. Tunable ultrafast valley excitations were demonstrated in monolayer systems\cite{Mrudul2021,Mitra2024,Weitz2024} and even in bulk inversion-symmetric layered materials\cite{Tyulnev2024} as long as the tailored laser drive itself breaks inversion symmetry\cite{Neufeld2019,Habibovic2024}. Theoretically, such effects can be described \textit{ab-initio} with time-dependent density functional theory (TDDFT)\cite{Tancogne-Dejean2020b}, or with other approaches\cite{ChangLee2024}. These allow making accurate predictions, but simultaneously make it difficult to uncover the potential role of interlayer interactions that allow electrons to hop between layers. Alternatively, semiconductor Bloch equations (SBE) formulations in density-matrix form are commonly employed\cite{Golde2008a,GaardeTutorialHHGSolids2021}. In these schemes one can also apply phenomenological dephasing that captures some extent of electron-phonon and electron-electron scattering processes, and the current can also be easily decomposed into various contributions for analyzing the physical mechanism. However, typically only few band models are utilized while interlayer hopping is neglected. A couple of works studied HHG from bilayers including interlayer coupling\cite{LeBreton2018,Mrudul2021a,Lee2021,Kim2022sy,Molinero2024}, but physical investigations of bulk 3D layered solids are lacking. Moreover, the general contribution of interlayer interactions to ultrafast laser-driven dynamics remain unclear, and it is also not obvious how the coupling can be extracted from measurements to potentially track dynamical processes\cite{Bionta2021,Alcala2022,Tyulnev2025,Liu2026}.

Here we explore HHG from a set of layered systems including graphite, hexagonal boron nitride (hBN), and the transition metal dichalcogenide (TMD) WS$_2$. We develop an SBE formalism based on an analytic $4\times4$ tight-binding (TB) model that incorporates first-order interlayer hopping. We numerically use this model to calculate the HHG dependence on the coupling parameter in typical strong-field conditions when the driving field is polarized within the layered planes. We find that sufficiently strong interlayer coupling can impact characteristic HHG features such as its ellipticity or angular dependence. Depending on the laser-matter regime, this effect can be pronounced for both low-order and high-order harmonics. To understand this behavior, we formulate an analytic perturbative expansion of the ultrafast current in the coupling parameter. This analysis uncovers that harmonic yields evolve as a general-form 4'th-order polynomial in the interlayer coupling, which we validate in numerical simulations. Our work presents an intuitive picture for understanding the role of interlayer coupling in ultrafast dynamics, which should motivate experimental spectroscopies of out-of-equilibrium layered solids.

\begin{figure}[!h]
    \centering
    
    \newlength{\fullcolwidth}
    \setlength{\fullcolwidth}{0.48\columnwidth}
    \begin{subfigure}{0.33\columnwidth}
        
        \setlength{\unitlength}{\fullcolwidth} 
        \begin{picture}(0.5,0.61) 

            \put(0.05,0.05){\includegraphics[width=0.538\fullcolwidth]{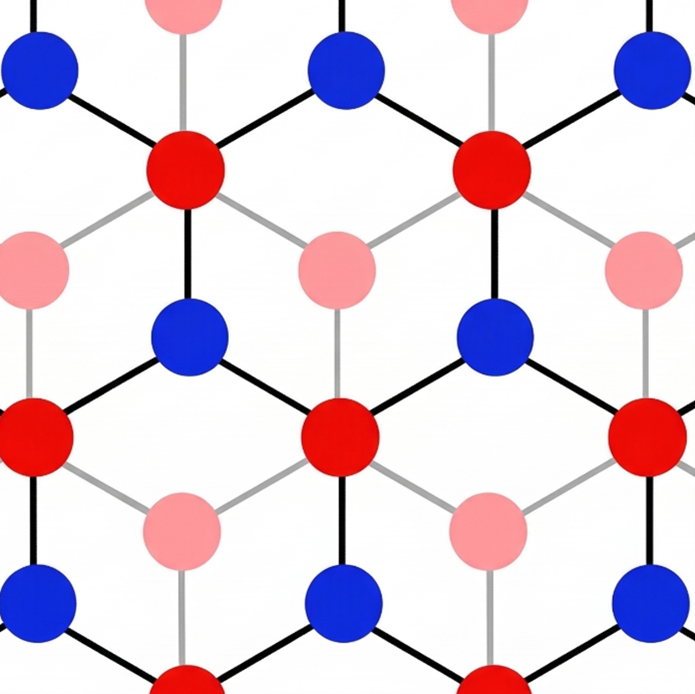}}

            \refstepcounter{subfigure}
            \put(-0.1,0.45){(\alph{subfigure})}

            \put(0.18,0.23){\textcolor{red}{A1}}
            \put(0.19,0.1){\textcolor{blue}{B1}}
            \put(0.33,0.36){\textcolor{red}{A2}}
            \put(0.36,0.225){\textcolor{blue}{B2}}
            
            \put(0,0){\vector(1,0){0.3}} 
            \put(0.32,-0.02){$x$}
    
            \put(0,0){\vector(0,1){0.3}} 
            \put(-0.02,0.33){$y$}
        \end{picture}

        \label{fig:LatticeTopView}
    \end{subfigure}%
    \begin{subfigure}{0.33\columnwidth}
        
        \setlength{\unitlength}{\fullcolwidth} 
        \begin{picture}(0.5,0.6)
            \put(0.05,0.05){\scalebox{-1}[1]{\includegraphics[width=0.538\fullcolwidth, angle=0]{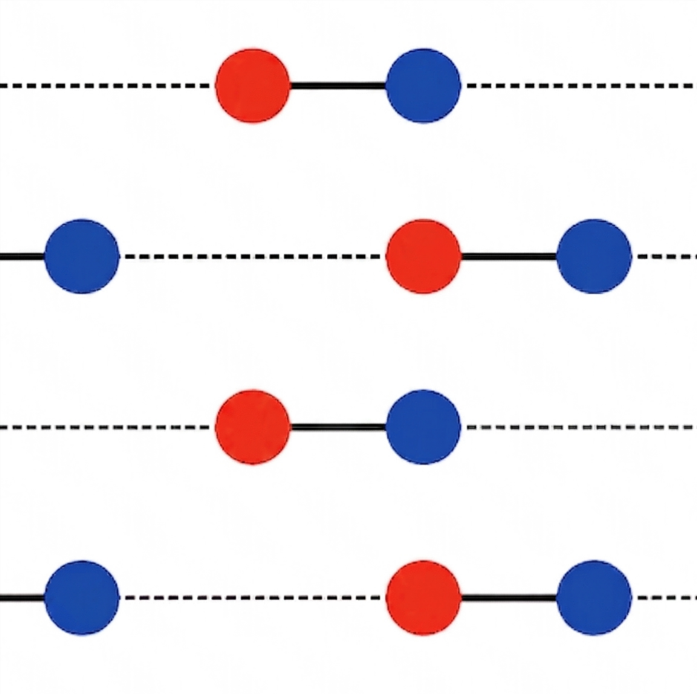}}
            }

            \refstepcounter{subfigure}
            \put(-0.1,0.45){(\alph{subfigure})}

            \put(0.35,0.29){\textcolor{red}{A2}}
            \put(0.21,0.29){\textcolor{blue}{B2}}
            \put(0.21,0.42){\textcolor{red}{A1}}
            \put(0.08,0.42){\textcolor{blue}{B1}}

            \put(0,0){\vector(1,0){0.3}} 
            \put(0.32,-0.02){$y$}
    
            \put(0,0){\vector(0,1){0.3}} 
            \put(-0.02,0.33){$z$}
        \end{picture}

        \label{fig:LatticeSideView}
    \end{subfigure}%
    \begin{subfigure}{0.33\columnwidth} 
        \raggedright
        \setlength{\unitlength}{\columnwidth}
        \begin{picture}(0.3,0.4) 
            \put(0.2,0.12){\includegraphics[width=0.7\columnwidth]{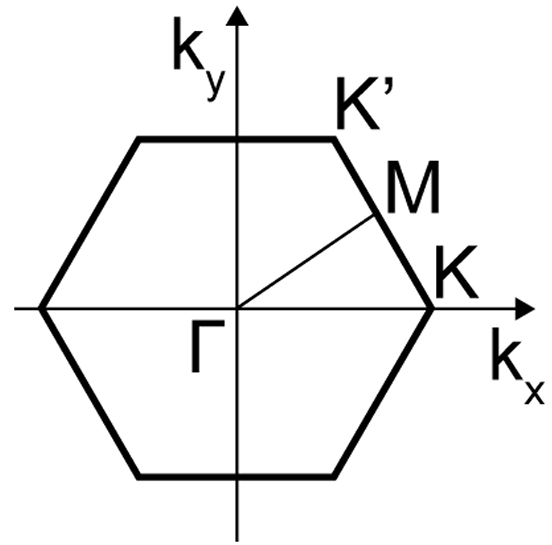}}

            \setlength{\unitlength}{\fullcolwidth} 
            
            \refstepcounter{subfigure}
            \put(0.1,0.45){(\alph{subfigure})}
        \end{picture}

        \label{fig:First Brillouin Zone}
    \end{subfigure}

    \begin{subfigure}{0.5\columnwidth}
        \raggedright
         \setlength{\unitlength}{\columnwidth}
        \begin{picture}(0.7,0.62) 
            \put(0,0){\includegraphics[width=\columnwidth]{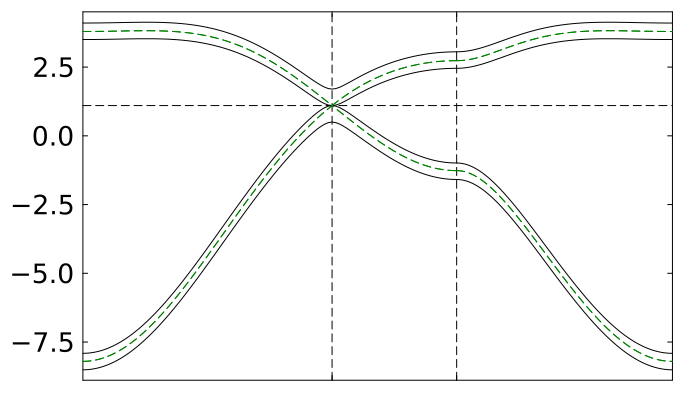}}
    
            \refstepcounter{subfigure}
            \put(0.17,0.3){(\alph{subfigure})}

            \put(0.1,-0.04){$\Gamma$}
            \put(0.45,-0.05){K}
            \put(0.63,-0.05){M}
            \put(0.95,-0.04){$\Gamma$}

            \put(-0.07,0.1){\rotatebox{90}{Energy [eV]}} 

            \put(0.8,0.26){\includegraphics[width=0.12\columnwidth]{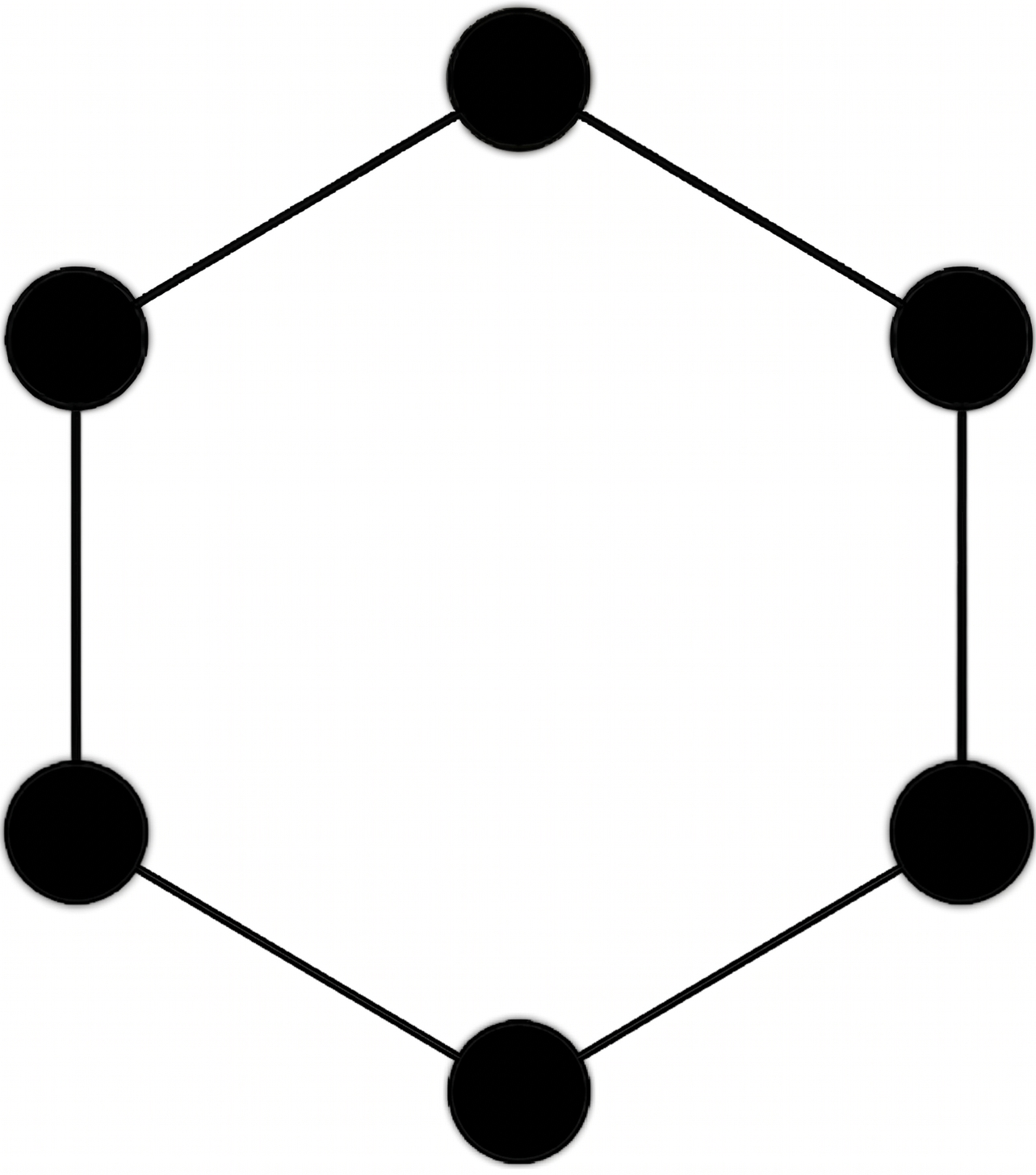}}
        \end{picture}

        \label{fig:Graphite Band Structure}
    \end{subfigure}%
    \begin{subfigure}{0.5\columnwidth}
        \raggedright
        \setlength{\unitlength}{\columnwidth}
        \begin{picture}(0.7,0.57) 
            \put(0,0){\includegraphics[width=\columnwidth]{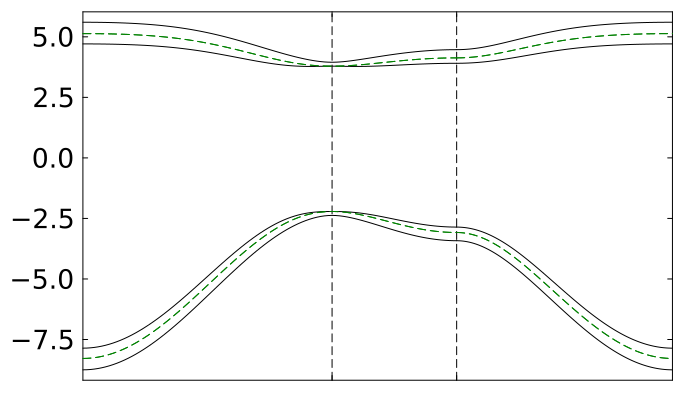}}
    
            \refstepcounter{subfigure}
            \put(0.17,0.3){(\alph{subfigure})}
            
            \put(0.1,-0.04){$\Gamma$}
            \put(0.45,-0.04){K}
            \put(0.63,-0.04){M}
            \put(0.95,-0.04){$\Gamma$}

            \put(0.8,0.26){\includegraphics[width=0.12\columnwidth]{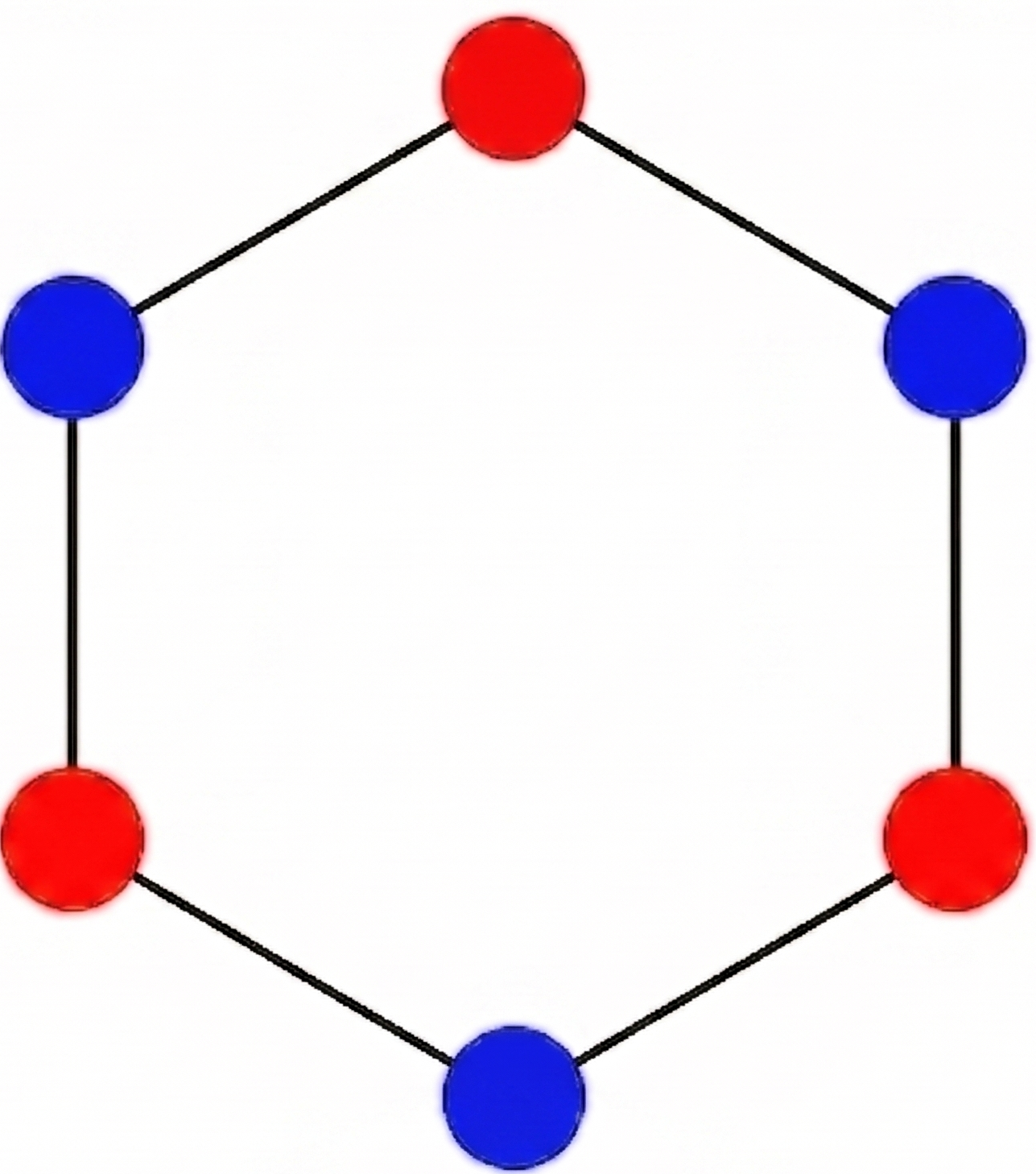}}
            
        \end{picture}

        \label{fig:hBN Band Structure}
    \end{subfigure}

    \begin{subfigure}{\columnwidth}
        \centering
        \setlength{\unitlength}{\columnwidth}
        \begin{picture}(1,0.46) 
            \put(0,0){\includegraphics[width=\columnwidth]{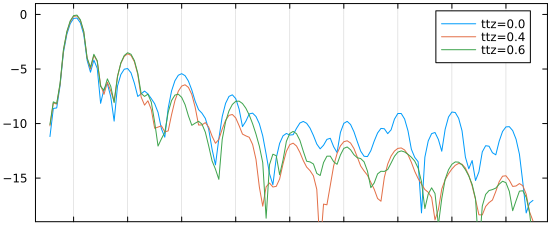}}

            \refstepcounter{subfigure}
            \put(0.495,0.34){\makebox(0,0)
            {\text{(\alph{subfigure})}}} 

            \put(0.54,0.3){\includegraphics[width=0.07\columnwidth]{black_hexagonal.png}}
        \end{picture}

        \label{fig:Graphite Yields}
    \end{subfigure}
    
    \begin{subfigure}{\columnwidth}
        \centering
        \setlength{\unitlength}{\columnwidth}
        \begin{picture}(1,0.44) 
            \put(0,0){\includegraphics[width=\columnwidth]{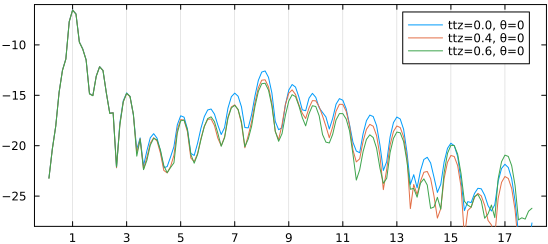}}

            \refstepcounter{subfigure}
            \put(0.5,0.37){\makebox(0,0){\text{(\alph{subfigure})}}} 
            
            \put(0.39,-0.05){ \textbf{$\mathrm{Harmonic}$ $\mathrm{Order}$}} 
            
            \put(-0.05,0.22){\rotatebox{90}{ \textbf{$\log_{10}(\mathrm{Intensitys})$ $\mathrm{[arb.}$ $\mathrm{units]}$}}} 

            \put(0.54,0.33){\includegraphics[width=0.07\columnwidth]{colored_hexagonal.png}}
            
        \end{picture}

        \label{fig:hBN Yields}
    \end{subfigure}

    \vspace{5mm}

    \caption{(a–b) Illustration of lattice geometry showing top view over \textit{xy} planes (a), and side view showing AB stacking configuration (b). (c) Hexagonal BZ with high symmetry points for $k_z=0$ plane. (d,e) TB band structures for graphene/graphite and monolayer/bulk hBN, where bands are indicated by green dashed lines for the monolayers, and black solid lines for bulks. (f–g) Exemplary HHG spectra for (f) graphene/graphite and (g) hBN, calculated for linear driving along $\Gamma$-K for various values of interlayer couplings (legends are in eV).}
    \label{fig:Lattice_Structure_Band_Structure_and_Yields}
\end{figure}

Let us begin by describing our methodological approach and employed formalism. The ground state of all hexagonal systems (see illustration in Figs.~\ref{fig:Lattice_Structure_Band_Structure_and_Yields}\subref{fig:LatticeTopView}-\subref{fig:LatticeSideView} for the stacked honeycomb lattice) is described by a $4\times4$ TB Hamiltonian. Here A and B represent in-equivalent sites that are coupled in-plane by a typical $2\times2$ Hamiltonian considering second-order nearest neighbor (NN) interactions:  
\begin{equation}
    H_{2D} = \begin{pmatrix}
        \alpha + \frac{\Delta}{2} & \beta_{12} \\
        \beta_{12}^* & \alpha - \frac{\Delta}{2}
    \end{pmatrix} .
\end{equation}
where $\Delta = \varepsilon_A - \varepsilon_B$ is the on-site energy difference, with $\varepsilon_{A/B}$ the on-site energies of A and B sites, respectively. The interaction between the NN sites has the form $\beta_{12} = t_1 \sum^{3}_{j=1} e^{i \mathbf{k} \cdot \mathbf{a}_j}$, where $t_1$ is the first-order NN hopping parameter that couples A and B sites in-plane, and $\mathbf{a}_j$ are the NN vectors (with a formalism similar to that employed in refs. \cite{Ofer2025SolidsNatCommun,Chen2025a}). The interaction between second order NN sites in-plane (A-A and B-B sites) has the form $\alpha = t_2 \sum^{6}_{j=1} e^{i \mathbf{k} \cdot \mathbf{b}_j}$, where $t_2$ is the second-order hopping parameter, and $\mathbf{b}_j$ are the second NN vectors. Here we assume only one orbital per lattice site. The resulting $H_{2D}$ is a function of only in-plane ($xy$ plane) Bloch momenta, such that $H_{2D}=H_{2D}(k_x,k_y)$.

The bulk Hamiltonian that includes interlayer hopping is given for AB-stacked layers as
\begin{equation}
    H_{3D} = \begin{pmatrix}
        \scalebox{1.3}{$H_{2D}$} & \begin{matrix}
        0 & \beta_{14} \\
        0 & 0
    \end{matrix}\\
    \begin{matrix}
        0 & 0 \\
        \beta_{14} & 0
    \end{matrix} & \scalebox{1.3}{$H_{2D}$}
    \end{pmatrix} ,
\end{equation}
\noindent , which is constructed from the individual layer 2D Hamiltonians with the added first-order interlayer coupling: $\beta_{14} = t_z \sum^{2}_{j=1} e^{i \mathbf{k} \cdot \mathbf{c}_j}$, where $t_z$ is the interlayer NN hopping parameter, and $\mathbf{c}_j$ are the interplanar NN vectors, which in this case are parallel to the crystal $c$-axis. Thus, $\beta_{14}$ is a function of $k_Z$, meaning $H_{3D}$ acts in the full 3D Brillouin zone (BZ) of the bulk solid. Note that we simulate AB-stacked bulks, which is not the most thermodynamically stable structure in gapped hexagonal systems such as hBN\cite{Gilbert2019,Su2024} (though can be synthesized in all cases, and has already been employed for HHG in graphite\cite{Chen2025a}). However, AB-stacking has a simpler functional form that  allows an easier analysis of the role of interlayer couplings. Moreover, AB-stacked gapped solids lack inversion symmetry, meaning that even harmonics will also be emitted in HHG\cite{LeBreton2018,Neufeld2019,Heide2023}, which continuously connects to the single layer dynamics where inversion symmetry is also broken (i.e. for $t_z\rightarrow0$ one obtains the monolayer limit). 

The band structure for this system is given analytically by $E(\mathbf{k}) = \alpha \pm |\beta_{14}| /2 \pm \sqrt{4|\beta_{12}|^2 + \beta^2_{14}}$. Importantly, since the Hamiltonian is $4\times4$, there exists an analytic (though cumbersome) solution to its eigenstates, from which we derive analytic forms of dipole transition matrix elements, $\textbf{d}_{mn}$ and $\textbf{p}_{nm}$, which are used in subsequent SBE simulations described below. Additional technical details are delegated to the Supplementary Information (SI).  

Using this ground state Hamiltonian, we solve the SBE in the Houston basis and length gauge, given in atomic units by\cite{GaardeTutorialHHGSolids2021}:
\begin{equation}
\begin{aligned}
    i \frac{d}{dt} \rho^{\mathbf{k}}_{mn} =& 
    ( E^{\mathbf{k+A}(t)/c}_{m} - E^{\mathbf{k+A}(t)/c}_{n} ) \rho^{\mathbf{k}}_{mn}
    + i \frac{1 - \delta_{mn}}{T_2} \rho^{\mathbf{k}}_{mn} \\
    &- \mathbf{F}(t) \cdot \sum_{l} [
    \mathbf{d}^{\mathbf{k+A}(t)/c}_{ml} \rho^{\mathbf{k}}_{ln} 
    - \mathbf{d}^{\mathbf{k+A}(t)/c}_{ln} \rho^{\mathbf{k}}_{ml}
    ],
\end{aligned}
\end{equation}
where $\mathbf{F}(t)$ and $\mathbf{A}(t)$ are the electric field and vector potential of the driving laser, respectively, and are related by $c\mathbf{F}(t) = - d \mathbf{A}(t) / dt$ (employing the dipole approximation). $E^{\mathbf{k}}_{m}$ is the band energy of band $m$ at $\textbf{k}$. $\mathbf{d}^{\mathbf{k}(t)}_{ln}$ is the dipole transition matrix element between $m$ and $n$ bands at $k$-point $\textbf{k}(t)=\textbf{k}_0+\textbf{A}(t)/c$ (in the Houston gauge), which is evaluated analytically as 
$\mathbf{d}^{\mathbf{k}}_{ln} = i \left\langle  u_{m,\mathbf{k}} \left|  \nabla_{\mathbf{k}}  \right|  u_{n,\mathbf{k}} \right\rangle$
, where $\left| u_{m,\mathbf{k}} \right\rangle$ is the periodic part of the Bloch state obtained by diagonalizing $H_{3D}$ analytically. A phenomenological term accounting for decoherence is added with a constant dephasing time $T_2$ (where we employ 25 fs for graphite\cite{Heide2021}, 20 fs for hBN\cite{Mitra2024,Lively2024}, and 5 fs for WS2\cite{Korolev2024,Kim2024,Ofer2025SolidsNatCommun}, consistent with recent reports).

We directly solve these equations of motion with a 4'th-order Runge-Kutta scheme with a converged time-step of 0.1 atomic units and a $k$-grid of $600\times 600 \times 9$ points in the BZ. At each time step and $\mathbf{k}$-point we evaluate the electric current:
\begin{equation}
    \mathbf{J}(\mathbf{k}, t) 
    = \sum_{m,n}   \mathbf{p}^{\mathbf{k+A}(t)}_{mn}   \rho^{\mathbf{k}}_{mn} .
\end{equation}
where, $\mathbf{p}^{\mathbf{k}}_{mn}= \left\langle  u_{m,\mathbf{k}} \left| \nabla_{\textbf{k}} H_{3D}({\mathbf{k}})  \right|  u_{n,\mathbf{k}} \right\rangle$ are the momentum matrix elements that are obtained analytically\cite{Pedersen2001}. HHG emission is calculated as the Fourier transform of the time derivative of the BZ summed current:
\begin{equation}
    \mathcal{I} (\omega)=\left |    \mathcal{FT} \left\{  \frac{d}{dt}  \left[     \int_{BZ}  \mathbf{J}(\mathbf{k}, t) \mathrm{d} \mathbf{k}  \right]  \right\}   \right| ^2 .
\end{equation}

\afterpage{
\begin{figure*}[!t]
    \centering

    \begin{tabular}{cc @{\hspace{10pt}} | @{\hspace{10pt}} cc}

        \begin{subfigure}{0.23\textwidth}
            \centering
            \setlength{\unitlength}{1mm} 
            \begin{picture}(\columnwidth,40) 
                \put(0,0){\includegraphics[width=\columnwidth]{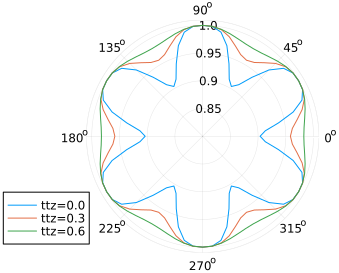}}

                \refstepcounter{subfigure}
                \put(0,30){(\alph{subfigure}) H3}

                 \put(0,18){\includegraphics[width=0.15\columnwidth]{black_hexagonal.png}}
                
            \end{picture}

            \label{Fig:Graphite H3}
        \end{subfigure} &

        \setcounter{subfigure}{3}
        \begin{subfigure}{0.23\textwidth}
            \centering
            \setlength{\unitlength}{1mm} 
            \begin{picture}(\columnwidth,40) 
                \put(0,0){\includegraphics[width=\columnwidth]{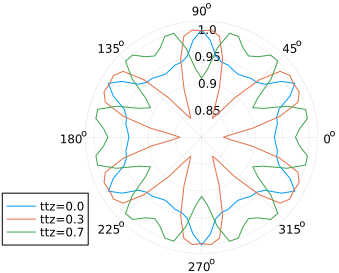}}

                \refstepcounter{subfigure}
                \put(0,30){(\alph{subfigure}) H11}
                
            \end{picture}

            \label{Fig:Graphite H11}
        \end{subfigure} &

        \setcounter{subfigure}{6}
        \begin{subfigure}{0.23\textwidth}
            \centering
            \setlength{\unitlength}{1mm} 
            \begin{picture}(\columnwidth,40) 
                \put(0,0){\includegraphics[width=\columnwidth]{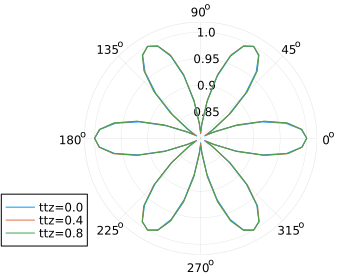}}

                \refstepcounter{subfigure}
                \put(0,30){(\alph{subfigure}) H6}

                \put(0,18){\includegraphics[width=0.15\columnwidth]{colored_hexagonal.png}}
                
            \end{picture}

            \label{Fig:hBN H6}
        \end{subfigure} &

        \setcounter{subfigure}{9}
        \begin{subfigure}{0.23\textwidth}
            \centering
            \setlength{\unitlength}{1mm} 
            \begin{picture}(\columnwidth,40) 
                \put(0,0){\includegraphics[width=\columnwidth]{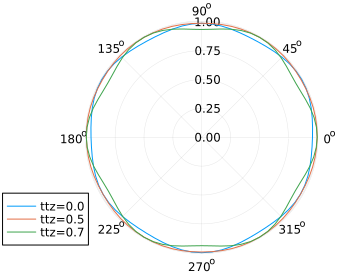}}

                \refstepcounter{subfigure}
                \put(0,30){(\alph{subfigure}) H5}
                
            \end{picture}

            \label{Fig:hBN H5}
        \end{subfigure} \\

        \setcounter{subfigure}{1}
        \begin{subfigure}{0.23\textwidth}
            \centering
            \setlength{\unitlength}{1mm} 
            \begin{picture}(\columnwidth,40) 
                \put(0,0){\includegraphics[width=\columnwidth]{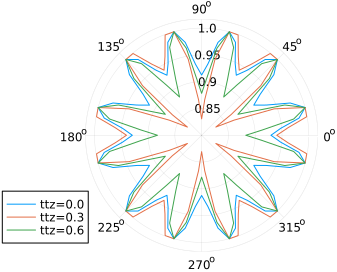}}

                \refstepcounter{subfigure}
                \put(0,30){(\alph{subfigure}) H5}
                
            \end{picture}

            \label{Fig:Graphite H5}
        \end{subfigure} &

        \setcounter{subfigure}{4}
        \begin{subfigure}{0.23\textwidth}
            \centering
            \setlength{\unitlength}{1mm} 
            \begin{picture}(\columnwidth,40) 
                \put(0,0){\includegraphics[width=\columnwidth]{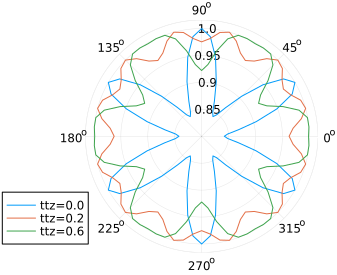}}

                \refstepcounter{subfigure}
                \put(0,30){(\alph{subfigure}) H13}
                
            \end{picture}

            \label{Fig:Graphite H13}
        \end{subfigure} &

        \setcounter{subfigure}{7}
        \begin{subfigure}{0.23\textwidth}
            \centering
            \setlength{\unitlength}{1mm} 
            \begin{picture}(\columnwidth,40) 
                \put(0,0){\includegraphics[width=\columnwidth]{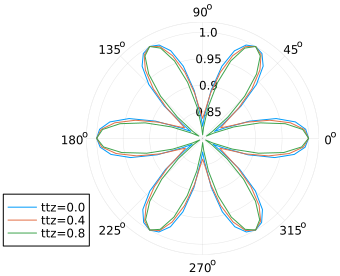}}

                \refstepcounter{subfigure}
                \put(0,30){(\alph{subfigure}) H10}
                
            \end{picture}

            \label{Fig:hBN H10}
        \end{subfigure}&

        \setcounter{subfigure}{10}
        \begin{subfigure}{0.23\textwidth}
            \centering
            \setlength{\unitlength}{1mm} 
            \begin{picture}(\columnwidth,40) 
                \put(0,0){\includegraphics[width=\columnwidth]{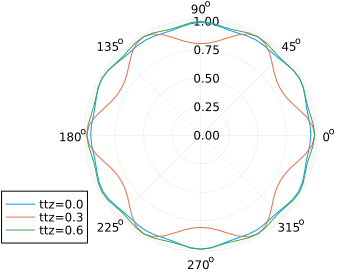}}

                \refstepcounter{subfigure}
                \put(0,30){(\alph{subfigure}) H9}
                
            \end{picture}

            \label{Fig:hBN H9}
        \end{subfigure} \\

        \setcounter{subfigure}{2}
        \begin{subfigure}{0.23\textwidth}
            \centering
            \setlength{\unitlength}{1mm} 
            \begin{picture}(\columnwidth,40) 
                \put(0,0){\includegraphics[width=\columnwidth]{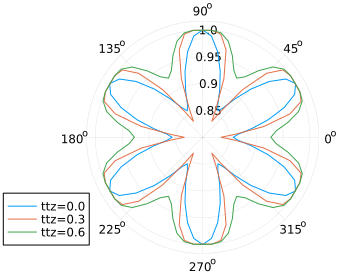}}

                \refstepcounter{subfigure}
                \put(0,30){(\alph{subfigure}) H7}
                
            \end{picture}

            \label{Fig:Graphite H7}
        \end{subfigure} &

        \setcounter{subfigure}{5}
        \begin{subfigure}{0.23\textwidth}
            \centering
            \setlength{\unitlength}{1mm} 
            \begin{picture}(\columnwidth,40) 
                \put(0,0){\includegraphics[width=\columnwidth]{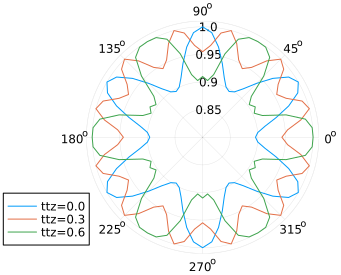}}

                \refstepcounter{subfigure}
                \put(0,30){(\alph{subfigure}) H15}
                
            \end{picture}

            \label{Fig:Graphite H15}
        \end{subfigure} &
        
        \setcounter{subfigure}{8}
        \begin{subfigure}{0.23\textwidth}
            \centering
            \setlength{\unitlength}{1mm} 
            \begin{picture}(\columnwidth,40) 
                \put(0,0){\includegraphics[width=\columnwidth]{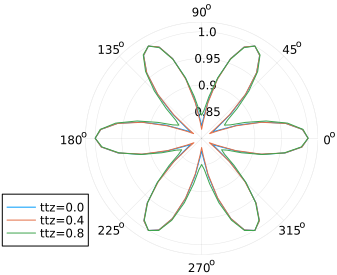}}

                \refstepcounter{subfigure}
                \put(0,30){(\alph{subfigure}) H12}
                
            \end{picture}

            \label{Fig:hBN H12}
        \end{subfigure}&

        \setcounter{subfigure}{11}
        \begin{subfigure}{0.23\textwidth}
            \centering
            \setlength{\unitlength}{1mm} 
            \begin{picture}(\columnwidth,40) 
                \put(0,0){\includegraphics[width=\columnwidth]{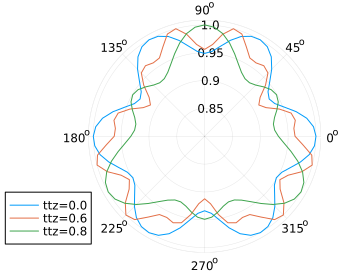}}

                \refstepcounter{subfigure}
                \put(0,30){(\alph{subfigure}) H14}
                
            \end{picture}

            \label{Fig:hBN H14}
        \end{subfigure}

    \end{tabular}
    
    \caption{Orientation dependence of normalized harmonic yields from layered solids for various interlayer coupling strengths. (a–f) Normalized HHG yields from graphite. (g–l) Normalized HHG yields from hBN. Select harmonics are presented that cover the full range of physical behaviors observed (all other harmonics orders are delegated to the SI).}
    \label{fig:Orientation Dependency Results}

\end{figure*}
}

\afterpage{
\begin{figure*}[!t]
    \centering

    \begin{tabular}{cc @{\hspace{5pt}} | @{\hspace{4pt}} cc}

        \begin{subfigure}{0.23\textwidth}
            \centering
            \setlength{\unitlength}{1mm} 
            \begin{picture}(\columnwidth,28) 
                \put(0,0){\includegraphics[width=\columnwidth]{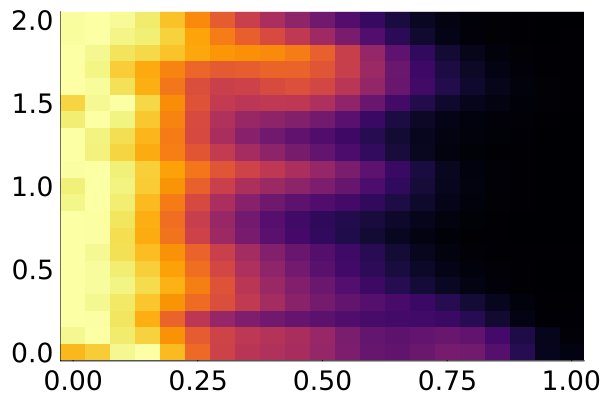}}

                \refstepcounter{subfigure}
                \put(29,23){\textcolor{white}{(\alph{subfigure}) H3}}

                \put(-4,10){\rotatebox{90}{\small $t_z$ [eV]}} 

                \put(32,12){\includegraphics[width=0.15\columnwidth]{black_hexagonal.png}}
                
            \end{picture}

            \label{Fig:Graphite H3}
        \end{subfigure} &
        
        \setcounter{subfigure}{3}
        \begin{subfigure}{0.23\textwidth}
            \centering
            \setlength{\unitlength}{1mm} 
            \begin{picture}(\columnwidth,28) 
                \put(0,0){\includegraphics[width=\columnwidth]{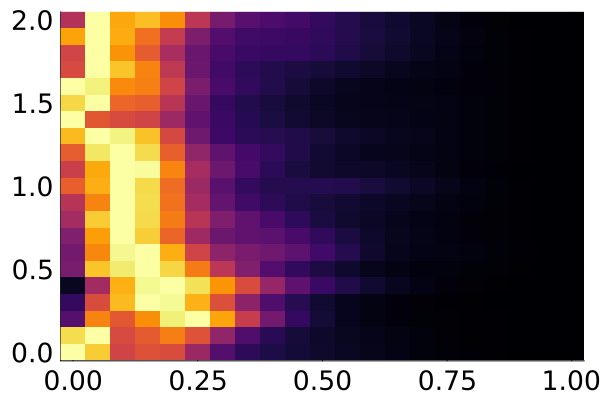}}

                \refstepcounter{subfigure}
                \put(29,23){\textcolor{white}{(\alph{subfigure}) H9}}

            \end{picture}

            \label{Fig:Graphite H9}
        \end{subfigure} &

        \setcounter{subfigure}{6}
            \begin{subfigure}{0.23\textwidth}
                \centering
                \setlength{\unitlength}{1mm} 
                \begin{picture}(\columnwidth,28) 
                    \put(0,0){\includegraphics[width=\columnwidth]{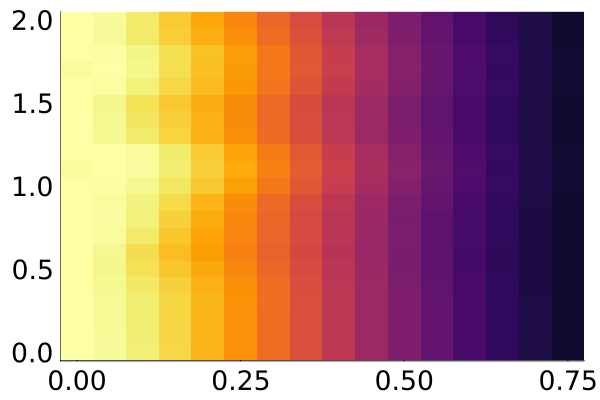}}

                    \refstepcounter{subfigure}
                    \put(29,23){\textcolor{white}{(\alph{subfigure}) H3}}

                    \put(32,12){\includegraphics[width=0.15\columnwidth]{colored_hexagonal.png}}
                    
                \end{picture}

                \label{Fig:hBN H3}
            \end{subfigure} &
            
            \setcounter{subfigure}{9}
            \begin{subfigure}{0.23\textwidth}
                \centering
                \setlength{\unitlength}{1mm} 
                \begin{picture}(\columnwidth,28) 
                    \put(0,0){\includegraphics[width=\columnwidth]{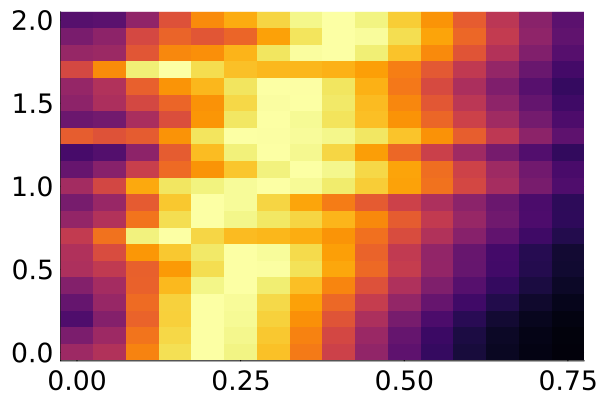}}

                    \refstepcounter{subfigure}
                    \put(29,23){\textcolor{white}{(\alph{subfigure}) H6}}

                    \put(42,1.9){\includegraphics[width=0.1425\columnwidth]{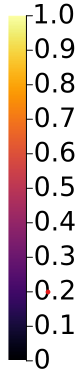}} 
                    
                \end{picture}

                \label{Fig:hBN H6}
            \end{subfigure} \\

        \setcounter{subfigure}{1}
        \begin{subfigure}{0.23\textwidth}
            \centering
            \setlength{\unitlength}{1mm} 
            \begin{picture}(\columnwidth,27) 
                \put(0,0){\includegraphics[width=\columnwidth]{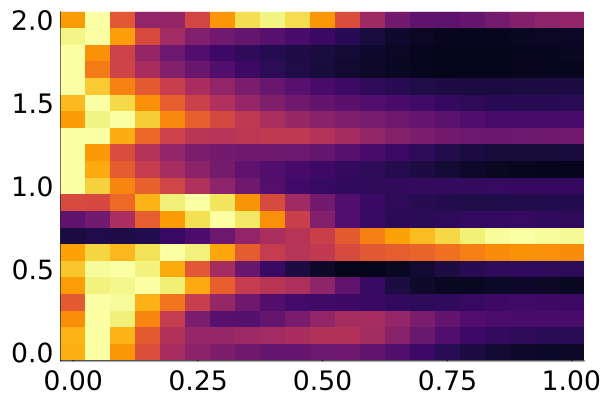}}

                \refstepcounter{subfigure}
                \put(29,23){\textcolor{white}{(\alph{subfigure}) H5}}

                \put(-4,10){\rotatebox{90}{\small $t_z$ [eV]}} 
                
            \end{picture}

            \label{Fig:Graphite H5}
        \end{subfigure} &
        
        \setcounter{subfigure}{4}
        \begin{subfigure}{0.23\textwidth}
            \centering
            \setlength{\unitlength}{1mm} 
            \begin{picture}(\columnwidth,27) 
                \put(0,0){\includegraphics[width=\columnwidth]{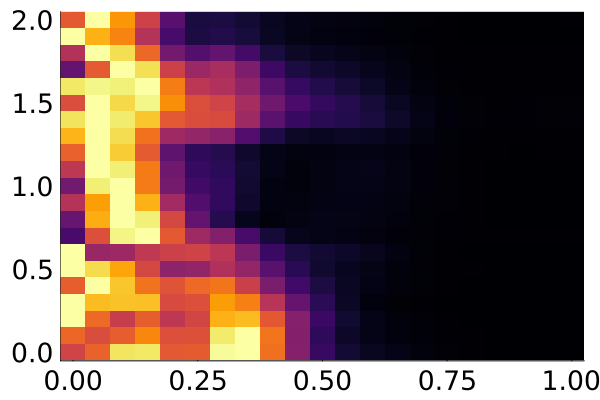}}

                \refstepcounter{subfigure}
                \put(29,23){\textcolor{white}{(\alph{subfigure}) H11}}
                
            \end{picture}

            \label{Fig:Graphite H11}
        \end{subfigure} &

        \setcounter{subfigure}{7}
            \begin{subfigure}{0.23\textwidth}
                \centering
                \setlength{\unitlength}{1mm} 
                \begin{picture}(\columnwidth,27) 
                    \put(0,0){\includegraphics[width=\columnwidth]{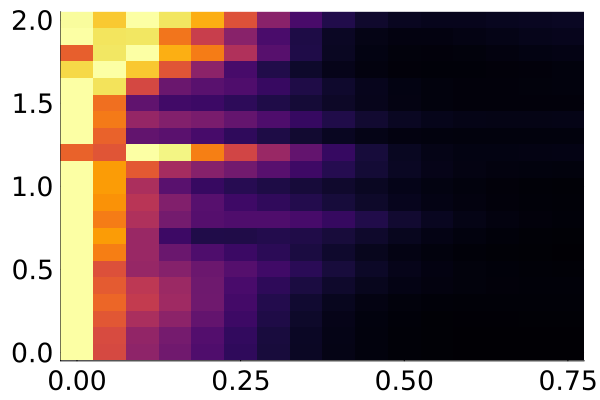}}

                    \refstepcounter{subfigure}
                    \put(29,23){\textcolor{white}{(\alph{subfigure}) H7}}
                    
                \end{picture}

                \label{Fig:hBN H7}
            \end{subfigure} &
            
            \setcounter{subfigure}{10}
            \begin{subfigure}{0.23\textwidth}
                \centering
                \setlength{\unitlength}{1mm} 
                \begin{picture}(\columnwidth,27) 
                    \put(0,0){\includegraphics[width=\columnwidth]{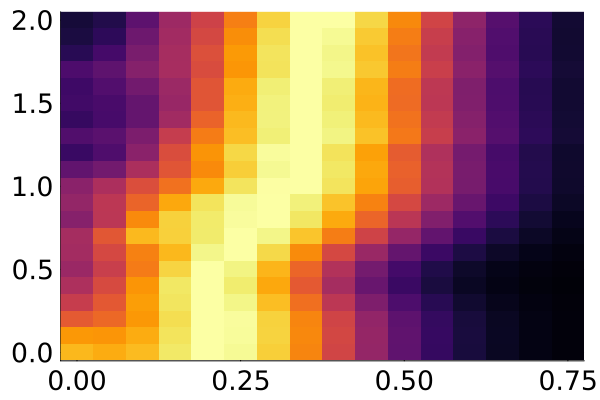}}
                    \put(29,23){\textcolor{white}{(\alph{subfigure}) H8}}

                    \put(42,1.9){\includegraphics[width=0.1425\columnwidth]{Colorbar.png}} 
                    
                \end{picture}

                \label{Fig:hBN H8}
            \end{subfigure} \\

        \setcounter{subfigure}{2}
        \begin{subfigure}{0.23\textwidth}
            \centering
            \setlength{\unitlength}{1mm} 
            \begin{picture}(\columnwidth,27) 
                \put(0,0){\includegraphics[width=\columnwidth]{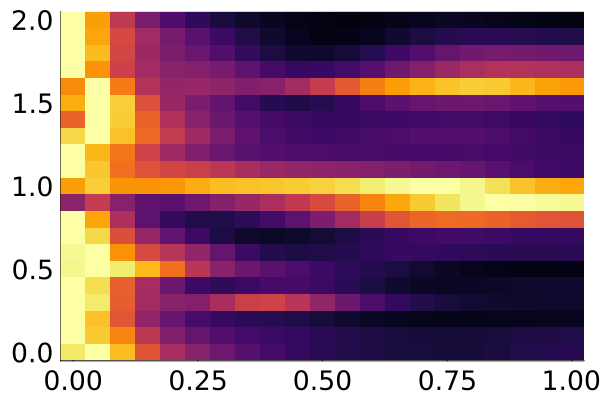}}

                \refstepcounter{subfigure}
                \put(29,23){\textcolor{white}{(\alph{subfigure}) H7}}

                \put(13,-3){{\normalsize $\varepsilon$} - Ellipticity} 

                \put(-4,10){\rotatebox{90}{\small $t_z$ [eV]}} 
                
            \end{picture}

            \label{Fig:Graphite H7}
        \end{subfigure} &
        
        \setcounter{subfigure}{5}
        \begin{subfigure}{0.23\textwidth}
            \centering
            \setlength{\unitlength}{1mm} 
            \begin{picture}(\columnwidth,27) 
                \put(0,0){\includegraphics[width=\columnwidth]{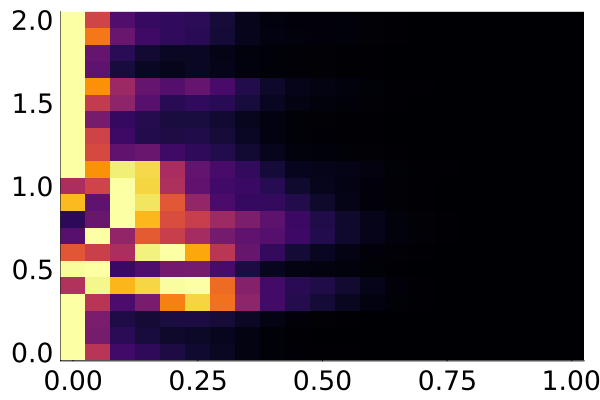}}

                \refstepcounter{subfigure}
                \put(29,23){\textcolor{white}{(\alph{subfigure}) H15}}

                \put(13,-3){{\normalsize $\varepsilon$} - Ellipticity} 
                
            \end{picture}

            \label{Fig:Graphite H15}
        \end{subfigure} &

        \setcounter{subfigure}{8}
        \begin{subfigure}{0.23\textwidth}
            \centering
            \setlength{\unitlength}{1mm} 
            \begin{picture}(\columnwidth,27) 
                \put(0,0){\includegraphics[width=\columnwidth]{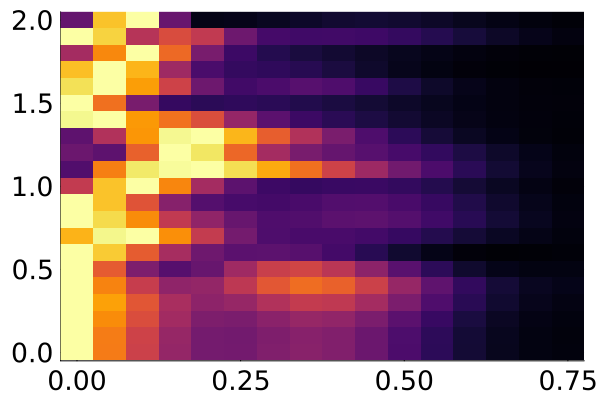}}

                \refstepcounter{subfigure}
                \put(29,23){\textcolor{white}{(\alph{subfigure}) H11}}

                \put(13,-3){{\normalsize $\varepsilon$} - Ellipticity} 
            \end{picture}

            \label{Fig:hBN H11}
        \end{subfigure} &
        
        \setcounter{subfigure}{11}
        \begin{subfigure}{0.23\textwidth}
            \centering
            \setlength{\unitlength}{1mm} 
            \begin{picture}(\columnwidth,27) 
                \put(0,0){\includegraphics[width=\columnwidth]{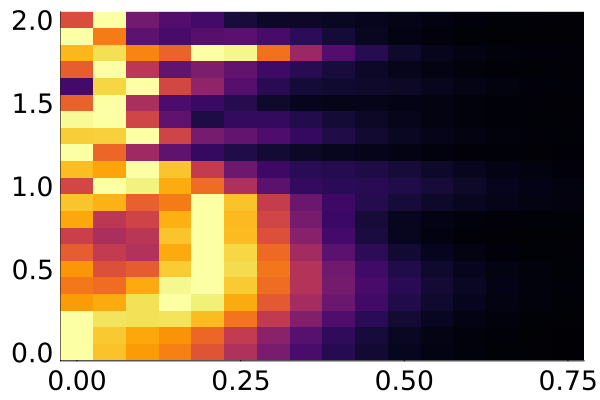}}

                \put(42,1.9){\includegraphics[width=0.1425\columnwidth]{Colorbar.png}} 

                \refstepcounter{subfigure}
                \put(29,23){\textcolor{white}{(\alph{subfigure}) H14}}

                \put(13,-3){{\normalsize $\varepsilon$} - Ellipticity} 
                
            \end{picture}

            \label{Fig:hBN H14}
        \end{subfigure}

    \end{tabular}
    
    \vspace{3mm}
    
    \caption{Ellipticity dependence of normalized harmonic yields from layered solids for varying interlayer coupling strength. (a–f) Select harmonics from graphite. (g–l) Select harmonics from hBN. Other harmonic orders, as well as inter/intra-band separation, is delegated to the SI.}
    \label{fig:Ellipticity Dependency Results}
\end{figure*}
}

\afterpage{
\begin{figure*}[!t]
\centering

\begin{subfigure}{0.33\textwidth}
    \centering
    \setlength{\unitlength}{\linewidth} 
    \begin{picture}(1,0.45)
        \put(0,0){\includegraphics[width=\unitlength]{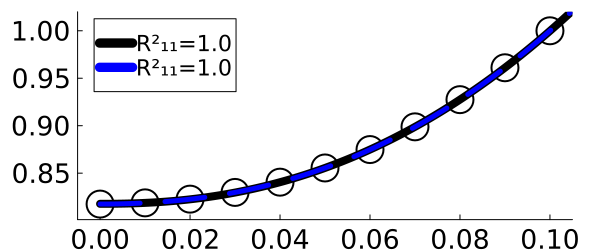}}

        \refstepcounter{subfigure}
        \put(0.42,0.25){
            \shortstack[l]{
                (\thesubfigure)\ H3 \\
                \hspace{1.7em}$\theta = 0^{\circ}$
            }
        }

        \put(0.82,0.1){\includegraphics[width=0.12\columnwidth]{black_hexagonal.png}}
    \end{picture}

    \label{Fig:Graphite A}
\end{subfigure}%
\begin{subfigure}{0.33\textwidth}
    \centering
    \setlength{\unitlength}{\linewidth}
    \begin{picture}(1,0.45)
        \put(0,0){\includegraphics[width=\unitlength]{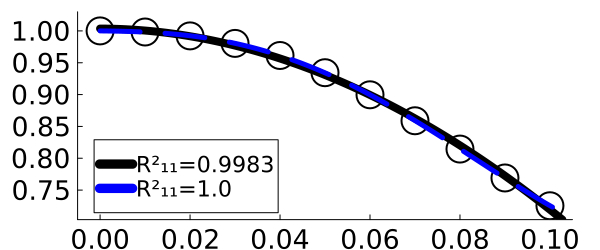}}

        \refstepcounter{subfigure}
        \put(0.65,0.23){
            \shortstack[l]{
                (\thesubfigure)\ H7 \\
                \hspace{1.7em}$\theta = 0^{\circ}$
            }
        }
    \end{picture}

    \label{Fig:Graphite B}
\end{subfigure}%
\begin{subfigure}{0.33\textwidth}
    \centering
    \setlength{\unitlength}{\linewidth}
    \begin{picture}(1,0.45)
        \put(0,0){\includegraphics[width=\unitlength]{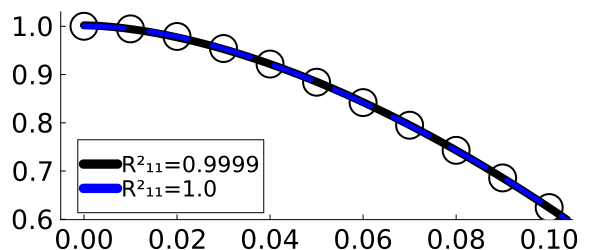}}

        \refstepcounter{subfigure}
        \put(0.65,0.23){
            \shortstack[l]{
                (\thesubfigure)\ H13 \\
                \hspace{1.7em}$\theta = 25^{\circ}$
            }
        }
    \end{picture}

    \label{Fig:Graphite C}
\end{subfigure}

\begin{subfigure}{0.33\textwidth}
    \centering
    \setlength{\unitlength}{\linewidth}
    \begin{picture}(1,0.45)
        \put(0,0){\includegraphics[width=\unitlength]{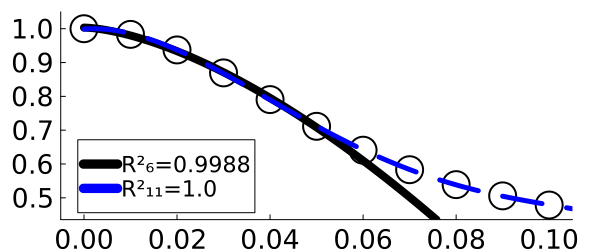}}

        \refstepcounter{subfigure}
        \put(0.65,0.23){
            \shortstack[l]{
                (\thesubfigure)\ H9 \\
                \hspace{1.7em}$\theta = 40^{\circ}$
            }
        }

         \put(-0.07,0.08){\rotatebox{90}{\footnotesize \shortstack{Normalized Intensity {[arb. units]} }}} 
    \end{picture}

    \label{Fig:Graphite D}
\end{subfigure}%
\begin{subfigure}{0.33\textwidth}
    \centering
    \setlength{\unitlength}{\linewidth}
    \begin{picture}(1,0.45)
        \put(0,0){\includegraphics[width=\unitlength]{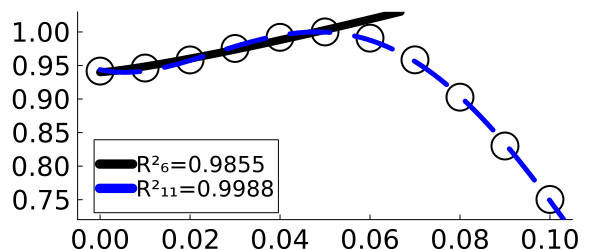}}

        \refstepcounter{subfigure}
        \put(0.7,0.27){
            \shortstack[l]{
                (\thesubfigure)\ H11 \\
                \hspace{1.7em}$\theta = 0^{\circ}$
            }
        }
    \end{picture}

    \label{Fig:Graphite E}
\end{subfigure}%
\begin{subfigure}{0.33\textwidth}
    \centering
    \setlength{\unitlength}{\linewidth}
    \begin{picture}(1,0.45)
        \put(0,0){\includegraphics[width=\unitlength]{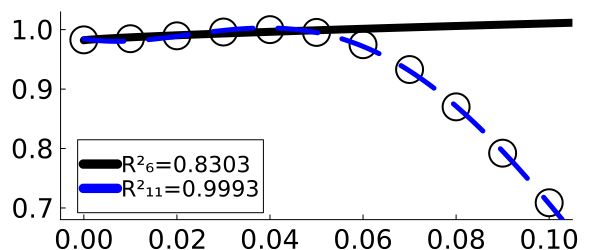}}

        \refstepcounter{subfigure}
        \put(0.7,0.24){
            \shortstack[l]{
                (\thesubfigure)\ H11 \\
                \hspace{1.7em}$\theta = 5^{\circ}$
            }
        }

    \end{picture}

    \label{Fig:Graphite F}
\end{subfigure}

\begin{subfigure}{0.33\textwidth}
    \centering
    \setlength{\unitlength}{\linewidth}
    \begin{picture}(1,0.45)
        \put(0,0){\includegraphics[width=\unitlength]{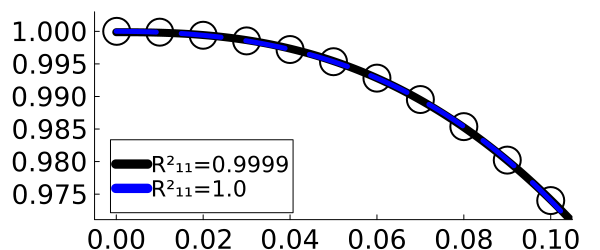}}

        \refstepcounter{subfigure}
        \put(0.7,0.23){
            \shortstack[l]{
                (\thesubfigure)\ H4 \\
                \hspace{1.7em}$\theta = 0^{\circ}$
            }
        }

        \put(0.52,0.08){\includegraphics[width=0.12\columnwidth]{colored_hexagonal.png}}
    \end{picture}

    \label{Fig:hBN G}
\end{subfigure}%
\begin{subfigure}{0.33\textwidth}
    \centering
    \setlength{\unitlength}{\linewidth}
    \begin{picture}(1,0.45)
        \put(0,0){\includegraphics[width=\unitlength]{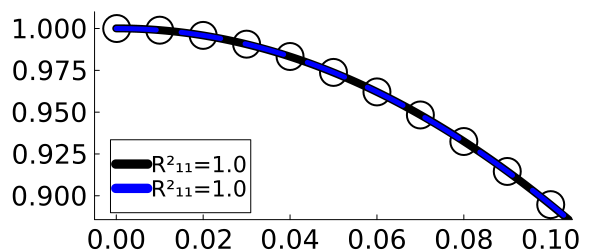}}

        \refstepcounter{subfigure}
        \put(0.65,0.23){
            \shortstack[l]{
                (\thesubfigure)\ H7 \\
                \hspace{1.7em}$\theta = 0^{\circ}$
            }
        }
    \end{picture}

    \label{Fig:hBN H}
\end{subfigure}%
\begin{subfigure}{0.33\textwidth}
    \centering
    \setlength{\unitlength}{\linewidth}
    \begin{picture}(1,0.45)
        \put(0,0){\includegraphics[width=\unitlength]{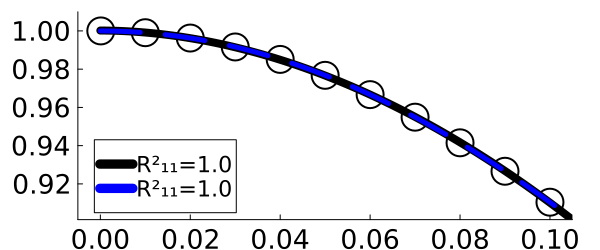}}

        \refstepcounter{subfigure}
        \put(0.65,0.23){
            \shortstack[l]{
                (\thesubfigure)\ H9 \\
                \hspace{1.7em}$\theta = 0^{\circ}$
            }
        }
    \end{picture}

    \label{Fig:hBN I}
\end{subfigure}

\begin{subfigure}{0.33\textwidth}
    \centering
    \setlength{\unitlength}{\linewidth}
    \begin{picture}(1,0.45)
        \put(0,0){\includegraphics[width=\unitlength]{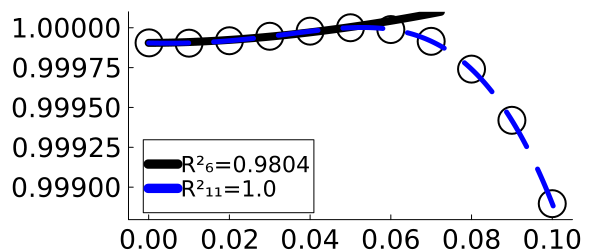}}

        \refstepcounter{subfigure}
        \put(0.55,0.13){
            \shortstack[l]{
                (\thesubfigure)\ H10 \\
                \hspace{1.7em}$\theta = 80^{\circ}$
            }
        }

        \put(0.47,-0.05){\small $t_z$ [eV]} 
        \put(-0.07,0.08){\rotatebox{90}{\footnotesize \shortstack{Normalized Intensity {[arb. units]} }}} 
    \end{picture}

    \label{Fig:hBN J}
\end{subfigure}%
\begin{subfigure}{0.33\textwidth}
    \centering
    \setlength{\unitlength}{\linewidth}
    \begin{picture}(1,0.45)
        \put(0,0){\includegraphics[width=\unitlength]{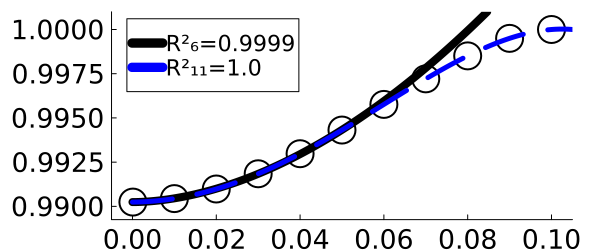}}

        \refstepcounter{subfigure}
        \put(0.7,0.13){
            \shortstack[l]{
                (\thesubfigure)\ H11 \\
                \hspace{1.7em}$\theta = 20^{\circ}$
            }
        }
        
        \put(0.47,-0.05){\small $t_z$ [eV]} 
    \end{picture}

    \label{Fig:hBN K}
\end{subfigure}%
\begin{subfigure}{0.33\textwidth}
    \centering
    \setlength{\unitlength}{\linewidth}
    \begin{picture}(1,0.45)
        \put(0,0){\includegraphics[width=\unitlength]{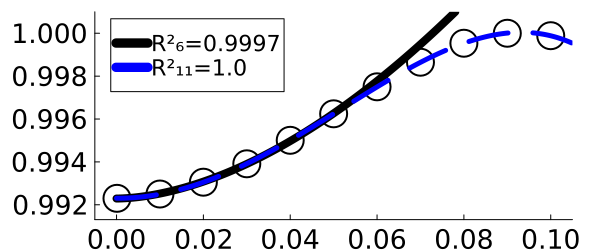}}

        \refstepcounter{subfigure}
        \put(0.7,0.13){
            \shortstack[l]{
                (\thesubfigure)\ H11 \\
                \hspace{1.7em}$\theta = 80^{\circ}$
            }
        }
        
        \put(0.47,-0.05){\small $t_z$ [eV]} 
    \end{picture}

    \label{Fig:hBN L}
\end{subfigure}

\vspace{3mm}

\caption{Harmonic yield dependence on the interlayer hopping parameter up to 100 meV in the perturbative regime. Exact numerical results from SBE simulations are given by the empty circle black markers. Black solid lines and blue dash lines are: generic power law fitted function, 4'th-order polynomial fitted function, respectively. Note that the $R^2$ values are given in the legend for each fit, and the sub-text in $R^2$ indicates the number of points over which the fit was made (generally the power-law dependence was fitted over a shorter domain since it cannot capture curvature changes that appear towards larger $t_z$ (see discussion in text)). (a–f) HHG in graphite. (g–l) HHG in hBN. }
\label{fig:Pertubition Theory Fits}
\end{figure*}
}

We drive the layered solids with a laser pulse of intensity $I_0=5\times10^{11} \mathrm{W / cm^2}$ and wavelength $\lambda=1600$nm, with the following vector potential:
\begin{equation}
\begin{aligned}
    \mathbf{A}(t) = &\frac{A_0 f(t)}{\sqrt{1+\epsilon^2}}  \{  [ \cos(\theta) \cos(\omega t) - \epsilon\sin(\theta) \sin(\omega t) ] \hat{\mathbf{e}}_x \\
    & + [ \sin(\theta) \cos(\omega t) + \epsilon\cos(\theta) \sin(\omega t)] \hat{\mathbf{e}}_y \}
\end{aligned}
\end{equation}
\noindent This expression describes an elliptically polarized laser pulse with ellipticity $\epsilon$, and a major elliptical axis at an angle $\theta$ above the \textit{x}-axis in the \textit{xy} layered planes. $f(t)$ represents the temporal envelope function taken as a super-sin form\cite{PhysRevLett.123.103202} with a total duration of 8 driving cycles ($\sim$21.35 full width half max).


Using this formalism, we now explore HHG in layered solids, starting with graphite. Figure~\ref{fig:Lattice_Structure_Band_Structure_and_Yields}\subref{fig:Graphite Yields} shows the band structure of graphite obtained with our 4-band Hamiltonian when setting the hopping amplitudes to optimally match density functional theory (DFT) calculations of graphite performed with octopus code\cite{Tancogne-Dejean2020b} at the experimental geometry. The interlayer hopping parameter causes a small gap opening around K/K' points that shifts the upper and bottom bands, while the valence and conduction bands remain fixed at the Dirac cone. $t_z\neq0$ also induces a weak dispersion along the $k_z$ axis. In standard conditions, the interlayer hopping in graphite is quite weak and on the order of $\sim$20 meV. However, it can in principle be manipulated by compression of the graphite layers, much like was recently explored in Diamond\cite{Su2026}, or possibly modulated dynamically by laser driving.

In what follows, we explore HHG from graphite while artificially tuning the interlayer hopping across various parameter regimes to investigate its influence on ultrafast dynamics in the bulk system. Note that we fix the laser polarization to the \textit{xy} plane within the monolayers, such that any impact of interlayer hopping is indirect (as opposed to being directly driven by a \textit{z}-polarized laser). This corresponds to typical experimental conditions. We note that under direct laser driving along the \textit{c}-axis, we expect similar ultrafast dynamics to take play as is usual in HHG, but with the interlayer hopping playing the main role, which is hence beyond our scope.


We start our analysis in HHG from graphite driven by a linearly-polarized laser ($\epsilon=0$). Figure~\ref{fig:Lattice_Structure_Band_Structure_and_Yields}\subref{fig:Graphite Yields} shows exemplary HHG spectra from graphite for $t_z=0$ (graphene) and $t_z>0$ on a scale of few hundred meV. Our simulations indicate changes to the HHG yields arise, especially towards higher harmonic orders and larger $t_z$ values. Angular HHG dependence is shown in Fig.~\ref{fig:Orientation Dependency Results}(\subref{Fig:Graphite H3}-\subref{Fig:Graphite H15}) for harmonic-integrated yields of select harmonic orders (full data sets are allocated to the SI). One main feature we observe is that the angle $\theta_{max}$ for which HHG yields maximize can shift for $t_z>0$. For perturbative harmonics, e.g. H3, H5 and H7 (Figs.~\ref{fig:Orientation Dependency Results}(\subref{Fig:Graphite H3}-\subref{Fig:Graphite H7}), this effect is non-existent and requires very high $t_z$ values to get yield modulation that inevitability do not shift $\theta_{max}$. However, in higher-order non-perturbative harmonics $\theta_{max}$ substantially shifts for large enough $t_z$ values. For instance, in H13 (H13, Fig.~\ref{fig:Orientation Dependency Results}\subref{Fig:Graphite H13}), $\theta_{max}$ shifts from $30^{\circ}$ to $45^{\circ}$ for $t_z\sim200$ meV. This still requires $t_z$ values larger than those arising naturally in graphite, but should be attainable in other materials with stronger Van der Waals bonding, or can otherwise be engineered. Overall, shifting of $\theta_{max}$ with $t_z$ is a generic effect in our simulations that can be utilized to probe interlayer coupling by comparing different material systems, or modulating the coupling by external stimuli.

Next, we extend our analysis to other material systems. Figures~\ref{fig:Orientation Dependency Results}(g-l) presents similar results in hBN, which is a wide-gap semiconductor that has been widely explored in HHG both at monolayer and bulk geometries. We generally observe more minor differences in HHG yield vs $t_z$ compared to those in graphite. This is attributed to the large gap in hBN (taken to be $\sim$6eV, much larger than the energy scales of $t_z$). Still, generic features such as shifting of $\theta_{max}$ are observed for select harmonics such as H9 (Fig.~\ref{fig:Orientation Dependency Results}\subref{Fig:hBN H9}) at $t_Z\sim$300 meV, and we further expect that they will become even more apparent at longer wavelength or more intense driving, where the HHG plateau is extended. Similar features are seen in the TMD WS$_2$ (these are delegated to the SI, Figs. S2). 

Moving forward, we allow the driving laser to be elliptically-polarized with $\epsilon\neq0$, while fixing the major elliptical axis to the \textit{y}-axis ($\theta=90^\circ$). In the past, the characteristic HHG response vs driving laser ellipticity was a major driver for understanding ultrafast dynamics in graphene\cite{Yoshikawa2017}, as well as in topological insulators\cite{Baykusheva2021a,Heide2022a,OferPRX2023}. Generically, anomalous behavior whereby HHG yields maximize at values $\epsilon_{max}\neq0$ (unlike in the gas phase\cite{MollerEllipticity2012PRA}) can be used as indicators of the material band structure characteristics. Figure~\ref{fig:Ellipticity Dependency Results} presents HHG yields per harmonic order for various driving ellipticity and interlayer coupling values in both graphite and hBN (with WS$_2$ delegated to the SI). Our simulations reveal two main effects of interest. First, as $t_z$ increases, the typical decay width of the HHG yield vs. $\epsilon$ can change, either increasing or decreasing depending on the harmonic order and conditions. For instance, in H11 in graphite (Fig.~\ref{fig:Ellipticity Dependency Results}\subref{Fig:Graphite H11}) the yield decreases substantially around $t_z\sim$0.5 eV, but increases again at $t_z\sim$1.2 eV. In H15 from graphite (Fig.~\ref{fig:Ellipticity Dependency Results}\subref{Fig:Graphite H15}) the effect is opposite. Second, $\epsilon_{max}$ for each harmonic order is continuously tuned with $t_z$, which can be a telling sign for probing interlayer coupling. A notable example is H14 in hBN (Fig.~\ref{fig:Ellipticity Dependency Results}\subref{Fig:hBN H14}) that shifts $\epsilon_{max}$ from 0 (typical gas-like behavior) up to $\sim$0.25 at $t_z\sim$0.5 eV. In the SI we also analyze the same dependence in inter/intra-band terms of HHG, and show that the features are general to both emission mechanisms. Similar features are robustly observed in all material platforms. This suggests that the ellipticity-dependence of HHG is highly sensitive to interlayer coupling. 

Having established robust signatures for interlayer couplings in HHG, we wish to analyze their physical origin and mechanism. Generically, HHG is a highly nonlinear and non-perturbative process. As such, it is usually extremely sensitive to minor changes in the underlying Hamiltonian and it is often difficult to analytically explore. In this case however, despite the overall process being non-perturbative in the laser drive, we can treat the interlayer coupling as a small perturbation to the laser-driven monolayer time-dependent Hamiltonian. That is, we assume that $\beta_{14}$ acts as off-diagonal perturbations for $H_{2D}$, which is block-diagonal. For rather small values of interlayer coupling (up to few hundred meV), this approximation should be valid, especially considering that at typical HHG conditions a single photon energy upholds $\omega>>t_Z$. In that respect, $\beta_{14}$ is negligible compared to the light-matter interaction term, as well as to the other intralayer hopping terms. 

We employ a perturbative approach and utilize the solutions of $H_{2D}$ as a basis for ultrafast laser-driven currents driven with the $t_z$ elements in $H_{3D}$. In practice, we analytically evaluate the form of the momentum matrix elements with the eigenstates expanded with perturbation theory to first order in $t_z$, utilizing the eigenstates of the block-diagonal $H_{2D}$. Since the current expectation value, $\textbf{J}(t)$, comprises a sum over $\textbf{p}_{nm}$ evaluated in different \textit{k}-points, if $\textbf{p}_{nm}$ ends up having a $k$-space universal analytical dependence in $t_z$, that should translate also to $\textbf{J}(t)$. We note of course that this approach is approximate since we directly neglect higher-order terms in $\textbf{p}_{nm}$, and since additional indirect dependence in $t_z$ enters also through the occupation of excited states ($\rho_{nm}(t)$). The full analytic derivation is delegated to the SI. Its end result uncovers that, to first order, $\textbf{p}_{nm}$ has a similar functional dependence in $t_z$ across the BZ, which leads to the generic form of the current: $\textbf{J}(t)\approx \textbf{a}_0(t) + \textbf{a}_1(t)t_z + \textbf{a}_2(t)t_z^2$ (with $\textbf{a}_n(t)$ being response functions that depend on the material and the laser-matter regime). The current can be directly connected to HHG yields that arise as the Fourier components of $\textbf{J}(t)$ (absolute value squared). Hence, individual harmonic yields are expected to behave as:
\begin{equation}
I^{(n)}\approx h^{(n)}_0 + h_1^{(n)}t_z + h_2^{(n)}(t)t_z^2 + h_3^{(n)}t_z^3 + h_4^{(n)}t_z^4
\end{equation}
\noindent , with $I^{(n)}$ the yield of the $n$'th harmonic and $h_j^{(n)}$ being $t_Z$-independent coefficients for the $n$'th harmonic. We can test this analytic result by comparing to our numeric simulations, limiting our analysis to the regime $t_Z<100$ meV where we expect the above expression to be valid. Figure~\ref{fig:Pertubition Theory Fits} presents numerically calculated integrated HHG yields in graphene and hBN in various select scenarios (different harmonics, orientations, $t_z$ values, etc.). Notably, these results follow the same numerical approach as utilized throughout the main text, directly solving $H_{3D}$ with full interlayer interactions. Therefore, comparison to the analytically derived form acts as a test to the theorem. Overlaid in each numeric case (indicated by empty circle markers) are two fitted lines: (i) A best-fit for typically expected power law $I^{(n)}\approx b_0^{(n)}+b_1^{(n)}(t_Z)^{k^{(n)}}$, where $b_0^{(n)}$, $b_1^{(n)}$, and $k^{(n)}$ are fitting parameters (black curves). (ii) A general 4'th order polynomial form with $h^{(n)}_j$ as fitting parameters (blue curves). In each case the quality of the fit is evaluated by an $R^2$ parameter over the domain of the fit.

The main result that emerges from Fig.~\ref{fig:Pertubition Theory Fits} is that dominantly the 4'th-order polynomial is a better fit (with higher $R^2$ values) for the functional behavior of HHG vs $t_Z$ than a simple naive power-law form. In most cases the simple power law is also a reasonably good approximation for the HHG physical behavior. However, it has a much shorter range of validity (e.g. see H9 in Fig.~\ref{fig:Pertubition Theory Fits}(d) with the black curve diverging for values $t_z>60 meV$). Moreover, in that case almost always the $k^{(n)}$ parameter that describes the power-law is non-integer, nor close to integer values. Often this suggests a highly non-perturbative behavior as is the case in HHG, but in this case we recall that the fitting is vs $t_Z$, not laser power, and hence the generic fit can be misleading for uncovering the role of $t_Z$ in the dynamics. We would also highlight that in some cases the HHG yield exhibits several curvature changes as in H10 in hBN (Fig.~\ref{fig:Pertubition Theory Fits}(j)), and H11 in graphite (Fig.~\ref{fig:Pertubition Theory Fits}(e)). These are quite robust features that dominantly arise i nour simulations, especially in higher harmonics. They cannot be captured by a simple power law, but are readily reproduced by the 4'th order polynomial derived analytically. 

Here we emphasize a main difference arising in our analysis between gapped and gapless solids --- In the gapless case our derivation reveals that the $\textbf{a}_1(t)$ coefficient that is proportional to $t_z$ in $J(t)$ vanishes for purely intraband emission, such that only even powers in $t_z$ appear in HHG yields (see SI). Numerical results in the SI support this and further validate the analytical model.

Overall, a physical picture emerges whereby interlayer coupling acts as an indirect perturbation to intralayer electron dynamics driven by the laser. The induced current `picks up' perturbative terms that naturally arise up to second order in $t_z$, indicating no more than two interlayer hopping events (yet higher order terms can become noticeable for larger $t_Z$). The resulting HHG yields follow a 4'th order polynomial, which in turn causes the unique physical behavior analyzed above vs driving angle and ellipticity. We expect this analysis to broadly apply to various laser-matter regimes.

To summarize, we numerically and analytically explored high harmonic generation from layered solids, including graphite, hBN, and WS$_2$. Our analysis uncovered the unique role of interlayer coupling in laser-driven dynamics dominantly within the monolayer planes (common experimental conditions). We showed that interlayer coupling can quantitatively modulate HHG yields, and also lead to qualitative shifts in spectral features such as: (i) Shifting HHG angular dependence, (ii) broadening (or narrowing) the HHG ellipticity dependent curve, (iii) shifting HHG yield maximizing ellipticity values. Using an analytic perturbation theory analysis, we showed that the current modulates up to a second order polynomial in interlayer coupling (indicating up to two hopping processes contribute), with HHG yields following a 4'th order polynomial. Slightly more intricate results were derived in the gapless case where some terms vanish in the intraband current, but overall the general behavior is similar across layered systems. Our numerical simulations validated the analytical theory. 

Our results can be used to develop spectroscopic schemes for probing interlayer coupling, as well as interlayer ultrafast electron dynamics. We expect that such predictions could also be tested by externally tuning the interlayer coupling through pressure or laser driving of coherent phonon dynamics and Floquet modes. Our results should also motivate caution in analysis of solid HHG spectroscopies that rely on angle- and ellipticity-dependent yields, illustrating that these values can vary depending on the quality level of the model. Lastly, we believe that dynamical tuning of the interlayer coupling in time could be employed as a unique novel tool to control HHG and subsequent attosecond pulse generation through gating schemes.

\noindent\textbf{Acknowledgments.} O.N. gratefully acknowledges the scientific support of Prof. Dr. Angel Rubio and the Young Faculty Award from the National Quantum Science and Technology program of Israel’s Council of Higher Education Planning and Budgeting Committee.

\bibliography{references}

\end{document}


\preprint{APS/123-QED}

\author{Eyal Uzner}
\affiliation{%
  Technion- Israel Institute of Technology, Schulich Faculty of Chemistry and Faculty of Physics, Haifa, 32000036, Israel
}%

\author{Ofer Neufeld}%
 \email{ofern@technion.ac.il}
\affiliation{%
  Technion- Israel Institute of Technology, Schulich Faculty of Chemistry, Haifa, 32000036, Israel
}%


\title{Supplementary information: Ultrafast spectroscopy and role of interlayer coupling in high harmonic generation from layered solids
}

\maketitle

\onecolumngrid

\noindent This supplementary information file contains additional technical details about simulations employed in the main text, as well as additional complementary results that support our conclusions and analysis. 

\section{Additional details of SBE simulations}

\noindent In the numerical simulation, we sample the first Brillouin zone using a grid with 600 $\times$ 600 $\times$ 9 points along the directions of the three basis vectors in the reciprocal space. We us the 4'th-order Runge-Kutta method to solve the time evolution of the density matrix. The step size of the time grid is 0.1 a.u. (17652 total steps). The lattice parameter, $a$, interlayer separation, $c$, hopping parameters, $t_1, t_2$, and band gap of graphite, hBN and WS$_2$ are summarized in Table~\ref{tab:Simulations Parameters}.

\begin{table}[H]
    \centering
    \resizebox{0.5\columnwidth}{!}{%
    \begin{tabular}{cccccc}
    \toprule
    Material & $a$ [\AA] & $c$ [\AA] & $t_1$ [eV] & $t_2$ [eV] & $\Delta$ [eV] \\
    \midrule
    Graphite & 2.46  & 3.85   & 2.0     & -0.367   & 0.0 \\
    hBN      & 2.505 & 3.33   & 1.99945 & -0.26244 & 6.0 \\
    WS$_2$   & 3.153 & 6.1615 & 1.48144 & 0.15882  & 1.3 \\
    \bottomrule
    \end{tabular}%
    }

    \caption{summary of material parameters employed in the Hamiltonians in our simulations.}
    \label{tab:Simulations Parameters}
\end{table}

\section{Additional data}
\begin{figure*}[!h]
    \centering

    \begin{tabular}{c c c c}

        \multicolumn{4}{c}{\textbf{Graphite}} \\[2mm]

            \begin{subfigure}{0.23\textwidth}
                \centering
                \setlength{\unitlength}{1mm} 
                \begin{picture}(\columnwidth,28) 
                    \put(0,0){\includegraphics[width=\columnwidth]{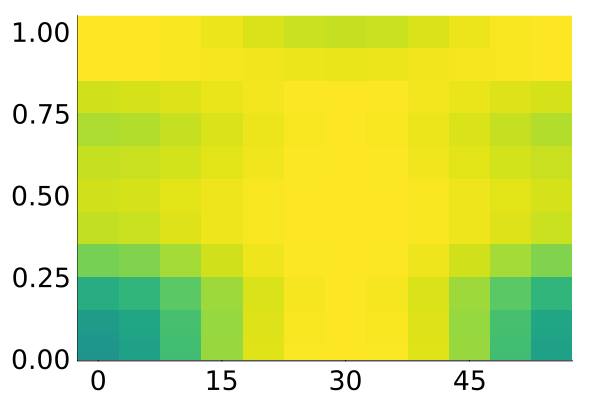}}

                    \refstepcounter{subfigure}
                    \put(29,23){\textcolor{white}{(\alph{subfigure}) H3}}

                    \put(-4,10){\rotatebox{90}{\small $t_z$ [eV]}} 
                    
                \end{picture}

                \label{fig:Graphite H3}
            \end{subfigure} &

            \begin{subfigure}{0.23\textwidth}
                \centering
                \setlength{\unitlength}{1mm} 
                \begin{picture}(\columnwidth,28) 
                    \put(0,0){\includegraphics[width=\columnwidth]{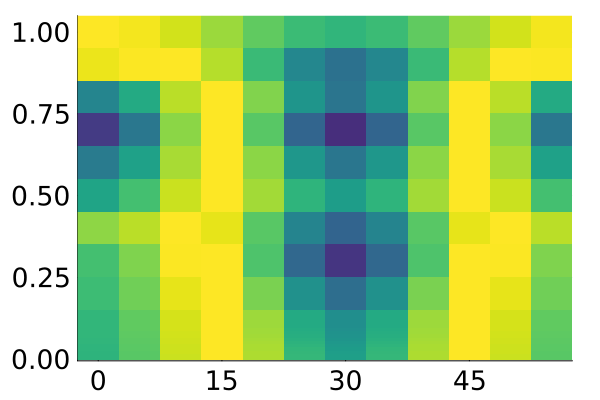}}

                    \refstepcounter{subfigure}
                    \put(29,23){\textcolor{white}{(\alph{subfigure}) H5}}
                    
                \end{picture}

                \label{fig:Graphite H5}
            \end{subfigure} &

            \begin{subfigure}{0.23\textwidth}
                \centering
                \setlength{\unitlength}{1mm} 
                \begin{picture}(\columnwidth,28) 
                    \put(0,0){\includegraphics[width=\columnwidth]{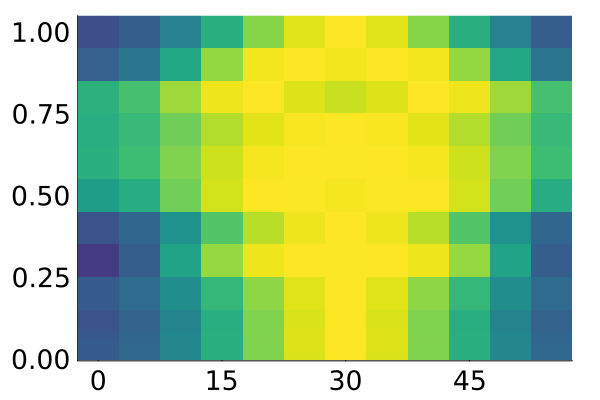}}

                    \refstepcounter{subfigure}
                    \put(29,23){\textcolor{white}{(\alph{subfigure}) H7}}
                    
                \end{picture}

                \label{fig:Graphite H7}
            \end{subfigure} &

            \begin{subfigure}{0.23\textwidth}
                \centering
                \setlength{\unitlength}{1mm} 
                \begin{picture}(\columnwidth,28) 
                    \put(0,0){\includegraphics[width=\columnwidth]{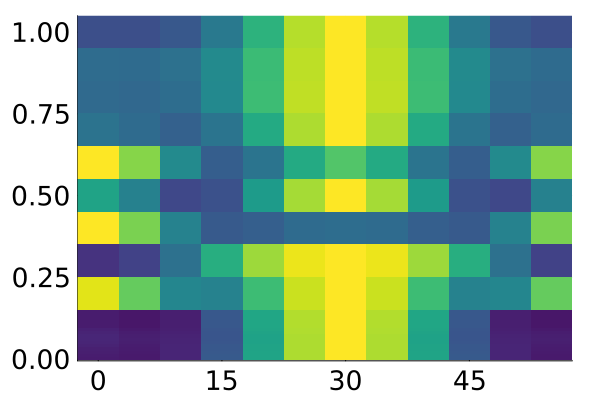}}

                    \refstepcounter{subfigure}
                    \put(29,23){\textcolor{white}{(\alph{subfigure}) H9}}

                    \put(42,1.7){\includegraphics[width=0.127\columnwidth]{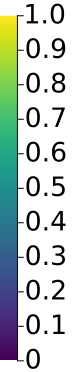}} 
                    
                \end{picture}

                \label{fig:Graphite H9}
            \end{subfigure} \\
    
            \begin{subfigure}{0.23\textwidth}
                \centering
                \setlength{\unitlength}{1mm} 
                \begin{picture}(\columnwidth,27) 
                    \put(0,0){\includegraphics[width=\columnwidth]{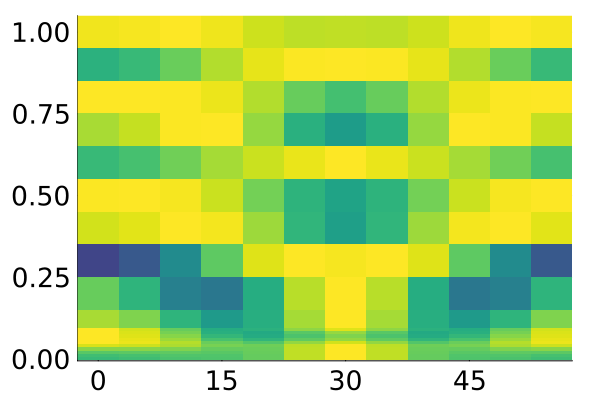}}

                    \refstepcounter{subfigure}
                    \put(27.2,23){\textcolor{white}{(\alph{subfigure}) H11}}

                    \put(-4,10){\rotatebox{90}{\small $t_z$ [eV]}} 

                    \put(19,-3){{\normalsize $\theta$} [\textdegree ]} 
                    
                \end{picture}

                \label{fig:Graphite H11}
            \end{subfigure} &
            
            \begin{subfigure}{0.23\textwidth}
                \centering
                \setlength{\unitlength}{1mm} 
                \begin{picture}(\columnwidth,27) 
                    \put(0,0){\includegraphics[width=\columnwidth]{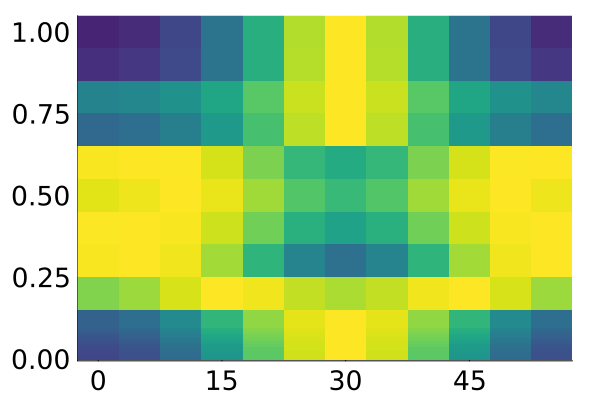}}

                    \refstepcounter{subfigure}
                    \put(27.2,23){\textcolor{white}{(\alph{subfigure}) H13}}

                    \put(19,-3){{\normalsize $\theta$} [\textdegree ]} 
                    
                \end{picture}

                \label{fig:Graphite H13}
            \end{subfigure}  &
            
            \begin{subfigure}{0.23\textwidth}
                \centering
                \setlength{\unitlength}{1mm} 
                \begin{picture}(\columnwidth,27) 
                    \put(0,0){\includegraphics[width=\columnwidth]{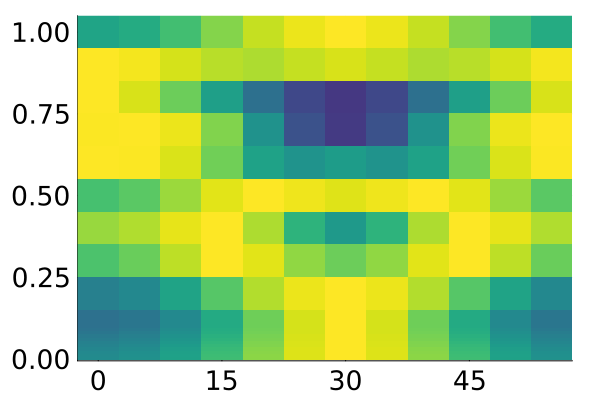}}

                    \refstepcounter{subfigure}
                    \put(27.2,23){\textcolor{white}{(\alph{subfigure}) H15}}

                    \put(19,-3){{\normalsize $\theta$} [\textdegree ]} 
                    
                \end{picture}

                \label{fig:Graphite H15}
            \end{subfigure} &
            
            \begin{subfigure}{0.23\textwidth}
                \centering
                \setlength{\unitlength}{1mm} 
                \begin{picture}(\columnwidth,27) 
                    \put(0,0){\includegraphics[width=\columnwidth]{Graphite_H15_Heatmap_Orientation_ttz.png}}

                    \refstepcounter{subfigure}
                    \put(27.2,23){\textcolor{white}{(\alph{subfigure}) H17}}

                    \put(19,-3){{\normalsize $\theta$} [\textdegree ]} 

                    \put(42,1.7){\includegraphics[width=0.127\columnwidth]{Colorbar_blue.png}} 
                    
                \end{picture}

                \label{fig:Graphite H17}
            \end{subfigure} \\ [7mm]

        \multicolumn{4}{c}{\textbf{hBN}} \\[2mm]

            \begin{subfigure}{0.23\textwidth}
                \centering
                \setlength{\unitlength}{1mm} 
                \begin{picture}(\columnwidth,28) 
                    \put(0,0){\includegraphics[width=\columnwidth]{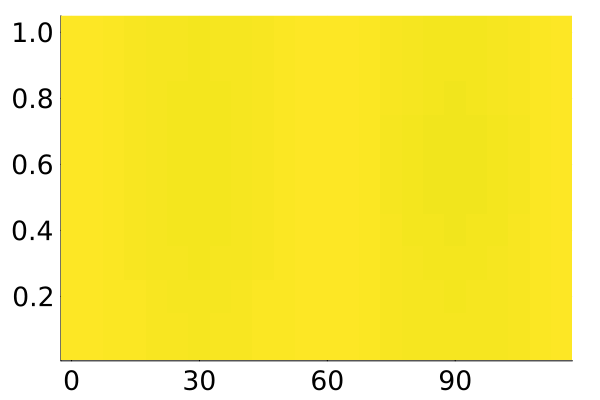}}

                    \refstepcounter{subfigure}
                    \put(29,23){\textcolor{white}{(\alph{subfigure}) H3}}

                    \put(-4,10){\rotatebox{90}{\small $t_z$ [eV]}} 
                    
                \end{picture}

                \label{fig:hBN H3}
            \end{subfigure} &
            
            \begin{subfigure}{0.23\textwidth}
                \centering
                \setlength{\unitlength}{1mm} 
                \begin{picture}(\columnwidth,28) 
                    \put(0,0){\includegraphics[width=\columnwidth]{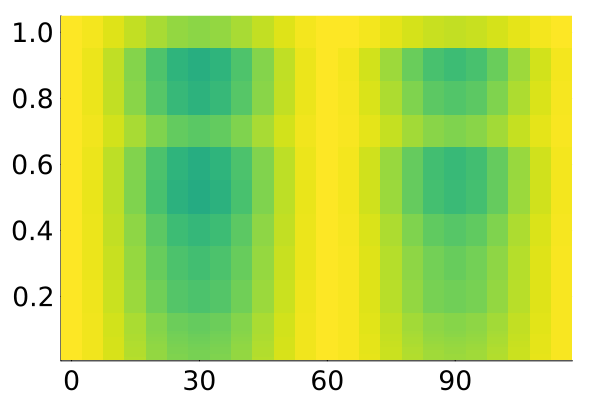}}

                    \refstepcounter{subfigure}
                    \put(29,23){\textcolor{white}{(\alph{subfigure}) H4}}
                    
                \end{picture}

                \label{fig:hBN H4}
            \end{subfigure} &

            \begin{subfigure}{0.23\textwidth}
                \centering
                \setlength{\unitlength}{1mm} 
                \begin{picture}(\columnwidth,28) 
                    \put(0,0){\includegraphics[width=\columnwidth]{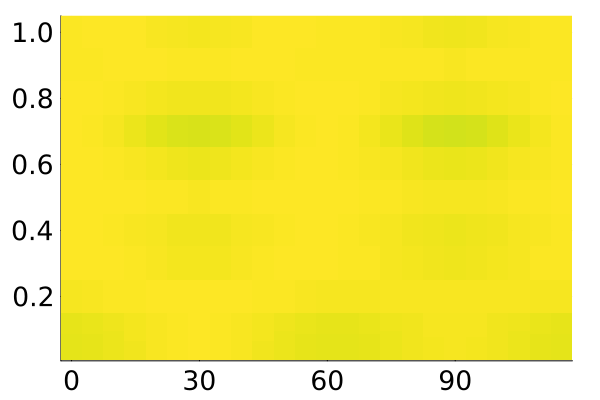}}

                    \refstepcounter{subfigure}
                    \put(29,23){\textcolor{white}{(\alph{subfigure}) H5}}
                    
                \end{picture}

                \label{fig:hBN H5}
            \end{subfigure} &
            
            \begin{subfigure}{0.23\textwidth}
                \centering
                \setlength{\unitlength}{1mm} 
                \begin{picture}(\columnwidth,28) 
                    \put(0,0){\includegraphics[width=\columnwidth]{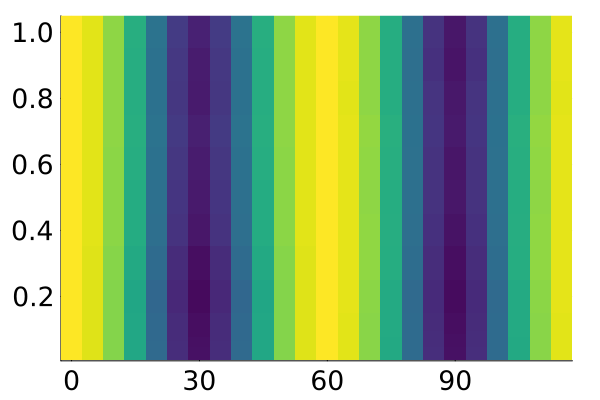}}

                    \refstepcounter{subfigure}
                    \put(29,23){\textcolor{white}{(\alph{subfigure}) H6}}

                    \put(42,1.7){\includegraphics[width=0.127\columnwidth]{Colorbar_blue.png}} 
                    
                \end{picture}

                \label{fig:hBN H6}
            \end{subfigure} \\
    
            \begin{subfigure}{0.23\textwidth}
                \centering
                \setlength{\unitlength}{1mm} 
                \begin{picture}(\columnwidth,27) 
                    \put(0,0){\includegraphics[width=\columnwidth]{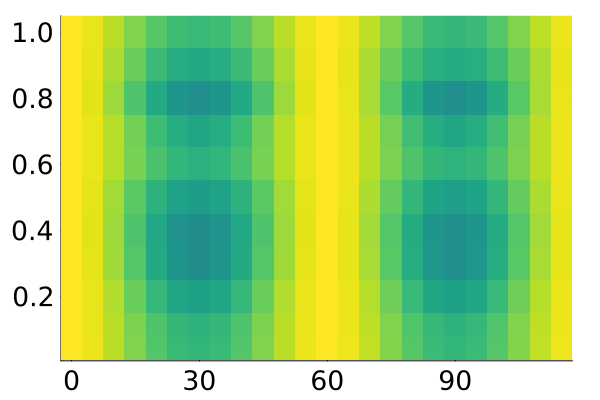}}

                    \refstepcounter{subfigure}
                    \put(27.2,23){\textcolor{white}{(\alph{subfigure}) H7}}

                    \put(-4,10){\rotatebox{90}{\small $t_z$ [eV]}} 
                    
                \end{picture}

                \label{fig:hBN H7}
            \end{subfigure} &
            
            \begin{subfigure}{0.23\textwidth}
                \centering
                \setlength{\unitlength}{1mm} 
                \begin{picture}(\columnwidth,27) 
                    \put(0,0){\includegraphics[width=\columnwidth]{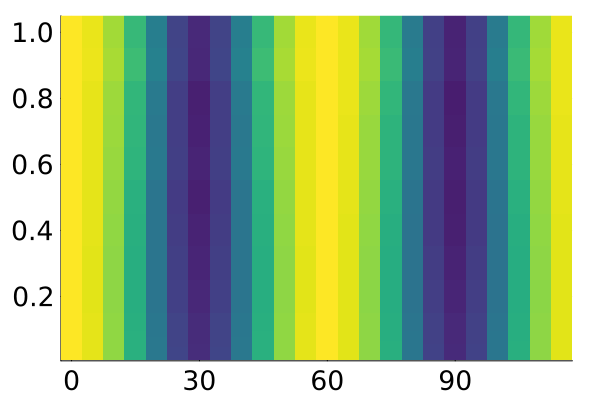}}

                    \refstepcounter{subfigure}
                    \put(29,23){\textcolor{white}{(\alph{subfigure}) H8}}
                    
                \end{picture}

                \label{fig:hBN H8}
            \end{subfigure}  &
            
            \begin{subfigure}{0.23\textwidth}
                \centering
                \setlength{\unitlength}{1mm} 
                \begin{picture}(\columnwidth,27) 
                    \put(0,0){\includegraphics[width=\columnwidth]{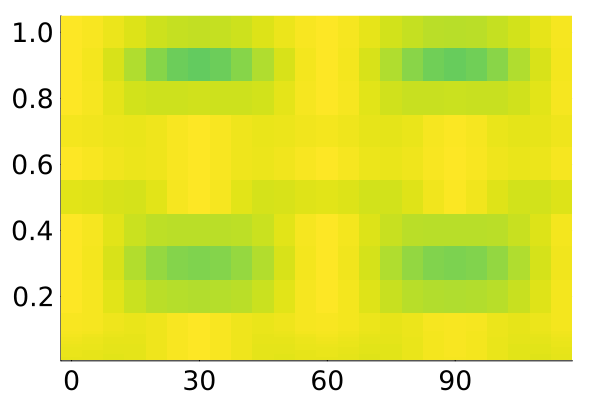}}

                    \refstepcounter{subfigure}
                    \put(29,23){\textcolor{white}{(\alph{subfigure}) H9}}
                    
                \end{picture}

                \label{fig:hBN H9}
            \end{subfigure} &
            
            \begin{subfigure}{0.23\textwidth}
                \centering
                \setlength{\unitlength}{1mm} 
                \begin{picture}(\columnwidth,27) 
                    \put(0,0){\includegraphics[width=\columnwidth]{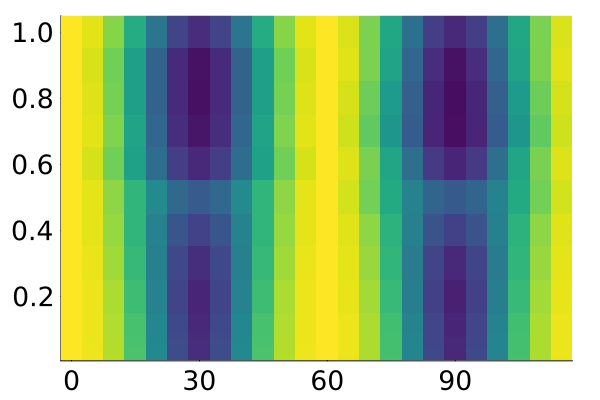}}

                    \refstepcounter{subfigure}
                    \put(27.2,23){\textcolor{white}{(\alph{subfigure}) H10}}

                    \put(42,1.7){\includegraphics[width=0.127\columnwidth]{Colorbar_blue.png}} 
                    
                \end{picture}

                \label{fig:hBN H10}
            \end{subfigure} \\
    
            \begin{subfigure}{0.23\textwidth}
                \centering
                \setlength{\unitlength}{1mm} 
                \begin{picture}(\columnwidth,27) 
                    \put(0,0){\includegraphics[width=\columnwidth]{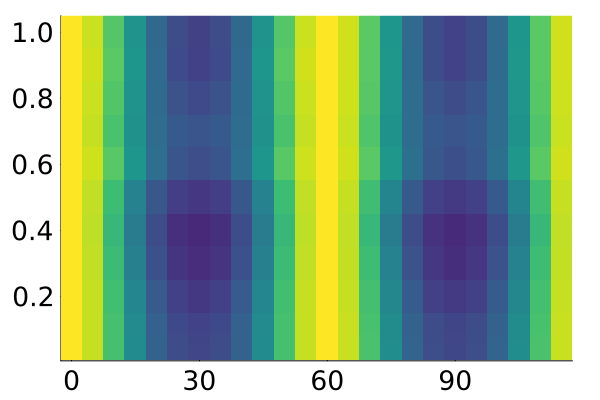}}

                    \refstepcounter{subfigure}
                    \put(27.2,23){\textcolor{white}{(\alph{subfigure}) H11}}

                    \put(19,-3){{\normalsize $\theta$} [\textdegree ]} 

                    \put(-4,10){\rotatebox{90}{\small $t_z$ [eV]}} 
                    
                \end{picture}

                \label{fig:hBN H11}
            \end{subfigure} &
            
            \begin{subfigure}{0.23\textwidth}
                \centering
                \setlength{\unitlength}{1mm} 
                \begin{picture}(\columnwidth,27) 
                    \put(0,0){\includegraphics[width=\columnwidth]{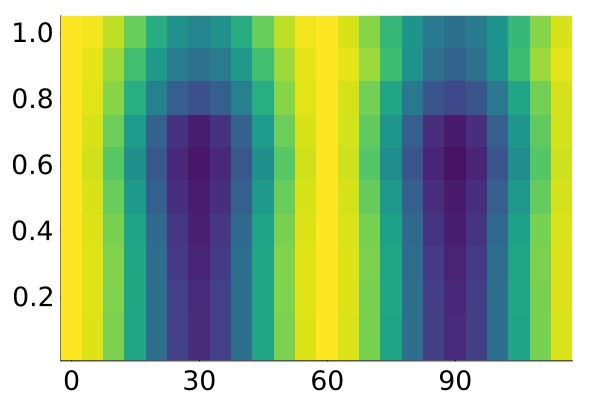}}

                    \refstepcounter{subfigure}
                    \put(27.2,23){\textcolor{white}{(\alph{subfigure}) H12}}

                    \put(19,-3){{\normalsize $\theta$} [\textdegree ]} 
                    
                \end{picture}

                \label{fig:hBN H12}
            \end{subfigure} &
        
            \begin{subfigure}{0.23\textwidth}
                \centering
                \setlength{\unitlength}{1mm} 
                \begin{picture}(\columnwidth,27) 
                    \put(0,0){\includegraphics[width=\columnwidth]{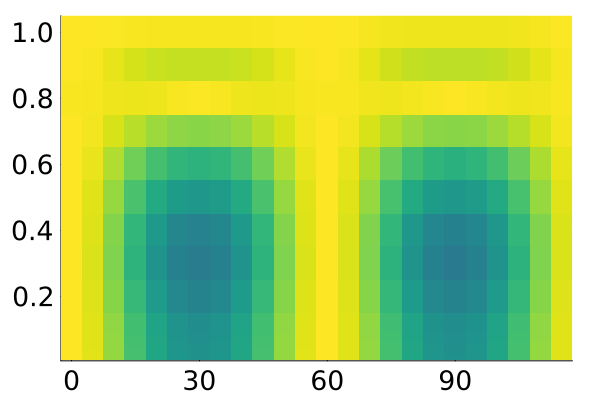}}

                    \refstepcounter{subfigure}
                    \put(27.2,23){\textcolor{white}{(\alph{subfigure}) H13}}

                    \put(19,-3){{\normalsize $\theta$} [\textdegree ]} 
                    
                \end{picture}

                \label{fig:hBN H13}
            \end{subfigure} &
            
            \begin{subfigure}{0.23\textwidth}
                \centering
                \setlength{\unitlength}{1mm} 
                \begin{picture}(\columnwidth,27) 
                    \put(0,0){\includegraphics[width=\columnwidth]{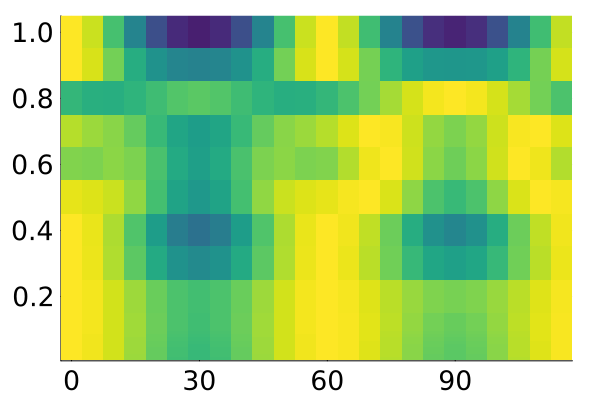}}

                    \refstepcounter{subfigure}

                    \put(42,1.7){\includegraphics[width=0.127\columnwidth]{Colorbar_blue.png}} 
                    
                    \put(27.2,23){\textcolor{white}{(\alph{subfigure}) H14}}

                    \put(19,-3){{\normalsize $\theta$} [\textdegree ]} 
                    
                \end{picture}

                \label{fig:hBN H14}
            \end{subfigure} \\[7mm]

            \multicolumn{4}{c}{\textbf{WS$_2$}} \\[2mm]

            \begin{subfigure}{0.23\textwidth}
                \centering
                \setlength{\unitlength}{1mm} 
                \begin{picture}(\columnwidth,27) 
                    \put(0,0){\includegraphics[width=\columnwidth]{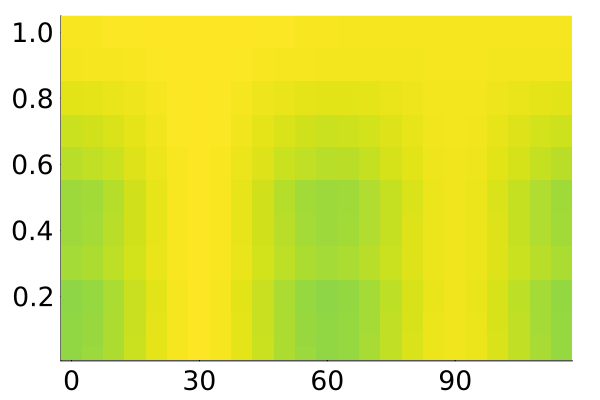}}

                    \refstepcounter{subfigure}
                    \put(28,23){\textcolor{white}{(\alph{subfigure}) H2}}

                    \put(-4,10){\rotatebox{90}{\small $t_z$ [eV]}} 

                    \put(19,-3){{\normalsize $\theta$} [\textdegree ]} 
                    
                \end{picture}

                \label{fig:WS2 H2}
            \end{subfigure} &
            
            \begin{subfigure}{0.23\textwidth}
                \centering
                \setlength{\unitlength}{1mm} 
                \begin{picture}(\columnwidth,27) 
                    \put(0,0){\includegraphics[width=\columnwidth]{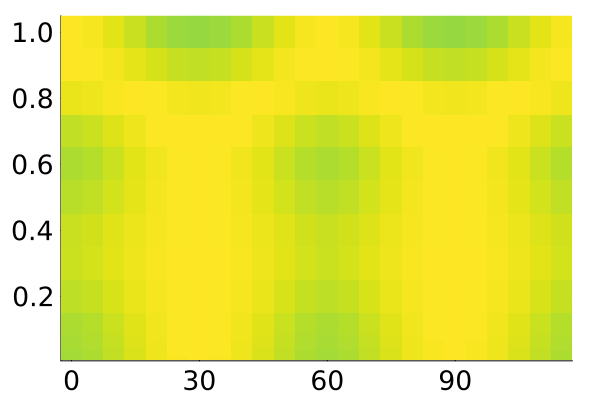}}

                    \refstepcounter{subfigure}
                    \put(28,23){\textcolor{white}{(\alph{subfigure}) H3}}

                    \put(19,-3){{\normalsize $\theta$} [\textdegree ]} 
                    
                \end{picture}

                \label{fig:WS2 H3}
            \end{subfigure}  &
            
            \begin{subfigure}{0.23\textwidth}
                \centering
                \setlength{\unitlength}{1mm} 
                \begin{picture}(\columnwidth,27) 
                    \put(0,0){\includegraphics[width=\columnwidth]{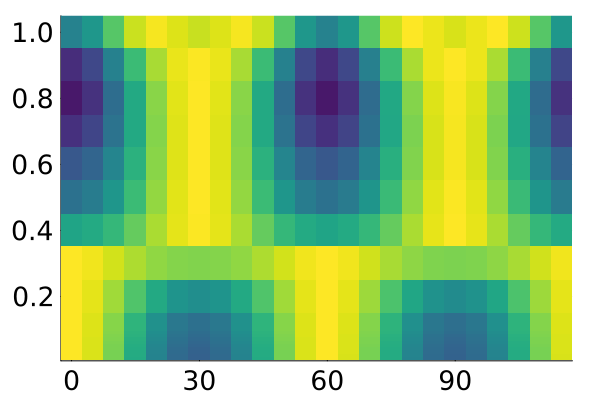}}

                    \refstepcounter{subfigure}
                    \put(28,23){\textcolor{white}{(\alph{subfigure}) H4}}

                    \put(19,-3){{\normalsize $\theta$} [\textdegree ]} 
                    
                \end{picture}

                \label{fig:WS2 H4}
            \end{subfigure} &
            
            \begin{subfigure}{0.23\textwidth}
                \centering
                \setlength{\unitlength}{1mm} 
                \begin{picture}(\columnwidth,27) 
                    \put(0,0){\includegraphics[width=\columnwidth]{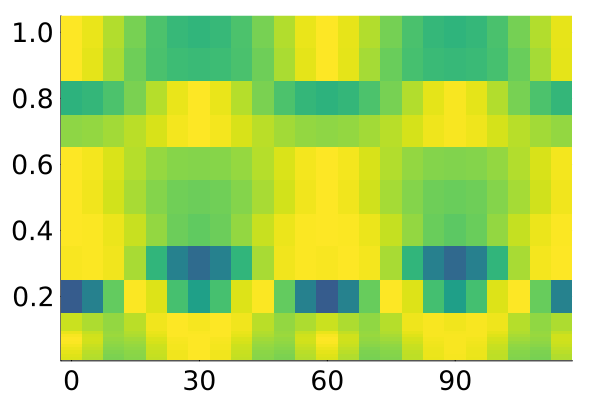}}

                    \refstepcounter{subfigure}
                    \put(28,23){\textcolor{white}{(\alph{subfigure}) H5}}

                    \put(19,-3){{\normalsize $\theta$} [\textdegree ]} 

                    \put(42,1.7){\includegraphics[width=0.127\columnwidth]{Colorbar_blue.png}} 
                    
                \end{picture}

                \label{fig:WS2 H5}
            \end{subfigure} \\

    \end{tabular}
    
    \vspace{3mm}
    
    \caption{Orientation-dependent HHG spectra for layered solids. Panels (a–h) show Graphite harmonics, while (i–t) show hBN results, and (u–x) show WS$_2$ results.}
    \label{fig:Total Current Orientation Dependency Heatmaps}
\end{figure*}

\begin{figure*}[!h]
    \centering

    \begin{tabular}{c c c c}

        \multicolumn{4}{c}{\textbf{Graphite}} \\[2mm]

            \begin{subfigure}{0.23\textwidth}
                \centering
                \setlength{\unitlength}{1mm} 
                \begin{picture}(\columnwidth,28) 
                    \put(0,0){\includegraphics[width=\columnwidth]{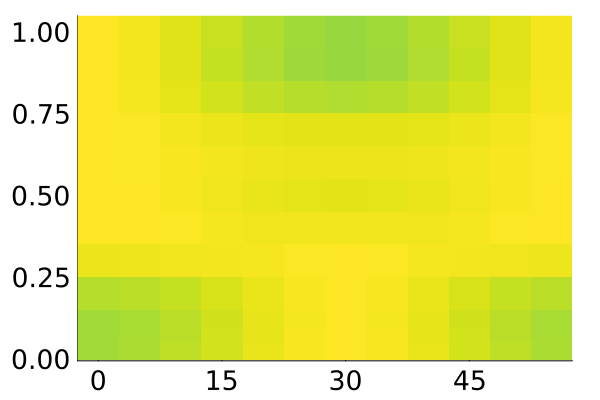}}
                    
                    \refstepcounter{subfigure}
                    \put(29,23){\textcolor{white}{(\alph{subfigure}) H3}}

                    \put(-4,10){\rotatebox{90}{\small $t_z$ [eV]}} 
                    
                \end{picture}
            \end{subfigure} &

            \begin{subfigure}{0.23\textwidth}
                \centering
                \setlength{\unitlength}{1mm} 
                \begin{picture}(\columnwidth,28) 
                    \put(0,0){\includegraphics[width=\columnwidth]{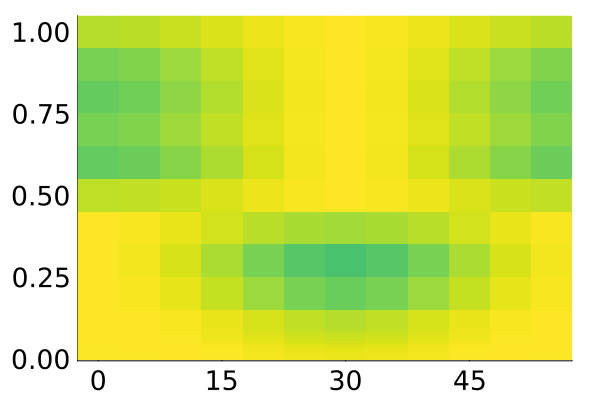}}
                    
                    \refstepcounter{subfigure}
                    \put(29,23){\textcolor{white}{(\alph{subfigure}) H5}}
                    
                \end{picture}
            \end{subfigure} &

            \begin{subfigure}{0.23\textwidth}
                \centering
                \setlength{\unitlength}{1mm} 
                \begin{picture}(\columnwidth,28) 
                    \put(0,0){\includegraphics[width=\columnwidth]{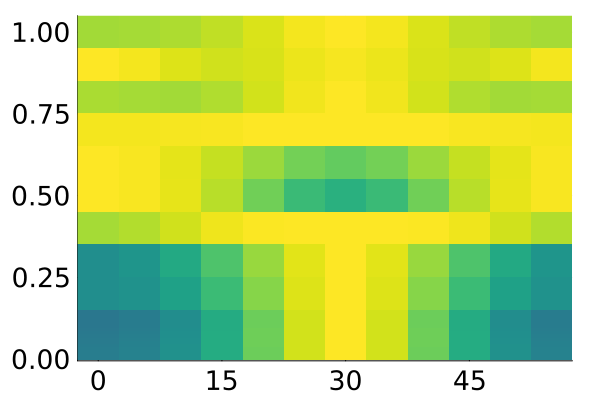}}
                    
                    \refstepcounter{subfigure}
                    \put(29,23){\textcolor{white}{(\alph{subfigure}) H7}}
                    
                \end{picture}
            \end{subfigure} &

            \begin{subfigure}{0.23\textwidth}
                \centering
                \setlength{\unitlength}{1mm} 
                \begin{picture}(\columnwidth,28) 
                    \put(0,0){\includegraphics[width=\columnwidth]{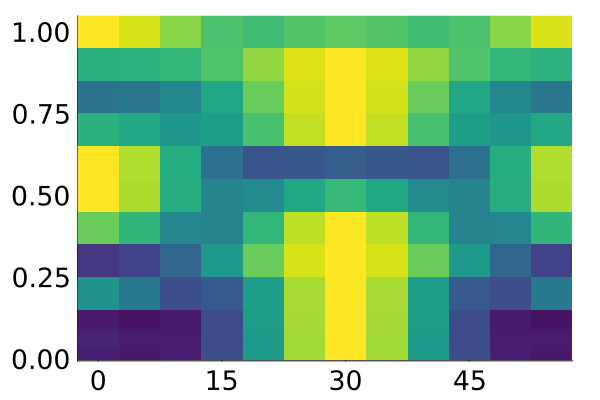}}
                    
                    \refstepcounter{subfigure}
                    \put(29,23){\textcolor{white}{(\alph{subfigure}) H9}}

                    \put(42,1.7){\includegraphics[width=0.127\columnwidth]{Colorbar_blue.png}} 
                    
                \end{picture}
            \end{subfigure} \\
    
            \begin{subfigure}{0.23\textwidth}
                \centering
                \setlength{\unitlength}{1mm} 
                \begin{picture}(\columnwidth,27) 
                    \put(0,0){\includegraphics[width=\columnwidth]{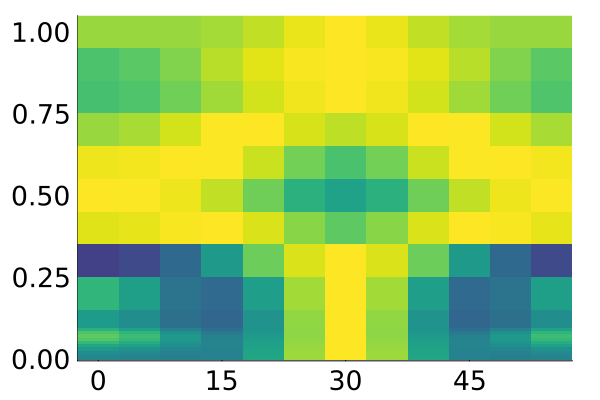}}
                    
                    \refstepcounter{subfigure}
                    \put(27.2,23){\textcolor{white}{(\alph{subfigure}) H11}}

                    \put(-4,10){\rotatebox{90}{\small $t_z$ [eV]}} 

                    \put(19,-3){{\normalsize $\theta$} [\textdegree ]} 
                    
                \end{picture}
            \end{subfigure} &
            
            \begin{subfigure}{0.23\textwidth}
                \centering
                \setlength{\unitlength}{1mm} 
                \begin{picture}(\columnwidth,27) 
                    \put(0,0){\includegraphics[width=\columnwidth]{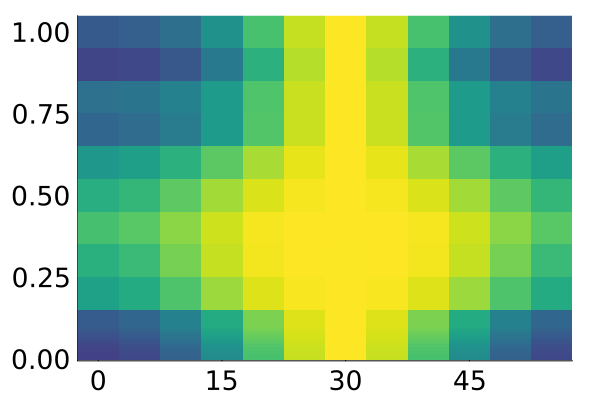}}
                    
                    \refstepcounter{subfigure}
                    \put(27.2,23){\textcolor{white}{(\alph{subfigure}) H13}}

                    \put(19,-3){{\normalsize $\theta$} [\textdegree ]} 
                    
                \end{picture}
            \end{subfigure}  &
            
            \begin{subfigure}{0.23\textwidth}
                \centering
                \setlength{\unitlength}{1mm} 
                \begin{picture}(\columnwidth,27) 
                    \put(0,0){\includegraphics[width=\columnwidth]{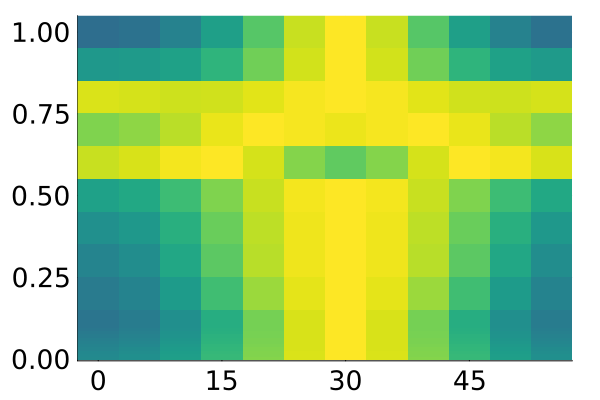}}
                    
                    \refstepcounter{subfigure}
                    \put(27.2,23){\textcolor{white}{(\alph{subfigure}) H15}}

                    \put(19,-3){{\normalsize $\theta$} [\textdegree ]} 
                    
                \end{picture}
            \end{subfigure} &
            
            \begin{subfigure}{0.23\textwidth}
                \centering
                \setlength{\unitlength}{1mm} 
                \begin{picture}(\columnwidth,27) 
                    \put(0,0){\includegraphics[width=\columnwidth]{Graphite_Intra_H15_Heatmap_Orientation_ttz.png}}
                    
                    \refstepcounter{subfigure}
                    \put(27.2,23){\textcolor{white}{(\alph{subfigure}) H17}}

                    \put(19,-3){{\normalsize $\theta$} [\textdegree ]} 

                    \put(42,1.7){\includegraphics[width=0.127\columnwidth]{Colorbar_blue.png}} 
                    
                \end{picture}
            \end{subfigure} \\ [7mm]

        \multicolumn{4}{c}{\textbf{hBN}} \\[2mm]

            \begin{subfigure}{0.23\textwidth}
                \centering
                \setlength{\unitlength}{1mm} 
                \begin{picture}(\columnwidth,28) 
                    \put(0,0){\includegraphics[width=\columnwidth]{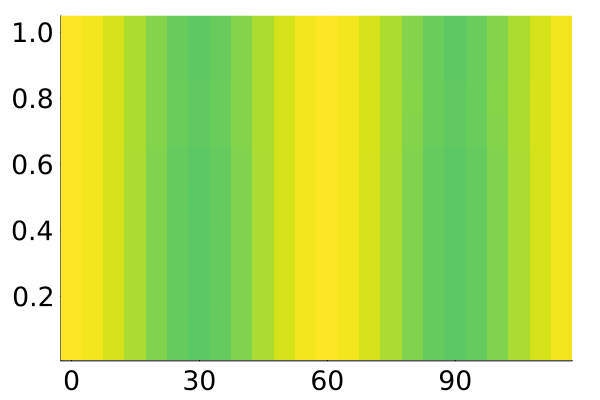}}
                    
                    \refstepcounter{subfigure}
                    \put(29,23){\textcolor{white}{(\alph{subfigure}) H3}}

                    \put(-4,10){\rotatebox{90}{\small $t_z$ [eV]}} 
                    
                \end{picture}
            \end{subfigure} &
            
            \begin{subfigure}{0.23\textwidth}
                \centering
                \setlength{\unitlength}{1mm} 
                \begin{picture}(\columnwidth,28) 
                    \put(0,0){\includegraphics[width=\columnwidth]{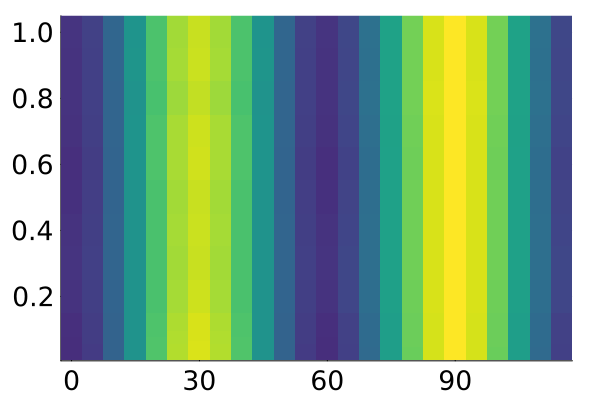}}
                    
                    \refstepcounter{subfigure}
                    \put(29,23){\textcolor{white}{(\alph{subfigure}) H4}}
                    
                \end{picture}
            \end{subfigure} &

            \begin{subfigure}{0.23\textwidth}
                \centering
                \setlength{\unitlength}{1mm} 
                \begin{picture}(\columnwidth,28) 
                    \put(0,0){\includegraphics[width=\columnwidth]{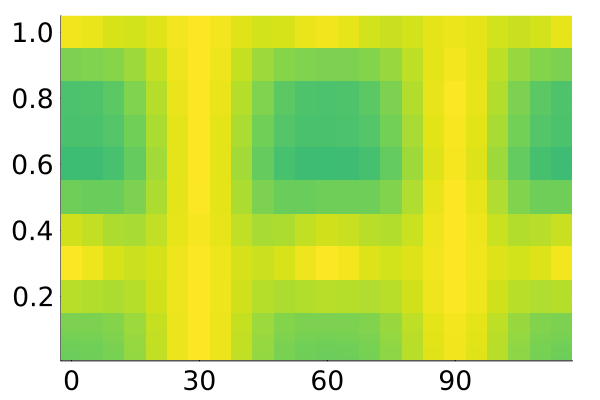}}
                    
                    \refstepcounter{subfigure}
                    \put(29,23){\textcolor{white}{(\alph{subfigure}) H5}}
                    
                \end{picture}
            \end{subfigure} &
            
            \begin{subfigure}{0.23\textwidth}
                \centering
                \setlength{\unitlength}{1mm} 
                \begin{picture}(\columnwidth,28) 
                    \put(0,0){\includegraphics[width=\columnwidth]{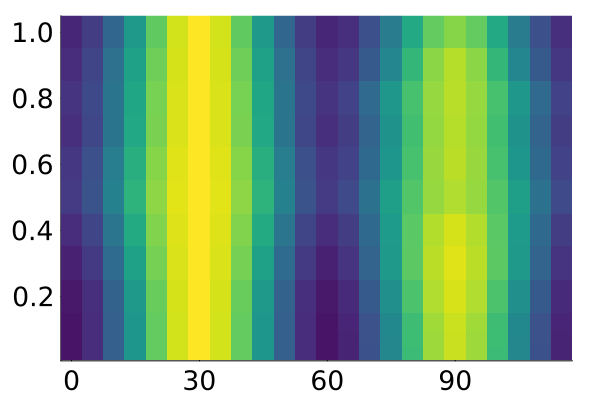}}
                    
                    \refstepcounter{subfigure}
                    \put(29,23){\textcolor{white}{(\alph{subfigure}) H6}}

                    \put(42,1.7){\includegraphics[width=0.127\columnwidth]{Colorbar_blue.png}} 
                    
                \end{picture}
            \end{subfigure} \\
    
            \begin{subfigure}{0.23\textwidth}
                \centering
                \setlength{\unitlength}{1mm} 
                \begin{picture}(\columnwidth,27) 
                    \put(0,0){\includegraphics[width=\columnwidth]{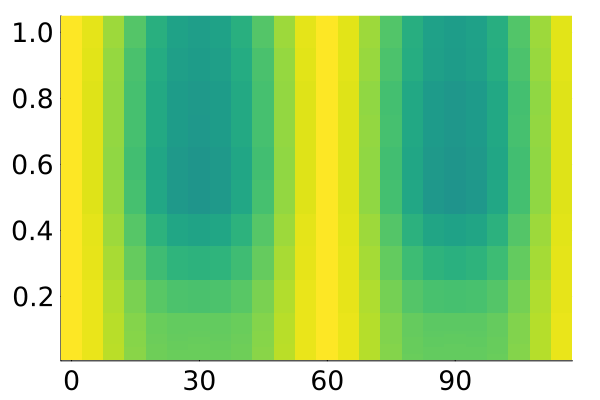}}
                    
                    \refstepcounter{subfigure}
                    \put(27.2,23){\textcolor{white}{(\alph{subfigure}) H7}}

                    \put(-4,10){\rotatebox{90}{\small $t_z$ [eV]}} 
                    
                \end{picture}
            \end{subfigure} &
            
            \begin{subfigure}{0.23\textwidth}
                \centering
                \setlength{\unitlength}{1mm} 
                \begin{picture}(\columnwidth,27) 
                    \put(0,0){\includegraphics[width=\columnwidth]{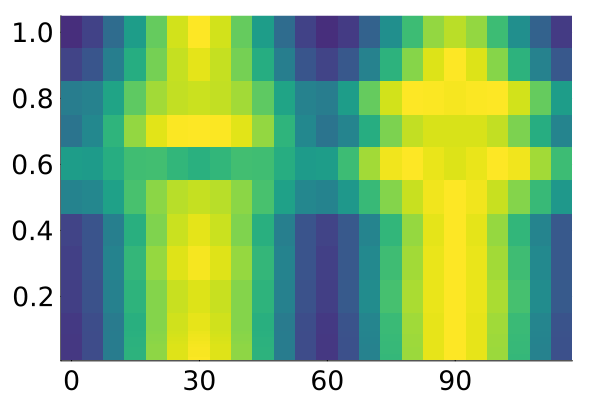}}
                    
                    \refstepcounter{subfigure}
                    \put(29,23){\textcolor{white}{(\alph{subfigure}) H8}}
                    
                \end{picture}
            \end{subfigure}  &
            
            \begin{subfigure}{0.23\textwidth}
                \centering
                \setlength{\unitlength}{1mm} 
                \begin{picture}(\columnwidth,27) 
                    \put(0,0){\includegraphics[width=\columnwidth]{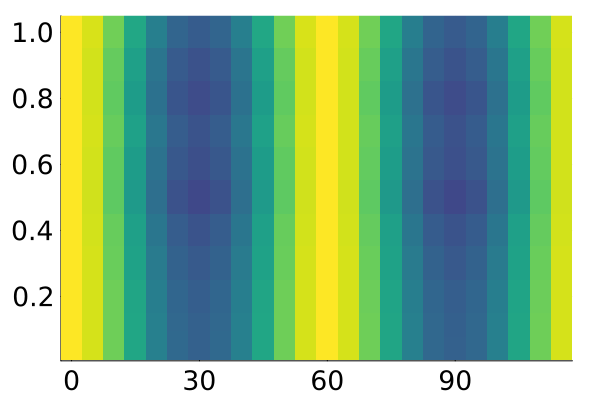}}
                    
                    \refstepcounter{subfigure}
                    \put(29,23){\textcolor{white}{(\alph{subfigure}) H9}}
                    
                \end{picture}
            \end{subfigure} &
            
            \begin{subfigure}{0.23\textwidth}
                \centering
                \setlength{\unitlength}{1mm} 
                \begin{picture}(\columnwidth,27) 
                    \put(0,0){\includegraphics[width=\columnwidth]{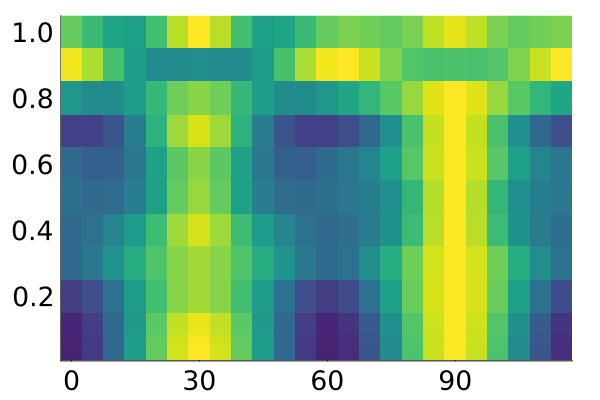}}
                    
                    \refstepcounter{subfigure}
                    \put(27.2,23){\textcolor{white}{(\alph{subfigure}) H10}}

                    \put(42,1.7){\includegraphics[width=0.127\columnwidth]{Colorbar_blue.png}} 
                    
                \end{picture}
            \end{subfigure} \\
    
            \begin{subfigure}{0.23\textwidth}
                \centering
                \setlength{\unitlength}{1mm} 
                \begin{picture}(\columnwidth,27) 
                    \put(0,0){\includegraphics[width=\columnwidth]{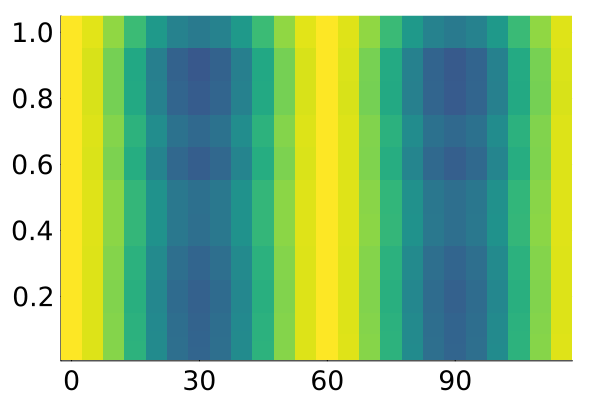}}
                    
                    \refstepcounter{subfigure}
                    \put(27.2,23){\textcolor{white}{(\alph{subfigure}) H11}}

                    \put(19,-3){{\normalsize $\theta$} [\textdegree ]} 

                    \put(-4,10){\rotatebox{90}{\small $t_z$ [eV]}} 
                    
                \end{picture}
            \end{subfigure} &
            
            \begin{subfigure}{0.23\textwidth}
                \centering
                \setlength{\unitlength}{1mm} 
                \begin{picture}(\columnwidth,27) 
                    \put(0,0){\includegraphics[width=\columnwidth]{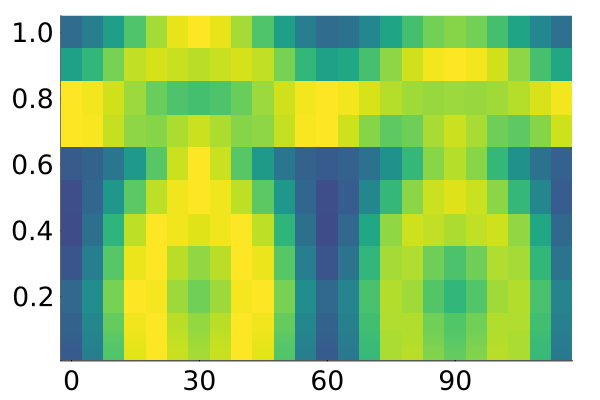}}
                    
                    \refstepcounter{subfigure}
                    \put(27.2,23){\textcolor{white}{(\alph{subfigure}) H12}}

                    \put(19,-3){{\normalsize $\theta$} [\textdegree ]} 
                    
                \end{picture}
            \end{subfigure} &
        
            \begin{subfigure}{0.23\textwidth}
                \centering
                \setlength{\unitlength}{1mm} 
                \begin{picture}(\columnwidth,27) 
                    \put(0,0){\includegraphics[width=\columnwidth]{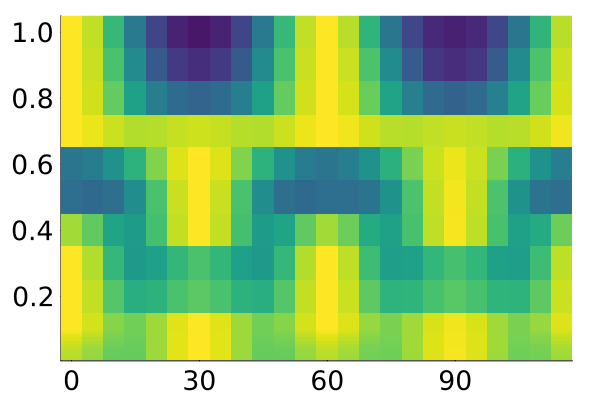}}
                    
                    \refstepcounter{subfigure}
                    \put(27.2,23){\textcolor{white}{(\alph{subfigure}) H13}}

                    \put(19,-3){{\normalsize $\theta$} [\textdegree ]} 
                    
                \end{picture}
            \end{subfigure} &
            
            \begin{subfigure}{0.23\textwidth}
                \centering
                \setlength{\unitlength}{1mm} 
                \begin{picture}(\columnwidth,27) 
                    \put(0,0){\includegraphics[width=\columnwidth]{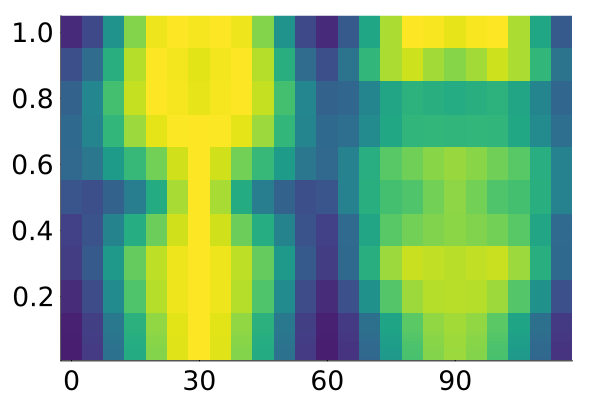}}

                    \put(42,1.7){\includegraphics[width=0.127\columnwidth]{Colorbar_blue.png}} 
                    
                    \refstepcounter{subfigure}
                    \put(27.2,23){\textcolor{white}{(\alph{subfigure}) H14}}

                    \put(19,-3){{\normalsize $\theta$} [\textdegree ]} 
                    
                \end{picture}
            \end{subfigure} \\[7mm]

            \multicolumn{4}{c}{\textbf{WS$_2$}} \\[2mm]

            \begin{subfigure}{0.23\textwidth}
                \centering
                \setlength{\unitlength}{1mm} 
                \begin{picture}(\columnwidth,27) 
                    \put(0,0){\includegraphics[width=\columnwidth]{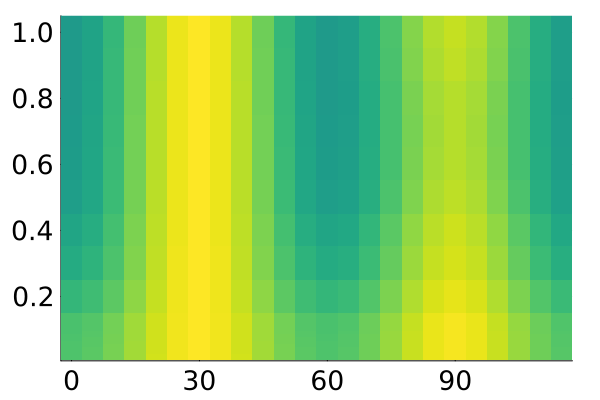}}
                    
                    \refstepcounter{subfigure}
                    \put(28,23){\textcolor{white}{(\alph{subfigure}) H2}}

                    \put(-4,10){\rotatebox{90}{\small $t_z$ [eV]}} 

                    \put(19,-3){{\normalsize $\theta$} [\textdegree ]} 
                    
                \end{picture}
            \end{subfigure} &
            
            \begin{subfigure}{0.23\textwidth}
                \centering
                \setlength{\unitlength}{1mm} 
                \begin{picture}(\columnwidth,27) 
                    \put(0,0){\includegraphics[width=\columnwidth]{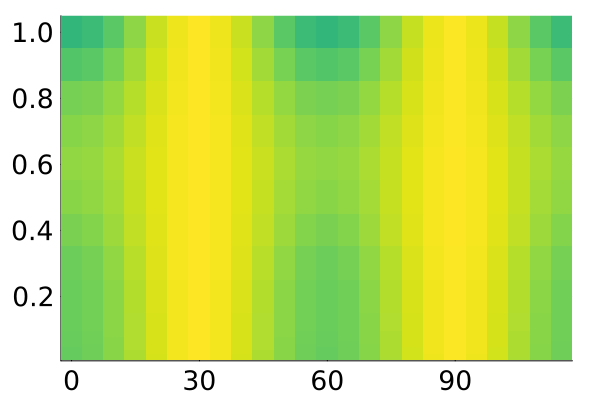}}
                    
                    \refstepcounter{subfigure}
                    \put(28,23){\textcolor{white}{(\alph{subfigure}) H3}}

                    \put(19,-3){{\normalsize $\theta$} [\textdegree ]} 
                    
                \end{picture}
            \end{subfigure}  &
            
            \begin{subfigure}{0.23\textwidth}
                \centering
                \setlength{\unitlength}{1mm} 
                \begin{picture}(\columnwidth,27) 
                    \put(0,0){\includegraphics[width=\columnwidth]{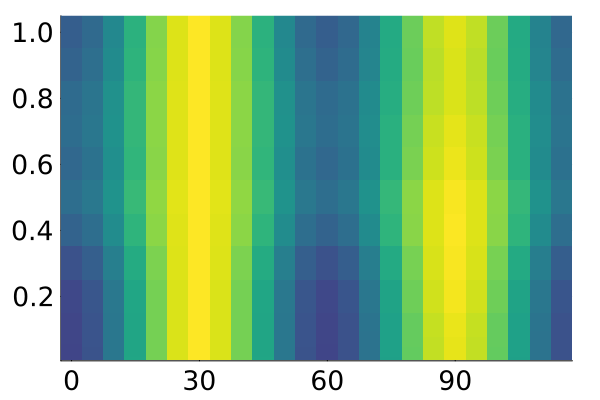}}
                    
                    \refstepcounter{subfigure}
                    \put(28,23){\textcolor{white}{(\alph{subfigure}) H4}}

                    \put(19,-3){{\normalsize $\theta$} [\textdegree ]} 
                    
                \end{picture}
            \end{subfigure} &
            
            \begin{subfigure}{0.23\textwidth}
                \centering
                \setlength{\unitlength}{1mm} 
                \begin{picture}(\columnwidth,27) 
                    \put(0,0){\includegraphics[width=\columnwidth]{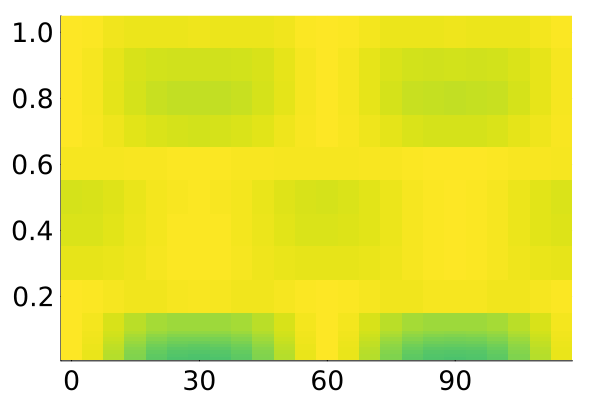}}
                    
                    \refstepcounter{subfigure}
                    \put(28,23){\textcolor{white}{(\alph{subfigure}) H5}}

                    \put(19,-3){{\normalsize $\theta$} [\textdegree ]} 

                    \put(42,1.7){\includegraphics[width=0.127\columnwidth]{Colorbar_blue.png}} 
                    
                \end{picture}
            \end{subfigure} \\

    \end{tabular}
    
    \vspace{3mm}
    
    \caption{Orientation-dependent HHG spectra of intraband harmonics from layered solids. Panels (a–h) show graphite harmonics, while (i–t) show hBN results, and (u–x) show WS$_2$ results.}
    \label{fig:Intra Current Orientation Dependency Heatmaps}
\end{figure*}

\begin{figure*}[!h]
    \centering

    \begin{tabular}{c c c c}

        \multicolumn{4}{c}{\textbf{Graphite}} \\[2mm]

            \begin{subfigure}{0.23\textwidth}
                \centering
                \setlength{\unitlength}{1mm} 
                \begin{picture}(\columnwidth,28) 
                    \put(0,0){\includegraphics[width=\columnwidth]{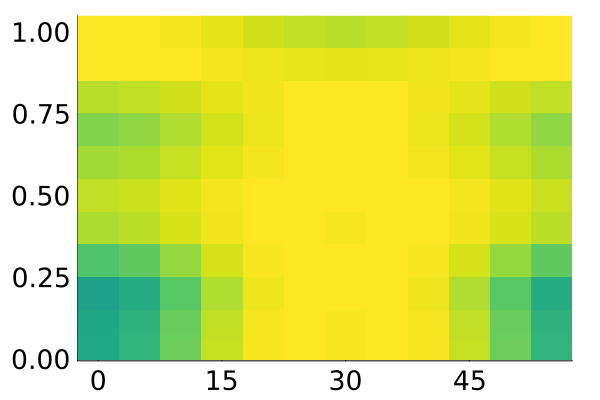}}
                    
                    \refstepcounter{subfigure}
                    \put(29,23){\textcolor{white}{(\alph{subfigure}) H3}}

                    \put(-4,10){\rotatebox{90}{\small $t_z$ [eV]}} 
                    
                \end{picture}
            \end{subfigure} &

            \begin{subfigure}{0.23\textwidth}
                \centering
                \setlength{\unitlength}{1mm} 
                \begin{picture}(\columnwidth,28) 
                    \put(0,0){\includegraphics[width=\columnwidth]{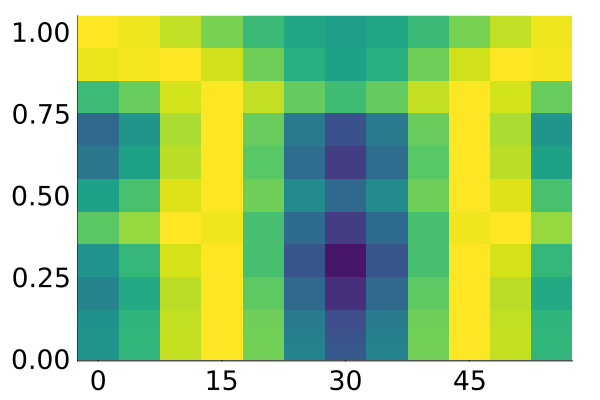}}
                    
                    \refstepcounter{subfigure}
                    \put(29,23){\textcolor{white}{(\alph{subfigure}) H5}}
                    
                \end{picture}
            \end{subfigure} &

            \begin{subfigure}{0.23\textwidth}
                \centering
                \setlength{\unitlength}{1mm} 
                \begin{picture}(\columnwidth,28) 
                    \put(0,0){\includegraphics[width=\columnwidth]{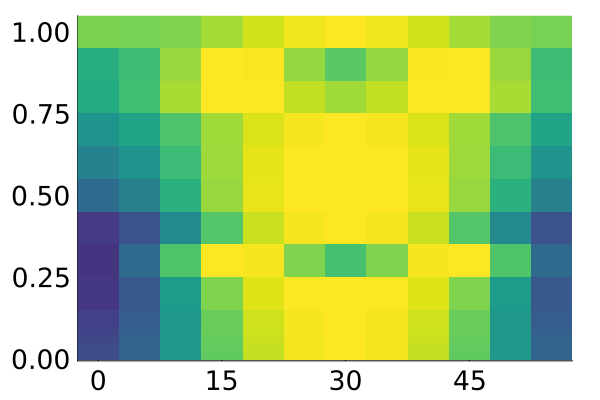}}
                    
                    \refstepcounter{subfigure}
                    \put(29,23){\textcolor{white}{(\alph{subfigure}) H7}}
                    
                \end{picture}
            \end{subfigure} &

            \begin{subfigure}{0.23\textwidth}
                \centering
                \setlength{\unitlength}{1mm} 
                \begin{picture}(\columnwidth,28) 
                    \put(0,0){\includegraphics[width=\columnwidth]{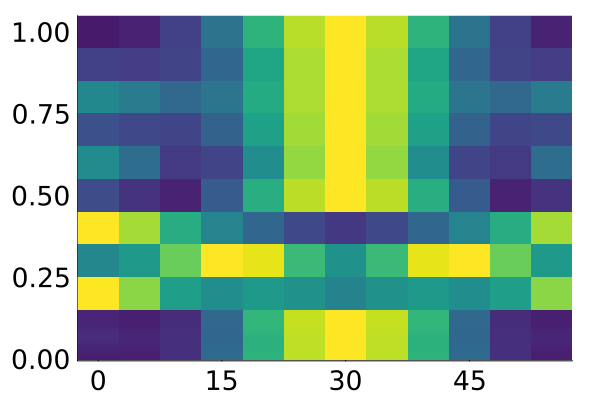}}
                    
                    \refstepcounter{subfigure}
                    \put(29,23){\textcolor{white}{(\alph{subfigure}) H9}}

                    \put(42,1.7){\includegraphics[width=0.127\columnwidth]{Colorbar_blue.png}} 
                    
                \end{picture}
            \end{subfigure} \\
    
            \begin{subfigure}{0.23\textwidth}
                \centering
                \setlength{\unitlength}{1mm} 
                \begin{picture}(\columnwidth,27) 
                    \put(0,0){\includegraphics[width=\columnwidth]{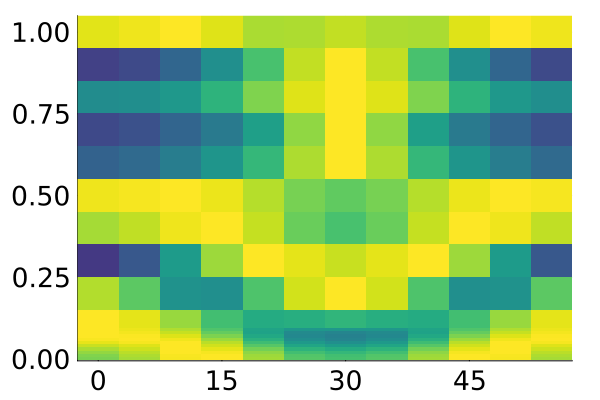}}
                    
                    \refstepcounter{subfigure}
                    \put(27.2,23){\textcolor{white}{(\alph{subfigure}) H11}}

                    \put(-4,10){\rotatebox{90}{\small $t_z$ [eV]}} 

                    \put(19,-3){{\normalsize $\theta$} [\textdegree ]} 
                    
                \end{picture}
            \end{subfigure} &
            
            \begin{subfigure}{0.23\textwidth}
                \centering
                \setlength{\unitlength}{1mm} 
                \begin{picture}(\columnwidth,27) 
                    \put(0,0){\includegraphics[width=\columnwidth]{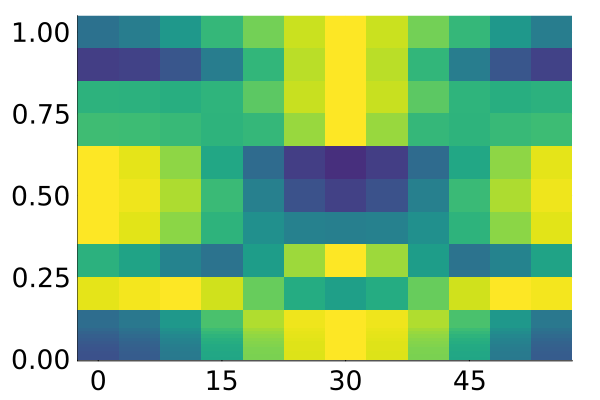}}
                    
                    \refstepcounter{subfigure}
                    \put(27.2,23){\textcolor{white}{(\alph{subfigure}) H13}}

                    \put(19,-3){{\normalsize $\theta$} [\textdegree ]} 
                    
                \end{picture}
            \end{subfigure}  &
            
            \begin{subfigure}{0.23\textwidth}
                \centering
                \setlength{\unitlength}{1mm} 
                \begin{picture}(\columnwidth,27) 
                    \put(0,0){\includegraphics[width=\columnwidth]{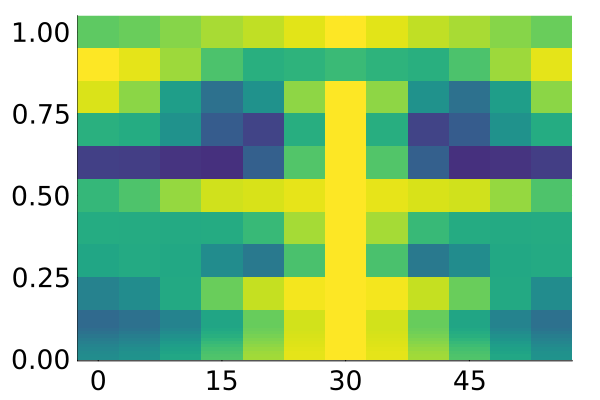}}
                    
                    \refstepcounter{subfigure}
                    \put(27.2,23){\textcolor{white}{(\alph{subfigure}) H15}}

                    \put(19,-3){{\normalsize $\theta$} [\textdegree ]} 
                    
                \end{picture}
            \end{subfigure} &
            
            \begin{subfigure}{0.23\textwidth}
                \centering
                \setlength{\unitlength}{1mm} 
                \begin{picture}(\columnwidth,27) 
                    \put(0,0){\includegraphics[width=\columnwidth]{Graphite_Inter_H15_Heatmap_Orientation_ttz.png}}
                    
                    \refstepcounter{subfigure}
                    \put(27.2,23){\textcolor{white}{(\alph{subfigure}) H17}}

                    \put(19,-3){{\normalsize $\theta$} [\textdegree ]} 

                    \put(42,1.7){\includegraphics[width=0.127\columnwidth]{Colorbar_blue.png}} 
                    
                \end{picture}
            \end{subfigure} \\ [7mm]

        \multicolumn{4}{c}{\textbf{hBN}} \\[2mm]

            \begin{subfigure}{0.23\textwidth}
                \centering
                \setlength{\unitlength}{1mm} 
                \begin{picture}(\columnwidth,28) 
                    \put(0,0){\includegraphics[width=\columnwidth]{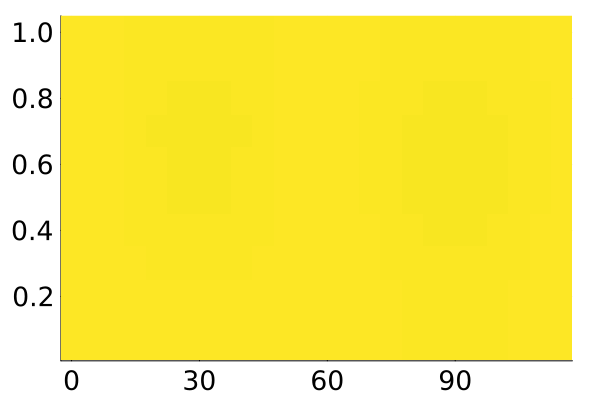}}
                    
                    \refstepcounter{subfigure}
                    \put(29,23){\textcolor{white}{(\alph{subfigure}) H3}}

                    \put(-4,10){\rotatebox{90}{\small $t_z$ [eV]}} 
                    
                \end{picture}
            \end{subfigure} &
            
            \begin{subfigure}{0.23\textwidth}
                \centering
                \setlength{\unitlength}{1mm} 
                \begin{picture}(\columnwidth,28) 
                    \put(0,0){\includegraphics[width=\columnwidth]{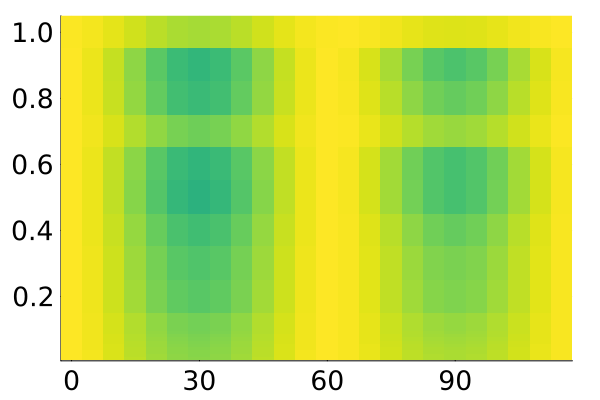}}
                    
                    \refstepcounter{subfigure}
                    \put(29,23){\textcolor{white}{(\alph{subfigure}) H4}}
                    
                \end{picture}
            \end{subfigure} &

            \begin{subfigure}{0.23\textwidth}
                \centering
                \setlength{\unitlength}{1mm} 
                \begin{picture}(\columnwidth,28) 
                    \put(0,0){\includegraphics[width=\columnwidth]{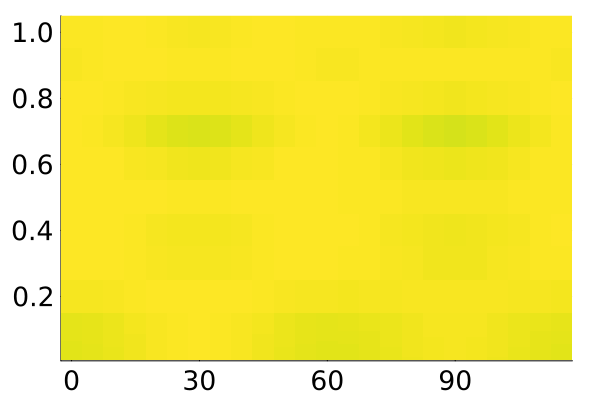}}
                    
                    \refstepcounter{subfigure}
                    \put(29,23){\textcolor{white}{(\alph{subfigure}) H5}}
                    
                \end{picture}
            \end{subfigure} &
            
            \begin{subfigure}{0.23\textwidth}
                \centering
                \setlength{\unitlength}{1mm} 
                \begin{picture}(\columnwidth,28) 
                    \put(0,0){\includegraphics[width=\columnwidth]{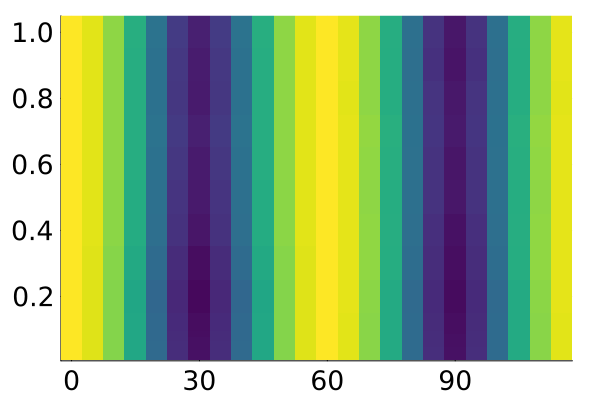}}
                    
                    \refstepcounter{subfigure}
                    \put(29,23){\textcolor{white}{(\alph{subfigure}) H6}}

                    \put(42,1.7){\includegraphics[width=0.127\columnwidth]{Colorbar_blue.png}} 
                    
                \end{picture}
            \end{subfigure} \\
    
            \begin{subfigure}{0.23\textwidth}
                \centering
                \setlength{\unitlength}{1mm} 
                \begin{picture}(\columnwidth,27) 
                    \put(0,0){\includegraphics[width=\columnwidth]{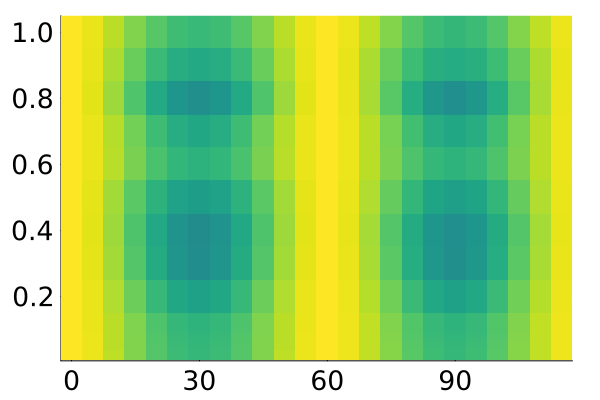}}
                    
                    \refstepcounter{subfigure}
                    \put(27.2,23){\textcolor{white}{(\alph{subfigure}) H7}}

                    \put(-4,10){\rotatebox{90}{\small $t_z$ [eV]}} 
                    
                \end{picture}
            \end{subfigure} &
            
            \begin{subfigure}{0.23\textwidth}
                \centering
                \setlength{\unitlength}{1mm} 
                \begin{picture}(\columnwidth,27) 
                    \put(0,0){\includegraphics[width=\columnwidth]{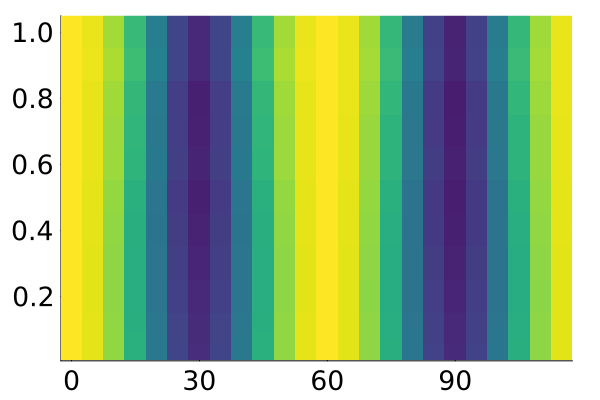}}
                    
                    \refstepcounter{subfigure}
                    \put(29,23){\textcolor{white}{(\alph{subfigure}) H8}}
                    
                \end{picture}
            \end{subfigure}  &
            
            \begin{subfigure}{0.23\textwidth}
                \centering
                \setlength{\unitlength}{1mm} 
                \begin{picture}(\columnwidth,27) 
                    \put(0,0){\includegraphics[width=\columnwidth]{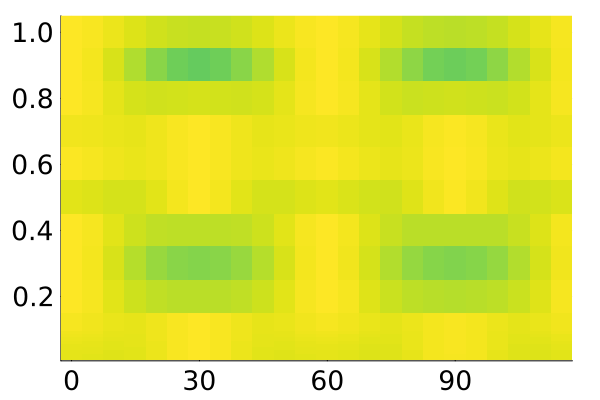}}
                    
                    \refstepcounter{subfigure}
                    \put(29,23){\textcolor{white}{(\alph{subfigure}) H9}}
                    
                \end{picture}
            \end{subfigure} &
            
            \begin{subfigure}{0.23\textwidth}
                \centering
                \setlength{\unitlength}{1mm} 
                \begin{picture}(\columnwidth,27) 
                    \put(0,0){\includegraphics[width=\columnwidth]{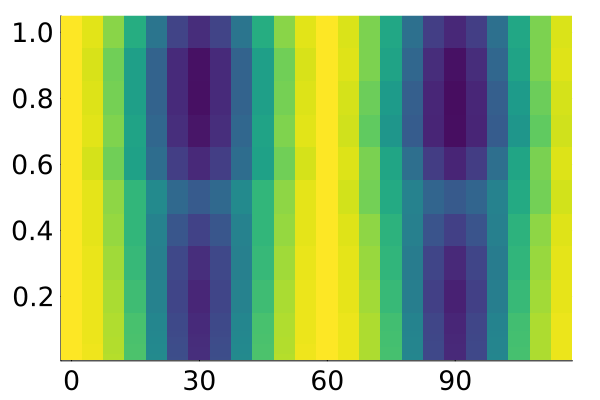}}
                    
                    \refstepcounter{subfigure}
                    \put(27.2,23){\textcolor{white}{(\alph{subfigure}) H10}}

                    \put(42,1.7){\includegraphics[width=0.127\columnwidth]{Colorbar_blue.png}} 
                    
                \end{picture}
            \end{subfigure} \\
    
            \begin{subfigure}{0.23\textwidth}
                \centering
                \setlength{\unitlength}{1mm} 
                \begin{picture}(\columnwidth,27) 
                    \put(0,0){\includegraphics[width=\columnwidth]{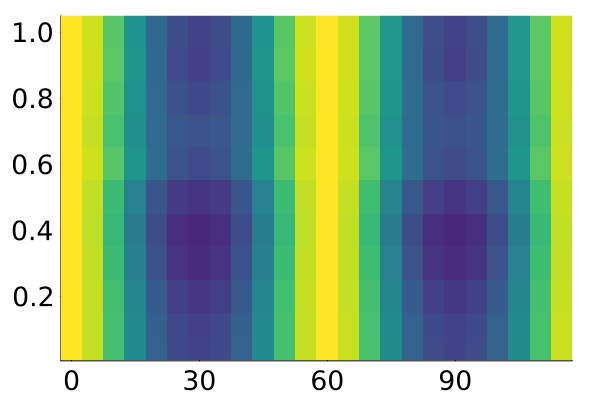}}
                    
                    \refstepcounter{subfigure}
                    \put(27.2,23){\textcolor{white}{(\alph{subfigure}) H11}}

                    \put(19,-3){{\normalsize $\theta$} [\textdegree ]} 

                    \put(-4,10){\rotatebox{90}{\small $t_z$ [eV]}} 
                    
                \end{picture}
            \end{subfigure} &
            
            \begin{subfigure}{0.23\textwidth}
                \centering
                \setlength{\unitlength}{1mm} 
                \begin{picture}(\columnwidth,27) 
                    \put(0,0){\includegraphics[width=\columnwidth]{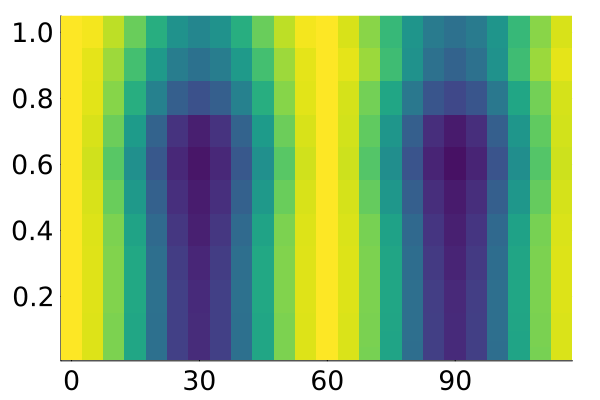}}
                    
                    \refstepcounter{subfigure}
                    \put(27.2,23){\textcolor{white}{(\alph{subfigure}) H12}}

                    \put(19,-3){{\normalsize $\theta$} [\textdegree ]} 
                    
                \end{picture}
            \end{subfigure} &
        
            \begin{subfigure}{0.23\textwidth}
                \centering
                \setlength{\unitlength}{1mm} 
                \begin{picture}(\columnwidth,27) 
                    \put(0,0){\includegraphics[width=\columnwidth]{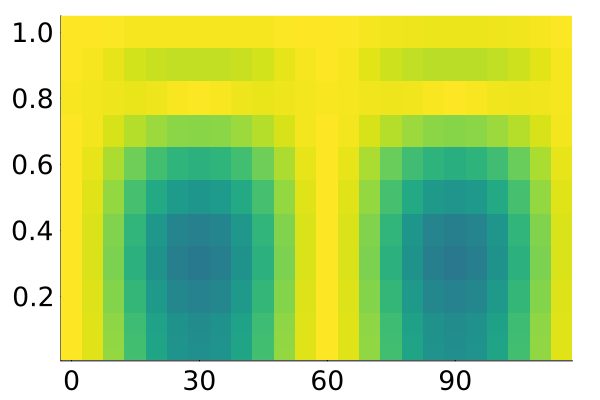}}
                    
                    \refstepcounter{subfigure}
                    \put(27.2,23){\textcolor{white}{(\alph{subfigure}) H13}}

                    \put(19,-3){{\normalsize $\theta$} [\textdegree ]} 
                    
                \end{picture}
            \end{subfigure} &
            
            \begin{subfigure}{0.23\textwidth}
                \centering
                \setlength{\unitlength}{1mm} 
                \begin{picture}(\columnwidth,27) 
                    \put(0,0){\includegraphics[width=\columnwidth]{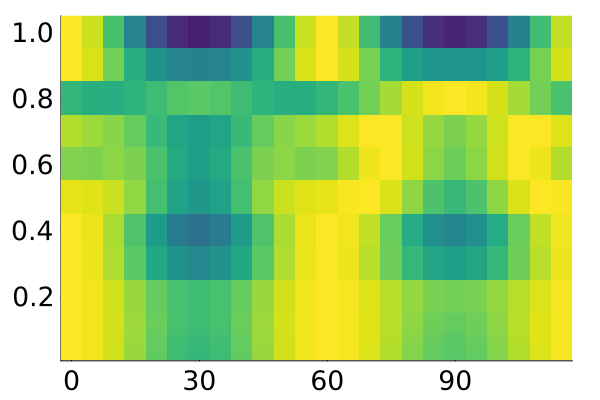}}
                    
                    \refstepcounter{subfigure}
                    \put(42,1.7){\includegraphics[width=0.127\columnwidth]{Colorbar_blue.png}} 
                    
                    \put(27.2,23){\textcolor{white}{(\alph{subfigure}) H14}}

                    \put(19,-3){{\normalsize $\theta$} [\textdegree ]} 
                    
                \end{picture}
            \end{subfigure} \\[7mm]

            \multicolumn{4}{c}{\textbf{WS$_2$}} \\[2mm]

            \begin{subfigure}{0.23\textwidth}
                \centering
                \setlength{\unitlength}{1mm} 
                \begin{picture}(\columnwidth,27) 
                    \put(0,0){\includegraphics[width=\columnwidth]{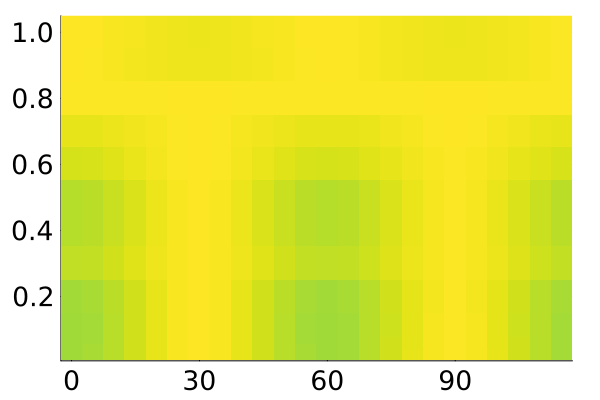}}
                    
                    \refstepcounter{subfigure}
                    \put(28,23){\textcolor{white}{(\alph{subfigure}) H2}}

                    \put(-4,10){\rotatebox{90}{\small $t_z$ [eV]}} 

                    \put(19,-3){{\normalsize $\theta$} [\textdegree ]} 
                    
                \end{picture}
            \end{subfigure} &
            
            \begin{subfigure}{0.23\textwidth}
                \centering
                \setlength{\unitlength}{1mm} 
                \begin{picture}(\columnwidth,27) 
                    \put(0,0){\includegraphics[width=\columnwidth]{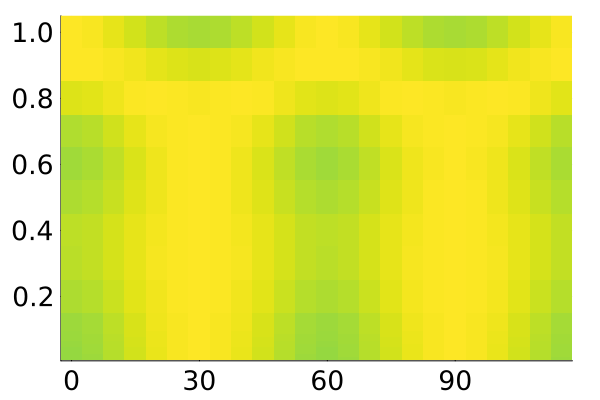}}
                    
                    \refstepcounter{subfigure}
                    \put(28,23){\textcolor{white}{(\alph{subfigure}) H3}}

                    \put(19,-3){{\normalsize $\theta$} [\textdegree ]} 
                    
                \end{picture}
            \end{subfigure}  &
            
            \begin{subfigure}{0.23\textwidth}
                \centering
                \setlength{\unitlength}{1mm} 
                \begin{picture}(\columnwidth,27) 
                    \put(0,0){\includegraphics[width=\columnwidth]{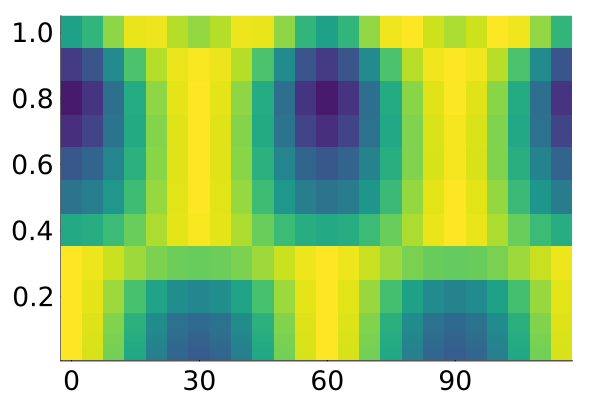}}
                    
                    \refstepcounter{subfigure}
                    \put(28,23){\textcolor{white}{(\alph{subfigure}) H4}}

                    \put(19,-3){{\normalsize $\theta$} [\textdegree ]} 
                    
                \end{picture}
            \end{subfigure} &
            
            \begin{subfigure}{0.23\textwidth}
                \centering
                \setlength{\unitlength}{1mm} 
                \begin{picture}(\columnwidth,27) 
                    \put(0,0){\includegraphics[width=\columnwidth]{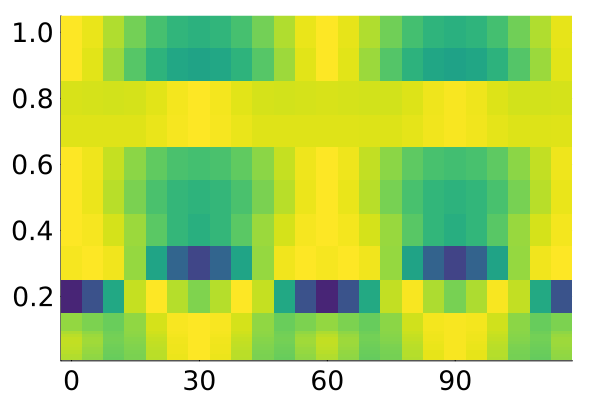}}
                    
                    \refstepcounter{subfigure}
                    \put(28,23){\textcolor{white}{(\alph{subfigure}) H5}}

                    \put(19,-3){{\normalsize $\theta$} [\textdegree ]} 

                    \put(42,1.7){\includegraphics[width=0.127\columnwidth]{Colorbar_blue.png}} 
                    
                \end{picture}
            \end{subfigure} \\

    \end{tabular}
    
    \vspace{3mm}
    
    \caption{Orientation-dependent HHG spectra of interband harmonics from layered solids. Panels (a–h) show graphite harmonics, while (i–t) show hBN results, and (u–x) show WS$_2$ results.}
    \label{fig:Inter Current Orientation Dependency Heatmaps}
\end{figure*}

\noindent Here we present additional complementary data to the main text analysis. First in Fig.~\ref{fig:Total Current Orientation Dependency Heatmaps} we present full orientation dependence of HHG yields from the three material systems. This complements data in Fig. 2 in the main text. In Fig.~\ref{fig:Intra Current Orientation Dependency Heatmaps}, \ref{fig:Inter Current Orientation Dependency Heatmaps}, we present similar data, but separated to intraband and interband, respectively, showing very similar physical features, indicating that both emission mechanisms are similarly susceptible to interlayer coupling. 

\begin{figure*}[!h]
    \centering

    \begin{tabular}{c c c c}

            \multicolumn{4}{c}{\textbf{WS$_2$}} \\[2mm]

            \begin{subfigure}{0.23\textwidth}
                \centering
                \setlength{\unitlength}{1mm} 
                \begin{picture}(\columnwidth,27) 
                    \put(0,0){\includegraphics[width=\columnwidth]{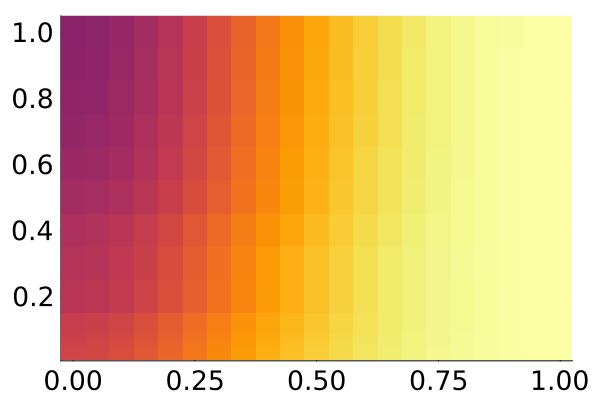}}
                    
                    \refstepcounter{subfigure}
                    \put(28,23){(\alph{subfigure}) H2}

                    \put(-4,10){\rotatebox{90}{\small $t_z$ [eV]}} 

                    \put(13,-3){{\normalsize $\varepsilon$} - Ellipticity} 
                    
                \end{picture}

                \label{fig:WS2 H2}
            \end{subfigure} &
            
            \begin{subfigure}{0.23\textwidth}
                \centering
                \setlength{\unitlength}{1mm} 
                \begin{picture}(\columnwidth,27) 
                    \put(0,0){\includegraphics[width=\columnwidth]{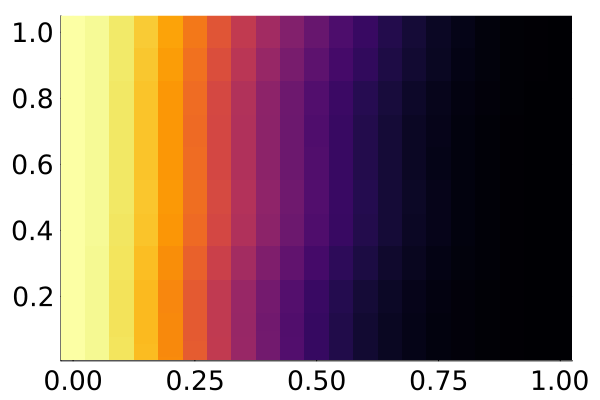}}
                    
                    \refstepcounter{subfigure}
                    \put(28,23){\textcolor{white}{(\alph{subfigure}) H3}}

                    \put(13,-3){{\normalsize $\varepsilon$} - Ellipticity} 
                    
                \end{picture}

                \label{fig:WS2 H3}
            \end{subfigure}  &
            
            \begin{subfigure}{0.23\textwidth}
                \centering
                \setlength{\unitlength}{1mm} 
                \begin{picture}(\columnwidth,27) 
                    \put(0,0){\includegraphics[width=\columnwidth]{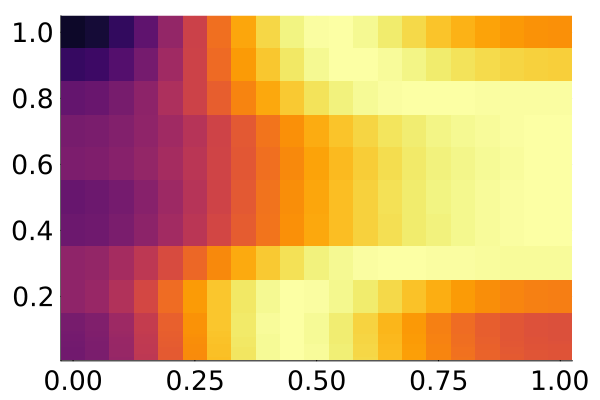}}
                    
                    \refstepcounter{subfigure}
                    \put(28,23){(\alph{subfigure}) H4}

                    \put(13,-3){{\normalsize $\varepsilon$} - Ellipticity} 
                    
                \end{picture}

                \label{fig:WS2 H4}
            \end{subfigure} &
            
            \begin{subfigure}{0.23\textwidth}
                \centering
                \setlength{\unitlength}{1mm} 
                \begin{picture}(\columnwidth,27) 
                    \put(0,0){\includegraphics[width=\columnwidth]{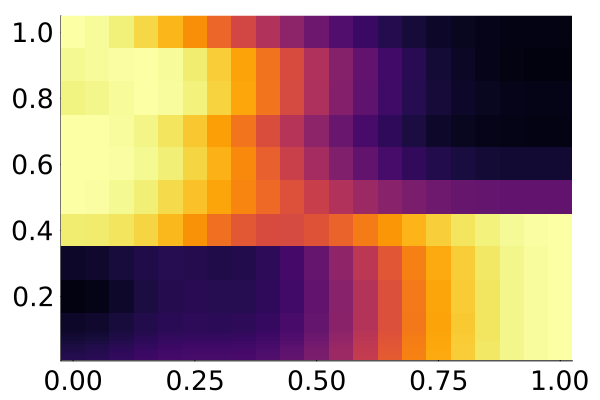}}
                    
                    \refstepcounter{subfigure}
                    \put(28,23){\textcolor{white}{(\alph{subfigure}) H5}}

                    \put(13,-3){{\normalsize $\varepsilon$} - Ellipticity} 

                    \put(42,1.7){\includegraphics[width=0.145\columnwidth]{Colorbar.png}} 
                    
                \end{picture}

                \label{fig:WS2 H5}
            \end{subfigure} \\

    \end{tabular}
    
    \vspace{3mm}
    
    \caption{}
    \label{fig:WS2 Ellipticity Dependency Results}
\end{figure*}

In Fig.~\ref{fig:WS2 Ellipticity Dependency Results} we present full ellipticity dependence of HHG yields from WS$_2$, providing additional dataset to the ellipticity studies of the other materials discussed in the primary text. Very similar physical effects are observed in HHG from WS$_2$, indicating the generality of our conclusions also to transition metal dichalcogenides and likely to other layered solids. 

\begin{figure*}[!h]
\centering

\begin{subfigure}{0.33\textwidth}
    \centering
    \setlength{\unitlength}{\linewidth} 
    \begin{picture}(1,0.7)
        \put(0,0){\includegraphics[width=\unitlength]{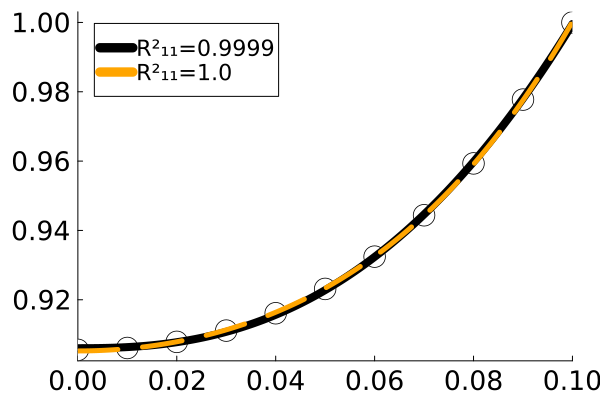}}
        
        \refstepcounter{subfigure}
        \put(0.17,0.3){
            \shortstack[l]{
                (\thesubfigure)\ H3 \\
                \hspace{1.7em}$\theta = 0^{\circ}$
            }
        }

        \put(-0.2,0.2){\rotatebox{90}{\footnotesize \shortstack{Normalized \\ Intensity \\ {[arb. units]} }}} 
    \end{picture}

    \label{fig:Intra A}
\end{subfigure}%
\begin{subfigure}{0.33\textwidth}
    \centering
    \setlength{\unitlength}{\linewidth}
    \begin{picture}(1,0.7)
        \put(0,0){\includegraphics[width=\unitlength]{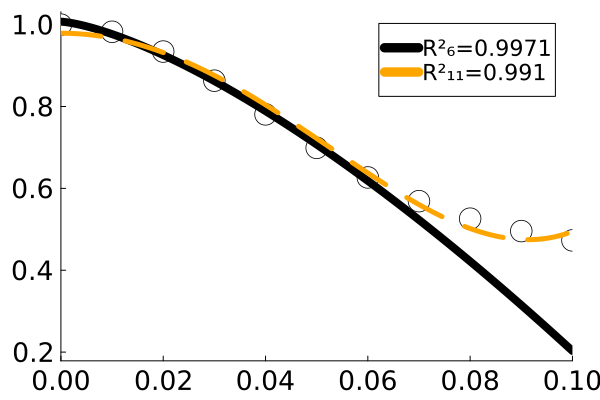}}
        
        \refstepcounter{subfigure}
        \put(0.14,0.3){
            \shortstack[l]{
                (\thesubfigure)\ H9 \\
                \hspace{1.7em}$\theta = 20^{\circ}$
            }
        }

        \put(0.39,0.75){Intra} 

        \put(0.46,-0.05){\small $t_z$ [eV]} 
    \end{picture}

    \label{fig:Intra B}
\end{subfigure}%
\begin{subfigure}{0.33\textwidth}
    \centering
    \setlength{\unitlength}{\linewidth}
    \begin{picture}(1,0.7)
        \put(0,0){\includegraphics[width=\unitlength]{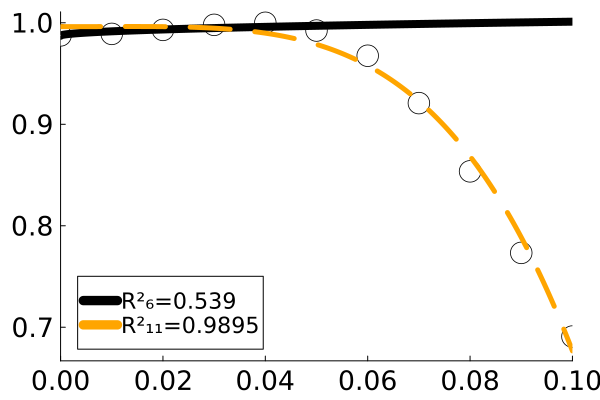}}
        
        \refstepcounter{subfigure}
        \put(0.14,0.3){
            \shortstack[l]{
                (\thesubfigure)\ H11 \\
                \hspace{1.7em}$\theta = 0^{\circ}$
            }
        }

        \put(0.46,-0.05){\small $t_z$ [eV]} 
    \end{picture}
    \label{fig:Intra C}
\end{subfigure}

\vspace{1.5cm}

\begin{subfigure}{0.33\textwidth}
    \centering
    \setlength{\unitlength}{\linewidth} 
    \begin{picture}(1,0.7)
        \put(0,0){\includegraphics[width=\unitlength]{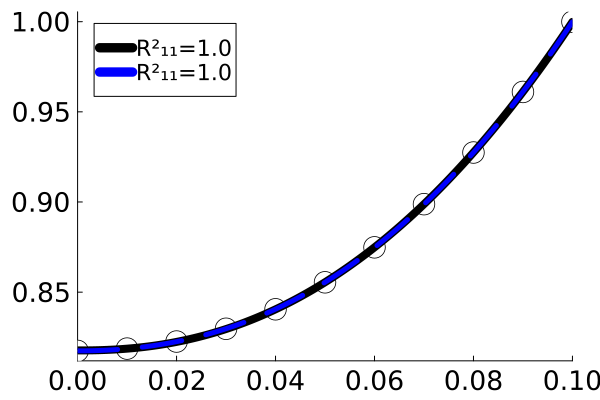}}
        
        \refstepcounter{subfigure}
        \put(0.16,0.3){
            \shortstack[l]{
                (\thesubfigure)\ H3 \\
                \hspace{1.7em}$\theta = 0^{\circ}$
            }
        }

        \put(-0.2,0.2){\rotatebox{90}{\footnotesize \shortstack{Normalized \\ Intensity \\ {[arb. units]} }}} 
    \end{picture}

    \label{fig:Inter D}
\end{subfigure}%
\begin{subfigure}{0.33\textwidth}
    \centering
    \setlength{\unitlength}{\linewidth}
    \begin{picture}(1,0.7)
        \put(0,0){\includegraphics[width=\unitlength]{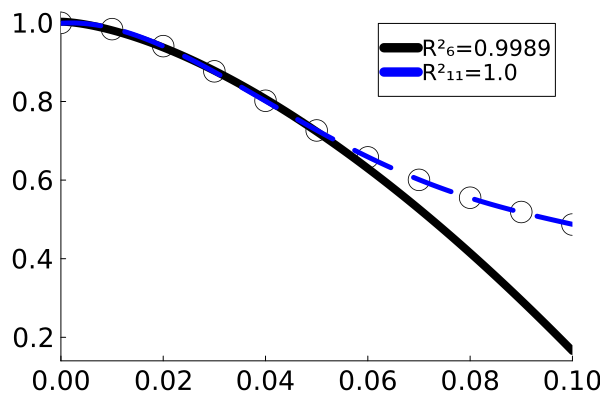}}
        
        \refstepcounter{subfigure}
        \put(0.13,0.3){
            \shortstack[l]{
                (\thesubfigure)\ H9 \\
                \hspace{1.7em}$\theta = 20^{\circ}$
            }
        }

        \put(0.39,0.75){Inter} 

        \put(0.46,-0.05){\small $t_z$ [eV]} 
    \end{picture}

    \label{fig:Inter E}
\end{subfigure}%
\begin{subfigure}{0.33\textwidth}
    \centering
    \setlength{\unitlength}{\linewidth}
    \begin{picture}(1,0.7)
        \put(0,0){\includegraphics[width=\unitlength]{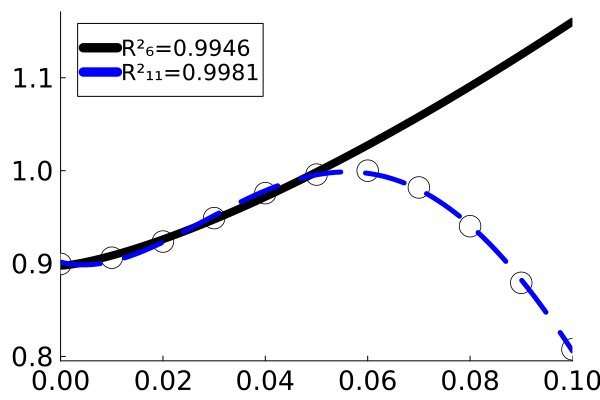}}
        
        \refstepcounter{subfigure}
        \put(0.12,0.35){
            \shortstack[l]{
                (\thesubfigure)\ H11 \\
                \hspace{1.7em}$\theta = 0^{\circ}$
            }
        }

        \put(0.46,-0.05){\small $t_z$ [eV]} 
    \end{picture}

    \label{fig:Inter F}
\end{subfigure}

\vspace{3mm}

\caption{HHG yield dependence on the interlayer hopping parameter from graphite. Black solid line represent generic power law fitted functions. Blue and yellow dashed lines are 4'th-order polynomial fitted lines, where the yellow lines in the intraband case include only even order in $t_z$. Notations of $R^2$ are similar to those employed in main-text figures}
\label{fig:Graphite Intra and Inter Numerical Fits}
\end{figure*}

Figure.~\ref{fig:Graphite Intra and Inter Numerical Fits} presents numerical fitting of the intra- and inter-band current yields in graphite vs the interlayer coupling strength, $t_z$. This evaluates the consistency of our results with theoretical predictions developed in the subsequent sections of this Supplementary Information (SI), which indicate that in the gapless graphene case ($\Delta=0)$, the intraband current should have some lower order terms in $t_z$ vanish, which changes the expected form of the HHG yields. As shown in Figures \ref{fig:Graphite Intra and Inter Numerical Fits}(\subref{fig:Intra A}–\subref{fig:Intra C}), the intraband data are fitted using a fourth-order polynomial of the form: $\mathcal{I}(t) \approx a_0(t) + a_2(t) \cdot t_z^2 + a_4(t) \cdot t_z^4$, avoiding odd-orders of $t_z$. This fitting procedure is employed to verify the agreement with the theoretical results derived from Eq.~(\ref{eq:Example Momentum Intra Elements}-\ref{eq:Momentum Intra Elements}) in the subsequent section.
This analysis further validates the results obtained via perturbation theory, and confirms the robustness of the modeled intraband dynamics. It further raises our level of confidence in the perturbation theory analysis.

\pagebreak

\section{Analytical derivation of perturbation theory}

\noindent We begin our analysis by obtaining perturbative expressions for the eigenstates of $H_{3D}$ utilized in the main text, where the eigenstates of the block-diagonal $H_{2D}$ (denoted here as $H_0$) are used as the starting point for perturbation (denoted here as $\ket{C^{(0)}_\pm}$ and $\ket{V^{(0)}_\pm}$). Note that in principle, we have the exact analytic form of the eigenstates of $H_{3D}$ including $t_z\neq0$, but their form is cumbersome and difficult to obtain insight from, motivating a perturbative treatment. The procedure follows by obtaining an approximation form for the eigenstates of $H_{3D}$ that includes the perturbative $t_z$ terms up to first order in perturbation theory. This is applied via a sum-of-states formula, summing over $\ket{C^{(0)}_\pm}$ and $\ket{V^{(0)}_\pm}$ with coefficients obtained including the perturbation. The end result for these states, up to first order in $t_z$, is given by:

\begin{equation}
\begin{gathered}
    \ket{C_{\pm}} = \frac{1}{\mathcal{N}} 
    \Bigl\{  \ket{C^{(0)}_\pm} \pm \frac{\beta_{14}}{2\sqrt{\Delta^2 + 4 | \beta_{12} |^2}} \ket{V^{(0)}_{\pm}}  
    \pm \frac{\Delta \beta_{14}}{2( \Delta^2 + 4 | \beta_{12} |^2 )} \ket{V^{(0)}_{\mp}} \Bigr\} \\[1ex]
    \ket{V_{\pm}} = \frac{1}{\mathcal{N}} 
    \Bigl\{  \ket{V^{(0)}_\pm} \pm \frac{\beta_{14}}{2\sqrt{\Delta^2 + 4 | \beta_{12} |^2}} \ket{C^{(0)}_{\pm}} 
    \mp \frac{\Delta \beta_{14}}{2( \Delta^2 + 4 | \beta_{12} |^2 )} \ket{C^{(0)}_{\mp}} \Bigr\}
\end{gathered}
\end{equation}
\noindent where $\mathcal{N}$ is a normalization factor that has the form $\sqrt{1+A \cdot \beta^2_{14}}$. Note that for simplicity, we do not include the normalization factor in this expansion, which has a largely weak dependence on $\beta_{14}$. Another important point is that all expressions above are \textit{k}-dependent (including the states themselves and several of the coefficient terms). For notational convenience, we have dropped the \textit{k}-dependence of all terms from this point on.

Next, we wish to evaluate a generic form for the time-dependent current, $\textbf{J}(t)$, and its functional behavior on $t_z$. In order to obtain this expression, we analyze the momentum matrix elements ($\textbf{p}_{nm}$) form on $t_z$, since they act as decisive factors in the current (see eq. (4) in main text). Notably, the full current arises as a summation of $\textbf{p}_{nm}$ terms in different \textit{k}-points. Hence, our hope is that each individual \textit{k}-point will inevitably lead to a similar functional dependence on $t_z$, such that the overall current dependence on $t_z$ could still be uncovered (despite the \textit{k}-summation). Note that we cannot in advance know if that will occur. We should also note that under this approximation we neglect additional dependencies of the occupations of bands on $t_z$, which we expect should have a higher-order form in $t_z$ since the dominant dynamics ours within the layered planes.  

We approximate the momentum matrix elements to first-order in $t_z$ by the form: $\textbf{p}_{CC}=\langle  C_{\pm}  | \hat{\textbf{{P}}} | C_{\pm}  \rangle $ (where $\textbf{P}=\partial_{\textbf{k}}H_{3D}$ is derived from the fully perturbed Hamiltonian), and similar forms are utilized for all other cross terms between valence and conduction bands. In this approximation the perturbation enters through the correction to the eigenstates, as well as in the Hamiltonian. As auxiliary quantities, we define $D$ and $C$ to simplify the matrix elements (this will become apparent in the equations below):

\begin{equation}
\begin{gathered}
    \mathbf{D} = \frac{ \beta_{12} \nabla_\mathbf{k} \beta^*_{12} + \beta^*_{12} \nabla_\mathbf{k} \beta_{12} }{\sqrt{\Delta^2 + 4 | \beta_{12} |^2}} \\
    \mathbf{\tilde{C}} = \frac{ \beta_{12} \nabla_\mathbf{k} \beta^*_{12} - \beta^*_{12} \nabla_\mathbf{k} \beta_{12} }{| \beta_{12} |}
\end{gathered}
\end{equation}

The $\mathbf{D}$ elements are real-valued, and $\mathbf{\tilde{C}}$ elements are purely imaginary. To further simplify the notations, we define $\mathbf{C} = \mathrm{Im} \big[ \mathbf{\tilde{C}}   \big]$ (note the tilde notation).

Moving forward, we being by evaluating the momentum matrix elements of the unperturbed system:
\begin{equation}
\begin{gathered}
    \langle C^{(0)}_{\pm} | \nabla_{\mathbf{k}} \hat{H}_0 | C^{(0)}_{\pm} \rangle
    = \nabla_{\mathbf{k}} \alpha + 2 \mathbf{D} \\[6pt]
    \langle V^{(0)}_{\pm} | \nabla_{\mathbf{k}} \hat{H}_0 | V^{(0)}_{\pm} \rangle 
    = \nabla_{\mathbf{k}} \alpha - 2 \mathbf{D} \\[6pt]
    \langle C^{(0)}_{\pm} | \nabla_{\mathbf{k}} \hat{H}_0 | V^{(0)}_{\mp} \rangle
    = \frac{\Delta}{| \beta_{12} |}  \mathbf{D} - i \mathbf{C}
\end{gathered}
\end{equation}

\noindent Note that terms not appearing above identically vanish. The perturbed system momentum expectation values involve the gradient of the full Hamiltonian $\hat{H}_{3D} = \hat{H}_0 + \hat{H}'$. We observe that the perturbation term $\hat{H}'(k_z)$ arises from interlayer hopping, proportional to $t_z$. In general notation, the perturbed system momentum matrix elements can be separated to the following parts:
\begin{equation}
\begin{aligned}
    \left\langle \phi  \left| \hat{p} \right| \psi   \right\rangle 
    =& \left\langle \phi \left| \nabla_{\mathbf{k}} \hat{H}_{3D}   \right| \psi  \right\rangle
    = \left\langle \phi  \left| \nabla_{\mathbf{k}} \hat{H}_0  \right| \psi  \right\rangle
    + \left\langle \phi  \left| \nabla_{\mathbf{k}} \hat{H'}   \right| \psi  \right\rangle
\end{aligned}
\end{equation}
Where $\phi$ and $\psi$ can be any of the conduction valence states, and the term in involving the perturbation $H'$ takes the form
\begin{equation}
\begin{aligned}
     \left\langle \phi  \left| \nabla_{\mathbf{k}} \hat{H'}   \right| \psi  \right\rangle
     = - c \cdot t_z \sin{\left( \frac{c}{2} \cdot k_z  \right)} \left\langle \phi  \left| 
     \begin{pmatrix}
        0 & 0 & 0 & 1 \\
        0 & 0 & 0 & 0 \\
        0 & 0 & 0 & 0 \\
        1 & 0 & 0 & 0 \\
    \end{pmatrix} .
    \right| \psi  \right\rangle
    \hat{\mathbf{z}}
\end{aligned}
\end{equation}
with $c$ the \textit{c}-axis lattice parameter of the unit cell.

Now comes a delicate and important point. Eventually, we will have to perform \textit{k}-space summation over the momentum expressions. These also inevitably include $k_z$ summation. We now argue that due to symmetry arguments, the terms involving $H'$ in the equation above will vanish upon $k_z$ summation. Indeed, note that since the basis states ( $\ket{C_{\pm}}$ and $\ket{V_{\pm}, \mathbf{k}}$ ) depend on $k_z$ via $\cos(ck_z/2)$, which is $k_z$-even, while expression above contain an odd function of $k_z$. Hence, overall the matrix element above behave as odd functions with respect to $k_z$. Hence, an integral of the following form vanishes:
\begin{equation}
    \int^{c/2}_{-c/2}  c^*_n ( \mathbf{k}, t) c_m ( \mathbf{k}, t) \left\langle u_n ( \mathbf{k} ) \left| \vec{\nabla}_{\mathbf{k}} \hat{H'}  \left( k_z  \right) \right| u_m ( \mathbf{k} ) \right\rangle \mathrm{dk_z} = 0
\end{equation}
\noindent Note that here we assumed that the $k_z$ occupation coefficients are also symmetric functions, which is applicable in our conditions due to the symmetric initial conditions and the Hamiltonian symmetrically occupying states with positive/negative $k_z$. Consequently, the momentum expectation values simplify to the contribution from the monolayer Hamiltonian $\hat{H}_0$ acting on the perturbed states alone: $\left\langle u_n ( \mathbf{k} ) \left| \vec{\nabla}_{\mathbf{k}} \hat{H}_0  \left( k_x,k_y  \right) \right| u_m ( \mathbf{k} ) \right\rangle$.

The next stage follows analytically evaluating these expressions. We obtain:

\begin{equation}
\begin{aligned}
    \left\langle C_{+} \left| \nabla_{\mathbf{k}} \hat{H}_0 \right| C_{+} \right\rangle
    =& \left\langle C_{+}^{0} \left| \nabla_{\mathbf{k}} \hat{H}_0  \right| C_{+}^{0} \right\rangle \\
    &
    \begin{aligned}
        + \frac{\beta_{14}}{2\sqrt{\Delta^{2} + 4|\beta_{12}|^{2}}}
        \Bigl| &
        \left\langle C_{+}^{0} \left| \nabla_{\mathbf{k}} \hat{H}_0  \right| V_{+}^{0} \right\rangle
        + \left\langle V_{+}^{0} \left| \nabla_{\mathbf{k}} \hat{H}_0  \right| C_{+}^{0} \right\rangle\\
        &+ \frac{\Delta}{\sqrt{\Delta^{2} + 4|\beta_{12}|^{2}}}
        \Bigl[
        \left\langle C_{+}^{0} \left| \nabla_{\mathbf{k}} \hat{H}_0  \right| V_{-}^{0} \right\rangle
        + \left\langle V_{-}^{0} \left| \nabla_{\mathbf{k}} \hat{H}_0  \right| C_{+}^{0} \right\rangle
        \Bigr]
        \Bigr\}
    \end{aligned}\\
    &
    \begin{aligned}
    + \frac{\beta_{14}^{2}}{4(\Delta^{2} + 4|\beta_{12}|^{2})}
        \Bigl\{ &
        \left\langle V_{+}^{0} \left| \nabla_{\mathbf{k}} \hat{H}_0  \right| V_{+}^{0} \right\rangle
        + \frac{\Delta}{\sqrt{\Delta^{2} + 4|\beta_{12}|^{2}}}
        \Bigl[
        \left\langle V_{+}^{0} \left| \nabla_{\mathbf{k}} \hat{H}_0  \right| V_{-}^{0} \right\rangle \\
        &+ \left\langle V_{-}^{0} \left| \nabla_{\mathbf{k}} \hat{H}_0  \right| V_{+}^{0} \right\rangle
        \Bigr]
        + \frac{\Delta^{2}}{\Delta^{2} + 4|\beta_{12}|^{2}}
        \left\langle V_{-}^{0} \left| \nabla_{\mathbf{k}} \hat{H}_0  \right| V_{-}^{0} \right\rangle
         \Bigr\}\\
    \end{aligned}\\
    =&  \left\langle C_{+}^{0} \left| \nabla_{\mathbf{k}} \hat{H}_0  \right| C_{+}^{0} \right\rangle +
    \frac{\Delta \beta_{14}}{2 \left( \Delta^{2} + 4|\beta_{12}|^{2} \right)}
    \Bigl[
    \left\langle C_{+}^{0} \left| \nabla_{\mathbf{k}} \hat{H}_0  \right| V_{-}^{0} \right\rangle
    + \left\langle V_{-}^{0} \left| \nabla_{\mathbf{k}} \hat{H}_0  \right| C_{+}^{0} \right\rangle
    \Bigr] \\
    & + \frac{\beta_{14}^{2}}{4 \left( \Delta^{2} + 4|\beta_{12}|^{2} \right)}
    \Bigl\{
    \left\langle V_{+}^{0} \left| \nabla_{\mathbf{k}} \hat{H}_0  \right| V_{+}^{0} \right\rangle
    + \frac{\Delta^{2}}{\Delta^{2} + 4|\beta_{12}|^{2}}
    \left\langle V_{-}^{0} \left| \nabla_{\mathbf{k}} \hat{H}_0  \right| V_{-}^{0} \right\rangle
    \Bigr\} \\
    =& \nabla_{\mathbf{k}} \alpha + 2 \mathbf{D}
    + \frac{\Delta \mathbf{D}}{\Delta^{2} + 4|\beta_{12}|^{2}} \cdot \beta_{14}
    + \frac{  \nabla_{\mathbf{k}} \alpha - 2 \mathbf{D}}{4\left(\Delta^{2} + 4\left|\beta_{12}\right|^{2}\right)}
    \Bigl[ 1
    + \frac{\Delta^{2}}{\Delta^{2}+4\left|\beta_{12}\right|^{2}} 
    \Bigr]
    \cdot \beta_{14}^{2}
    \label{eq:Example Momentum Intra Elements}
\end{aligned}
\end{equation}

Similarly, for the other cross terms we derive:

\begin{equation}
\begin{aligned}
    \left\langle C_{-} \left| \nabla_{\mathbf{k}} \hat{H}_0  \right| C_{-} \right\rangle  
    =& \nabla_{\mathbf{k}} \alpha + 2 \mathbf{D}
    - \frac{\Delta \mathbf{D}}{ \Delta^{2} + 4\left|\beta_{12}\right|^{2} } \cdot \beta_{14}
    + \frac{  \nabla_{\mathbf{k}} \alpha - 2 \mathbf{D}}{4\left(\Delta^{2} + 4\left|\beta_{12}\right|^{2}\right)}
    \Bigl[ 1
    + \frac{\Delta^{2}}{\Delta^{2}+4\left|\beta_{12}\right|^{2}} 
    \Bigr]
    \cdot \beta_{14}^{2}\\
    \left\langle V_{+} \left| \nabla_{\mathbf{k}} \hat{H}_0  \right| V_{+} \right\rangle 
    =& \nabla_{\mathbf{k}} \alpha - 2 \mathbf{D}
    - \frac{\Delta \mathbf{D}}{ \Delta^{2} + 4\left|\beta_{12}\right|^{2} } \cdot \beta_{14}
    + \frac{  \nabla_{\mathbf{k}} \alpha + 2 \mathbf{D}}{4\left(\Delta^{2} + 4\left|\beta_{12}\right|^{2}\right)}
    \Bigl[ 1
    + \frac{\Delta^{2}}{\Delta^{2}+4\left|\beta_{12}\right|^{2}} 
    \Bigr]
    \cdot \beta_{14}^{2}\\
    \left\langle V_{-} \left| \nabla_{\mathbf{k}} \hat{H}_0  \right| V_{-} \right\rangle 
    =& \nabla_{\mathbf{k}} \alpha - 2 \mathbf{D}
    + \frac{\Delta \mathbf{D}}{ \Delta^{2} + 4\left|\beta_{12}\right|^{2} } \cdot \beta_{14}
    + \frac{  \nabla_{\mathbf{k}} \alpha + 2 \mathbf{D}}{4\left(\Delta^{2} + 4\left|\beta_{12}\right|^{2}\right)}
    \Bigl[ 1
    + \frac{\Delta^{2}}{\Delta^{2}+4\left|\beta_{12}\right|^{2}} 
    \Bigr]
    \cdot \beta_{14}^{2}
    \label{eq:Momentum Intra Elements}
\end{aligned}
\end{equation}

\begin{equation}
\begin{gathered}
    \left\langle C_{+} \left| \nabla_{\mathbf{k}} \hat{H}_0  \right| C_{-} \right\rangle 
    = -i \cdot \frac{\mathbf{C}}{ \Delta^{2} + 4\left|\beta_{12}\right|^{2} } \cdot \beta_{14} 
    - \frac{ \Delta \left( \nabla_{\mathbf{k}} \alpha - 2 \mathbf{D} \right)}{2\left(\Delta^{2} + 4\left|\beta_{12}\right|^{2}\right) ^{3/2}} \cdot \beta_{14}^{2}\\
    \left\langle C_{+} \left| \nabla_{\mathbf{k}} \hat{H}_0  \right| V_{+} \right\rangle 
    = \frac{  \nabla_{\mathbf{k}} \alpha - 2 \mathbf{D}}{\sqrt{\Delta^{2} + 4\left|\beta_{12}\right|^{2}} } \cdot \beta_{14}\\
    \left\langle C_{+} \left| \nabla_{\mathbf{k}} \hat{H}_0  \right| V_{-} \right\rangle 
    = \frac{\Delta}{2\left|\beta_{12}\right|} \mathbf{D} - i \mathbf{C}
    + \frac{\Delta \nabla_{\mathbf{k}} \alpha}{ \Delta^{2} + 4\left|\beta_{12}\right|^{2} } \cdot \beta_{14} 
     - \frac{\beta_{14}^{2}}{4\left(\Delta^{2} + 4\left|\beta_{12}\right|^{2}\right)}
    \Bigl[ 1
    - \frac{\Delta^{2}}{\Delta^{2}+4\left|\beta_{12}\right|^{2}} 
    \Bigr]
    \Bigl[  \frac{\Delta}{2\left|\beta_{12}\right|} \mathbf{D} + i \mathbf{C} \Bigr]\\
    \left\langle C_{-} \left| \nabla_{\mathbf{k}} \hat{H}_0  \right| V_{+} \right\rangle
    = \frac{\Delta}{2\left|\beta_{12}\right|} \mathbf{D} - i \mathbf{C}
    - \frac{\Delta \nabla_{\mathbf{k}} \alpha}{ \Delta^{2} + 4\left|\beta_{12}\right|^{2} } \cdot \beta_{14} 
     - \frac{\beta_{14}^{2}}{4\left(\Delta^{2} + 4\left|\beta_{12}\right|^{2}\right)}
    \Bigl[ 1
    - \frac{\Delta^{2}}{\Delta^{2}+4\left|\beta_{12}\right|^{2}} 
    \Bigr]
    \Bigl[  \frac{\Delta}{2\left|\beta_{12}\right|} \mathbf{D} + i \mathbf{C} \Bigr]\\
    \left\langle C_{-} \left| \nabla_{\mathbf{k}} \hat{H}_0  \right| V_{-} \right\rangle
    = - \frac{  \nabla_{\mathbf{k}} \alpha - 2 \mathbf{D}}{\sqrt{\Delta^{2} + 4\left|\beta_{12}\right|^{2}} } \cdot \beta_{14}\\ 
    \left\langle V_{+} \left| \nabla_{\mathbf{k}} \hat{H}_0  \right| V_{-} \right\rangle 
    = i \cdot \frac{\mathbf{C}}{ \Delta^{2} + 4\left|\beta_{12}\right|^{2} } \cdot \beta_{14} 
    - \frac{ \Delta \left( \nabla_{\mathbf{k}} \alpha - 2 \mathbf{D} \right)}{2\left(\Delta^{2} + 4\left|\beta_{12}\right|^{2}\right) ^{3/2}} \cdot \beta_{14}^{2}\\
\end{gathered}
\end{equation}

\noindent From these, the most important result that arises, is that the momentum matrix elements at each individual \textit{k}-point should follow a generic second order polynomial in $\beta_{14}$, i.e. in the $t_z$ perturbation. The coefficients for these terms are \textit{k}-dependent, but the total $t_z$ dependence is general, and should survive BZ integration. Hence, the general form for the time-dependent current should also follow a second-order polynomial, where the coefficients forms are complex and difficult to derive, but should be guaranteed from this order of perturbation theory. Note that coefficients can also be negative here, such that some harmonic yields can increase with $t_z$, while others decrease.

We note some finer details here, which have to do with vanishing elements in the equations above for the gapless case ($\Delta$=0). Indeed, in the case of graphite the intraband momentum matrix elements coefficients of the $t_z$-linear term vanish. This means that specifically in graphite, the intraband current should follow a reduced polynomial of pure quadratic form (without curvature changes as observed in the main text in the general case). We indeed validate all of these expressions by comparison to numerical results in the main text, and the purely quadratic form is validated in Fig. S5. 

Another noteworthy point is that in the gapless case of graphite ($\Delta$=0), the sates diverge at $K/K'$, which implies that we should employ a degenerate perturbation theory. We have ignored this in the above derivation, also since in numerical simulations we soften this divergence.

\bibliographystyle{apsrev4-2}
\bibliography{references}